\documentclass[11pt]{article}

\usepackage[T1]{fontenc}
\usepackage[cp1251]{inputenc}
\usepackage[russian,english]{babel}
\usepackage[centertags]{amsmath}
\usepackage{amsfonts}
\usepackage{amssymb}
\usepackage{axodraw4j}
\usepackage[dvips]{graphicx}
\usepackage{multirow}
\usepackage[numbers,sort&compress]{natbib}
\usepackage[hypertex]{hyperref}

\usepackage{paperinitial}

\paperinitialization{15mm}{15mm}{20mm}{10mm}{2pt}{10pt}

\DeclareMathOperator{\re}{Re}
\DeclareMathOperator{\im}{Im}
\DeclareMathOperator{\reg}{reg}
\DeclareMathOperator{\Tr}{Tr}
\DeclareMathOperator{\Texp}{Texp}

\DeclareMathOperator{\sgn}{sgn}
\DeclareMathOperator{\arcch}{arcch}
\DeclareMathOperator{\tnh}{th}
\DeclareMathOperator{\Li}{Li}
\newcommand{\lan}{\langle}
\newcommand{\ran}{\rangle}

\newcommand{\bnabla}{\bar{\nabla}}
\newcommand{\udg}[3][0.3em]{\ensuremath{\vcenter{\hbox{\hspace{#1}\scalebox{#2}{\usebox{#3}}}}}}

\newcommand{\e}{\varepsilon}
\newcommand{\vf}{\varphi}

\newcommand{\s}{\sigma}
\newcommand{\bs}{\bar{\sigma}}
\newcommand{\Si}{\Sigma}
\newcommand{\al}{\alpha}
\newcommand{\be}{\beta}
\newcommand{\ga}{\gamma}
\newcommand{\Ga}{\Gamma}
\newcommand{\de}{\delta}
\newcommand{\De}{\Delta}
\newcommand{\ka}{\varkappa}
\newcommand{\la}{\lambda}

\newcommand{\ups}{\upsilon}

\newcommand{\spx}{\mathbf{x}}
\newcommand{\spy}{\mathbf{y}}

\newcommand{\N}{\mathbb{N}}

\begin{document}
\setlength{\unitlength}{1pt}
\newcounter{local}
\allowdisplaybreaks[4]
\selectlanguage{russian}
\selectlanguage{english}

\newsavebox{\diagai}
\savebox{\diagai}{%
\begin{picture}(50,30)
    \DashLine(2,2)(25,15){2} \Vertex(2,2){2}
    \DashLine(2,28)(25,15){2} \Vertex(2,28){2}
    \Line(48,2)(25,15) \Vertex(48,2){2}
    \Line(48,28)(25,15) \Vertex(48,28){2}
    \Vertex(25,15){2}
\end{picture}}
\newsavebox{\diagaii}
\savebox{\diagaii}{%
\begin{picture}(50,30)
    \DashLine(2,2)(20,15){2} \Vertex(2,2){2}
    \DashLine(2,28)(20,15){2} \Vertex(2,28){2}
    \BCirc(35,15){15}
    \Vertex(20,15){2}
\end{picture}}
\newsavebox{\diagaiii}
\savebox{\diagaiii}{%
\begin{picture}(50,30)
    \Line(2,2)(20,15) \Vertex(2,2){2}
    \Line(2,28)(20,15) \Vertex(2,28){2}
    \DashCArc(35,15)(15,0,360){2}
    \Vertex(20,15){2}
\end{picture}}
\newsavebox{\diagaiv}
\savebox{\diagaiv}{%
\begin{picture}(50,30)
    \DashLine(2,2)(20,15){2} \Vertex(2,2){2}
    \Line(2,28)(20,15) \Vertex(2,28){2}
    \DashCArc(35,15)(15,0,180){2}
    \CArc(35,15)(15,180,360)
    \Vertex(20,15){2}
\end{picture}}
\newsavebox{\diagav}
\savebox{\diagav}{%
\begin{picture}(50,30)
    \DashCArc(13,15)(12,0,360){2}
    \CArc(37,15)(12,0,360)
    \Vertex(25,15){2}
\end{picture}}
\newsavebox{\diagavi}
\savebox{\diagavi}{%
\begin{picture}(50,30)
    \DashCArc(13,15)(12,0,180){2}
    \CArc(13,15)(12,180,360)
    \DashCArc(37,15)(12,180,360){2}
    \CArc(37,15)(12,0,180)
    \Vertex(25,15){2}
\end{picture}}
\newsavebox{\diagbi}
\savebox{\diagbi}{%
\begin{picture}(50,30)
    \DashLine(2,15)(25,15){2} \Vertex(2,15){2}
    \Line(48,2)(25,15) \Vertex(48,2){2}
    \Line(48,28)(25,15) \Vertex(48,28){2}
    \Vertex(25,15){2}
\end{picture}}
\newsavebox{\diagbii}
\savebox{\diagbii}{%
\begin{picture}(50,30)
    \DashLine(2,15)(20,15){2} \Vertex(2,15){2}
    \BCirc(35,15){15}
    \Vertex(20,15){2}
\end{picture}}
\newsavebox{\diagbiii}
\savebox{\diagbiii}{%
\begin{picture}(50,30)
    \Line(2,15)(20,15) \Vertex(2,15){2}
    \DashCArc(35,15)(15,0,180){2}
    \CArc(35,15)(15,180,360)
    \Vertex(20,15){2}
\end{picture}}
\newsavebox{\diagc}
\savebox{\diagc}{%
\begin{picture}(10,10)
    \GCirc(5,5){4}{0.5}
\end{picture}}
\newsavebox{\diagViip}
\savebox{\diagViip}{%
\begin{picture}(50,30)
    \DashLine(2,2)(25,15){2}
    \DashLine(2,28)(25,15){2}
    \Line(48,2)(25,15)
    \Line(48,28)(25,15)
    \Vertex(45,15){1}
    \Vertex(44,10){1}
    \Vertex(44,20){1}
    \Vertex(25,15){2}
\end{picture}}
\newsavebox{\diagViipw}
\savebox{\diagViipw}{%
\begin{picture}(50,30)
    \DashLine(2,2)(25,15){2}
    \DashLine(2,28)(25,15){2}
    \Line(48,2)(25,15)
    \Line(48,28)(25,15)
    \Vertex(25,15){2}
\end{picture}}
\newsavebox{\diagVip}
\savebox{\diagVip}{%
\begin{picture}(50,30)
    \DashLine(2,15)(25,15){2}
    \Line(48,2)(25,15)
    \Line(48,28)(25,15)
    \Vertex(45,15){1}
    \Vertex(44,10){1}
    \Vertex(44,20){1}
    \Vertex(25,15){2}
\end{picture}}
\newsavebox{\diagVipw}
\savebox{\diagVipw}{%
\begin{picture}(50,30)
    \DashLine(2,15)(25,15){2}
    \Line(48,2)(25,15)
    \Line(48,28)(25,15)
    \Vertex(25,15){2}
\end{picture}}
\newsavebox{\diagVx}
\savebox{\diagVx}{%
\begin{picture}(50,30)
    \Line(2,15)(25,15)
    \Line(48,2)(25,15)
    \Line(48,28)(25,15)
    \Vertex(45,15){1}
    \Vertex(44,10){1}
    \Vertex(44,20){1}
    \Vertex(25,15){2}
\end{picture}}
\newsavebox{\diagVo}
\savebox{\diagVo}{%
\begin{picture}(50,30)
    \Line(2,15)(25,15)
    \Line(48,2)(25,15)
    \Line(48,28)(25,15)
    \Vertex(45,15){1}
    \Vertex(44,10){1}
    \Vertex(44,20){1}
    \GCirc(25,15){4}{0.5}
\end{picture}}


\title{{\Large \textbf{Non-perturbative corrections to the one-loop free energy induced by a massive scalar field on a stationary slowly varying in space gravitational background}}}

\date{}

\author{I.~S. Kalinichenko\thanks{E-mail: \texttt{theo@sibmail.com}}\;  and P.~O. Kazinski\thanks{E-mail: \texttt{kpo@phys.tsu.ru}}\\[0.5em]
{\normalsize Physics Faculty, Tomsk State University, Tomsk 634050, Russia}}

\maketitle

\begin{abstract}

The explicit expressions for the one-loop non-perturbative corrections to the gravitational effective action induced by a scalar field on a stationary gravitational background are obtained both at zero and finite temperatures. The perturbative and non-perturbative contributions to the one-loop effective action are explicitly separated. It is proved that, after a suitable renormalization, the perturbative part of the effective action at zero temperature can be expressed in a covariant form solely in terms of the metric and its derivatives. This part coincides with the known large mass expansion of the one-loop effective action. The non-perturbative part of the renormalized one-loop effective action at zero temperature is proved to depend explicitly on the Killing vector defining the vacuum state of quantum fields. This part cannot be expressed in a covariant way through the metric and its derivatives alone. The implications of this result for the structure and symmetries of the effective action for gravity are discussed.

\end{abstract}

\section{Introduction}

A construction of self-consistent quantum theory of gravity remains elusive due to several conceptual and technical problems. The main technical problem is, of course, a high non-linearity of classical theory of gravity (general relativity) and, as a result, nonrenormalizability of its quantum counterpart. The major conceptual issue is related to the problem of time in quantum gravity (see for a review \cite{Isham} and also \cite{DeWQG,GriMaMos,MamMostStar,Kuchar}) or, put another way, it regards the problem of definition of a natural vacuum state and a representation of the algebra of observables. On the other hand, there is a widespread belief resting on perturbative calculations on a flat background that the second problem does not actually exist. According to this point of view, quantum gravity is a mere another one effective quantum field theory similar to the Fermi theory of weak interactions. The aim of the present paper is to show that such a viewpoint is somewhat naive as long as it does not take into account non-perturbative corrections to the effective action.

One of the most powerful methods to find the non-perturbative corrections to the effective action (the generating functional of the one-particle irreducible Green functions) is the celebrated background field method (see, e.g., \cite{DeWGAQFT,BuchOdinShap}) with its renormalization group improvements (see, e.g., \cite{BuchOdinShap}). It is, perhaps, the only one for many particle quantum systems in dimension $D\geq3$ without extra symmetries. We shall use this method to obtain the one-loop non-perturbative (in the gravitational constant) contributions to the effective action of gravity from a massive scalar field at zero and finite temperatures (for the renormalization group improvement of the perturbative contributions see, e.g., \cite{EliOdiRG}). We shall derive the explicit expressions for these corrections in the case of a stationary slowly varying in space gravitational background (the stationary infrared limit). As far as we know, this problem has not been solved yet for such a general formulation.

The fact that the effective action has to possess the non-perturbative terms of the form that we shall derive in this paper was repeatedly noticed in the literature (see, e.g., \cite{GriMaMos,ZelnProc,GavrGit}). However, neither the explicit form of these corrections for a sufficiently wide class of background metrics nor even their expression in terms of some integrals were given. Whereas it is that statement of the problem which is necessary to solve for a construction of the effective action functional. In the present paper, we restrict our consideration to the contributions from a massive scalar field with a mass $m$ on a stationary slowly varying in space gravitational background with the standard vacuum state for quantum fields on stationary backgrounds (see, e.g., \cite{DeWGAQFT}, Sections 17, 18). The treatment of the contributions of quantum fields with higher spins ($1/2$, $1$, and $2$) is analogous but bulkier. We shall explicitly separate the perturbative and non-perturbative corrections to the effective action. The perturbative corrections turn out to be expandable in an asymptotic Laurent series in $m^{-2}$ with a finite principal part, while the non-perturbative are not. This, obviously, implies that the non-perturbative contributions cannot be extracted from the large mass expansion.

We shall show that, for the odd-dimensional spacetimes, the coefficients of the asymptotic series in $m^{-2}$ of the finite part of the perturbative corrections are expressed in terms of covariant combinations of the metric $g_{\mu\nu}$ and its derivatives only. These coefficients do not depend on any external structure. As for the even-dimensional spacetimes, the coefficients of this series (for the finite part) at the negative powers of $m^2$ and the coefficient at the logarithmic divergence can be also written in a covariant form in terms of the metric alone, while the terms at the nonnegative powers of $m^2$ are not \cite{Page,BrOtPa,FrZel,AndHisSam,gmse,KalKaz}. They depend explicitly on the Killing vector field of the metric. This vector field defines the vacuum state and the unique representation of the algebra of observables, according to the Gelfand-Naimark-Segal (GNS) construction (see, e.g., \cite{Emch}). If we cancel out these ``noncovariant'' terms by the counterterms, as discussed in \cite{DeWQFTcspt,KalKaz}, then the perturbative corrections to the effective action become covariant and expressible through the metric only.

As far as the non-perturbative corrections are concerned, we shall see that they explicitly depend on the Killing vector field and cannot be expressed in terms of the metric. Though they are covariant combinations of the metric, the Killing vector, and their derivatives. This result is quite expectable since the generating functional of the Green functions must depend on the structures (the Killing vector, in our case) that distinguish the unique vacuum state with respect to which the Green functions are defined. The non-perturbative corrections prove to be very small for the gravitational fields occurring in nature out of the ergosphere. Nevertheless, the very existence of such corrections and the fact that they are not expressible via the metric alone are of paramount importance for our understanding of the structure and symmetries of the effective action for quantum gravity.

The main technical tool, that we shall use in addition to the background field method to derive the explicit expression for the one-loop correction to the effective action, is the relation mentioned in \cite{olopq} (see also \cite{KalKaz}) between the free energy, or the $\Omega$-potential, at high temperatures (the reciprocal temperature $\be\rightarrow0$) and the effective action at zero temperature ($\be\rightarrow\infty$). It turns out that, in order to find the one-loop correction to the effective action at zero temperature (the vacuum contribution), it is sufficient to know the divergent and finite parts of the high-temperature expansion of the one-loop correction to the free energy without the vacuum contribution. Therefore, at the beginning, we shall provide a more rigorous derivation of the general formula \cite{KalKaz} for the high-temperature expansion of the one-loop contribution to the $\Omega$-potential with the non-perturbative corrections included. This derivation is given in Sec. \ref{HTE}. It is found that the formula derived in \cite{KalKaz} keeps its form provided the exponentially suppressed contributions at $\be\rightarrow0$ are discarded from the high-temperature expansion. We shall have to dwell in Sec. \ref{HTE} on some mathematical aspects of the zeta functions of hyperbolic type operators on stationary (non-ultrastatic) backgrounds since a mathematical theory of this type of zeta functions is almost absent in the literature (see, however, \cite{Fursaev1,Fursaev2}).

Then, in Sec. \ref{DescForm}, we shall prove the so-called descent formulas \cite{Fursaev2} that relate the coefficients of the heat kernel expansion for the Laplace type operator (see for a review \cite{VasilHeatKer}) in the space dimension $(d+1)$, but with the coefficients at derivatives independent of some coordinate $x^0$, with the coefficients of the heat kernel expansion for the same Laplace type operator in the space dimension $d$. The dependence on $x^0$ of the latter operator is separated by the standard means. These formulas will allow us to prove in this section that the coefficients at the negative powers of $m^2$ in the perturbative finite part of the high-temperature expansion are independent of the Killing vector and coincide with the standard large mass expansion of the effective action (see, e.g., \cite{BirDav}). For the odd-dimensional spacetime, the coefficients at the nonnegative powers of $m^2$ in the perturbative finite part are independent of the Killing vector as well. Besides, using the descent formulas, it is easy to show that the coefficient at the logarithmic divergence of the high-temperature expansion, which is the conformal anomaly and the logarithmic part of the energy-time anomaly \cite{KalKaz}, is expressed solely in terms of the metric and coincides with the standard expression for the conformal anomaly \cite{DowKen,NakFuk,DowSch,DowSch1,Fursaev1,Fursaev2,KalKaz}.

Section \ref{NonpertContrScalField} is the heart of the paper, where all the general results of the preceding sections are collected together in order to obtain the high-temperature expansion of the free energy of a scalar field and describe its properties. Also, in this section, we heavily rely on the results of the papers \cite{KalKaz} and \cite{prop}. In fact, it is the combination of the results of these papers that made it possible to derive a complete high-temperature expansion with the non-perturbative contributions. In Sec. \ref{PertTheor_sec}, we shall develop a perturbative procedure for the heat kernel that allows us to deduce systematically non-perturbative corrections to the effective action. Using the procedure elaborated, we shall find the first correction to the leading (Gaussian) contribution to the heat kernel obtained in \cite{prop} (for the Euclidean case see \cite{HuOCon,Avramid}). The counting scheme, which sorts an infinite set of Feynman diagrams for the heat kernel and distinguishes the most relevant ones, will be also presented there. Section \ref{NonPertCorr} is devoted to evaluation of the explicit expressions for non-perturbative contributions to the high-temperature expansion. From the technical point of view, this problem is reduced to a calculation of certain double integrals in the complex planes: one integral is over the proper-time and another one is over the energy of a mode. This issue is solved for both massive and massless cases in the weak field limit.

In conclusion we shall outline some implications of the results obtained and the possible further directions of research. In Appendix \ref{Zeta0}, we prove a certain property for a variant of the zeta function of the Laplace type operator coming from the hyperbolic type operator Fourier transformed over the time variable. This property is used in Sec. \ref{HTE}. In Appendix \ref{Pert_Theor}, the general formulas of perturbation theory are gathered and the first correction to the leading contribution to the heat kernel is explicitly calculated.

Since, in the present text, we shall try to reduce the duplication of formulas from \cite{KalKaz} and \cite{prop} to a minimum, the reader is strongly encouraged to have these papers close at hand. In the course of the discussion, several misprints made in \cite{KalKaz} and \cite{prop} will be also corrected. Besides, in Sec. \ref{HTE}, we shall extensively employ the analytic regularization technique for singular integrals \cite{GSh}. The acquaintance with Sec. 3 of \cite{GSh} will be required. The knowledge of the notion of heat kernel and the heat kernel expansion technique \cite{VasilHeatKer} is also desired in Sections \ref{HTE} and \ref{DescForm}. In order to realize better the main idea of the procedure developed in Sec. \ref{NonpertContrScalField} and the structure of density of states considered in Sections \ref{HTE} and \ref{NonpertContrScalField}, it is recommended to know the results of \cite{BalBloch,BalBlochSch} and especially \cite{BalBlochSch}. As for the construction of quantum field theory (QFT) on a curved stationary background, we shall assume that the reader is familiar with the paper \cite{DeWQFTcspt} and Sections 17, 18 of \cite{DeWGAQFT}. Of course, the other references we provide in the paper are also advised for reading.

We shall use the conventions adopted in \cite{DeWGAQFT}
\begin{equation}
    R^\al_{\ \be\mu\nu}=\partial_{[\mu}\Ga^\al_{\nu]\beta}+\Ga^\al_{[\mu\ga}\Ga^\ga_{\nu]\beta},\qquad R_{\mu\nu}=R^\al_{\ \mu\al\nu},\qquad R=R^\mu_\mu,
\end{equation}
for the curvatures and the other structures appearing in the heat kernel expansion. The square and round brackets at a pair of indices denote antisymmetrization and symmetrization without $1/2$, respectively. The Greek indices are raised and lowered by the metric $g_{\mu\nu}$ which has the signature $-2$. Also we assume that the metric possesses the timelike Killing vector $\xi^\mu$:
\begin{equation}
    \mathcal{L}_\xi g_{\mu\nu}=0,\qquad\xi^2=g_{\mu\nu}\xi^\mu\xi^\nu>0,
\end{equation}
that allows us to make the decomposition (\cite{LandLifshCTF}, Sec. 84; \cite{Zelman,Vladimir,Mitskev,FrZel})
\begin{equation}\label{metric_decomp}
    ds^2=g_{\mu\nu}dx^\mu dx^\nu=:\xi^2(g_\mu dx^\mu)^2-\bar{g}_{\mu\nu}dx^\mu dx^\nu,
\end{equation}
where $g_\mu=\xi_\mu/\xi^2$ is a one-form dual to the Killing vector (the Tolman temperature one-form). Notice that this decomposition is not the Arnowitt-Deser-Misner (ADM) one, but in some sense dual to it. The decomposition \eqref{metric_decomp} is constructed by the use of the vector field, while the ADM one is associated with the system of hypersurfaces or, equivalently, with the integrable one-from. In case of a static spacetime, these decompositions coincide so long as the family of hypersurfaces is identified with the integral manifolds of the one-form $g_\mu$. In the system of coordinates, where $\xi^\mu=(1,0,0,0)$, we have the relations
\begin{equation}\label{gbar_rels}
\begin{gathered}
    \bar{g}_{ik}g^{kj}=-\de_i^j,\qquad \xi^2\det(-\bar{g}_{ij})=g,\qquad g^{00}+\bar{g}_{ij}g^{0i}g^{0j}=(g_{00})^{-1},\qquad g_i=\bar{g}_{ij}g^{j0},\\
    -\bar{g}_{\mu\nu}=\begin{bmatrix}
                       0 & 0 \\
                       0 & g_{ij}-\frac{g_{i0}g_{j0}}{g_{00}} \\
                     \end{bmatrix},\qquad
    -\bar{g}^{\mu\nu}=g^{\mu\nu}-\xi^2g^\mu g^\nu=\begin{bmatrix}
                       g^{00}-(g_{00})^{-1} & g^{0j} \\
                       g^{i0} & g^{ij} \\
                     \end{bmatrix}.
\end{gathered}
\end{equation}
We have changed the sign of the metric $\bar{g}_{ij}$ in comparison with \cite{KalKaz}. The Latin indices corresponding to the space are raised and lowered by the positive-definite metric $\bar{g}_{ij}$. The curvatures associated with this metric will be distinguished by the overbars, e.g., $\bar{R}$. Note that we consider a general stationary spacetime, i.e., the Tolman temperature one-form is supposed to be non-integrable (\cite{Wald}, App. C; \cite{LandLifshCTF}, Sec. 88; \cite{Zelman,FrZel,Vladimir,Mitskev}),
\begin{equation}\label{fmunu}
    f_{\mu\nu}:=\partial_{[\mu}g_{\nu]}\neq0,
\end{equation}
in general. The system of units is chosen such that $c=\hbar=1$.

\section{High-temperature expansion}\label{HTE}

\subsection{General formulas}

Let $H(\omega)$ be a Fourier transform of a kernel of the wave operator on a stationary background (see, for instance, \eqref{KG_eq}). Suppose the corresponding operator is of a Laplacian type, depends analytically on $\omega$, has the spectrum bounded from above at fixed $\omega$ in the Hilbert space of square-integrable functions, and there is no accumulation points in the discrete spectrum. Consider the operator (cf. \cite{Gilkey1}, Sec. 1.10)
\begin{equation}\label{zeta+_nu}
    H_+^{-\nu}(\omega):=\int_C\frac{d\tau\tau^{\nu-1}}{(e^{2\pi i\nu}-1)\Ga(\nu)}e^{-\tau H(\omega)},
\end{equation}
where the contour $C$ runs along the imaginary axis from top to bottom and encircles the origin from the left. For the special case, $\nu=0$, $H_+^0(\omega)=\theta(H(\omega))$ is the projector to the subspace of the total Hilbert space. This subspace is spanned on the eigenvectors of $H(\omega)$ corresponding to the positive eigenvalues.

If the system is placed in a sufficiently large ``box'' in space so that the spectrum of $H(\omega)$ is discrete, then $H^{-\nu}_+(\omega)$ is trace-class for any $\nu\in\mathbb{C}$. Consequently, there exists the entire function of $\nu$,
\begin{equation}
    \zeta_+(\nu,\omega)=\Tr H_+^{-\nu}(\omega).
\end{equation}
The investigation of passing to an infinitely large ``box'' can be found, for example, in \cite{Dew_scat}. If the operator $H(\omega)$ also possesses a continuous spectrum, then
\begin{equation}\label{zeta+}
    \zeta_+(\nu,\omega)=\sum_k\theta(\e_k(\omega))\e_k^{-\nu}(\omega)+\int_0^{\e_c(\omega)}d\e n(\e,\omega)\e^{-\nu},\qquad\re\nu<1,
\end{equation}
where $\e_k(\omega)$ are the eigenvalues of $H(\omega)$ and $n(\e,\omega)d\e$ is the number of states of the continuous spectrum in the interval $[\e,\e+d\e]$ (usually, it is proportional to the volume of a system). The quantity $\e_c(\omega)$ specifies the boundary of the continuous spectrum. The generalization of formula \eqref{zeta+} and the following ones to the case of several zones is quite obvious. As an example, we present the density of states for a free scalar particle in a $(d+1)$ dimensional Minkowski space
\begin{equation}\label{dens_stat}
    n(\e,\omega)=V_d\frac{\Ga(D/2)}{2\pi^{D/2}\Ga(d)}\theta(\omega^2-m^2-\e)(\omega^2-m^2-\e)^{d/2-1},\qquad\e_c(\omega)=\omega^2-m^2,
\end{equation}
where $D:=d+1$. The quasiclassical approximation for the Laplacian type operators we study gives for $\zeta_+(\nu,\omega)$ at large $\omega$ (see \cite{olopq,KalKaz,Fursaev1,Fursaev2,DowKen,DowSch,DowSch1,NakFuk} and below),
\begin{equation}\label{asympt_UV}
    \zeta_+(\nu,\omega)\sim \frac{\Ga(1-\nu)|\omega|^{d-2\nu}}{\Ga(d/2-\nu+1)},\qquad \omega\rightarrow\pm\infty.
\end{equation}
In fact, to derive this asymptotics, one needs to substitute \eqref{dens_stat} to \eqref{zeta+} and evaluate the integral. For those $H(\omega)$, which depend nonquadratically on $\omega$,  for example, for the wave operator in a dispersive media, it is also reasonable to expect the asymptotic behavior \eqref{asympt_UV} so long as a media becomes transparent in the ultraviolet regime.

Consider, in general, that $\e_c'(\omega)>0$ for $\omega>0$ and there exists $\omega_c>0$ such that $\e_c(\omega_c)=0$. Also suppose that
\begin{enumerate}
  \item $n(\e,\omega)$ is a smooth function of $\omega$ and $\e$ for $\omega>0$, $\e\in(0,\e_c(\omega))$;
  \item The integral $\int_a^{\e_c(\omega)}d\e n(\e,\omega)$ converges for $a>0$ and $\omega>0$;
  \item $\partial^k_\e n(\e,\omega)$ is finite for $\e=0$ and $\omega>\omega_c$.
\end{enumerate}
The first condition is rather technical and may be weaken. It makes it possible not to take care of the existence of derivatives. The second condition says that the integral over $\e$ in \eqref{zeta+} converges on the upper integration limit. The last condition is necessary for the Gelfand-Shilov analytical regularization (in its standard form) \cite{GSh} of the integral over $\e$ in the neighborhood of $\e=0$. If these conditions are met, the function \eqref{zeta+} can be analytically continued to the region $\re\nu\geq1$. So, the integral in \eqref{zeta+} is understood as analytically regularized \cite{GSh} in the case when zero belongs to the continuous spectrum of $H(\omega)$. As follows from the general procedure \cite{GSh}, the function \eqref{zeta+} has simple poles at $\nu\in\mathbb{N}$ under the above restrictions on the density of states $n(\e,\omega)$.

Using the function $\zeta_+(\nu,\omega)$, it is easy to obtain the expression for the one-loop correction to the $\Omega$-potential. Assuming the condition of the vacuum stability is fulfilled for the eigenvalues $\e_k(\omega)$ of $H(\omega)$ at the points, where $\e_k(\omega)=0$, (see, e.g., \cite{Migdal})
\begin{equation}\label{vac_stab}
    \e'_k(\omega)>0\; \text{for}\; \omega>0 \quad\text{and}\quad \e'_k(\omega)<0\; \text{for}\; \omega<0,
\end{equation}
we have (see for details, e.g., \cite{KalKaz})
\begin{equation}\label{omega_pot_1}
    \mp\beta\Omega=\int_0^\infty d\omega\big[\partial_\omega\zeta_+(0,\omega)\ln(1\pm e^{-\be(\omega-\mu)})+\partial_\omega\zeta_+(0,-\omega)\ln(1\pm e^{-\be(\omega+\mu)})\big].
\end{equation}
The contributions from both particles and antiparticles are taken into account in this expression.

If the vacuum is stable \eqref{vac_stab} then $\zeta_+(\nu,0)=0$, i.e., $H(0)$ has no positive eigenvalues. In Appendix \ref{Zeta0}, we shall prove this statement deforming the wave operator of free fields, which possesses this property, into the wave operator with interaction $H(\omega)$. The proof given in Appendix \ref{Zeta0} requires, additionally, the condition of a ``smooth deformability'' of the eigenvalues of a family of operators under consideration. In what follows, we put $\zeta_+(\nu,0)=0$.

Integrating by parts in \eqref{omega_pot_1} and taking into account that $\zeta_+(\nu,0)=0$, we get
\begin{equation}\label{omeg_pot}
    \Omega=-\int_0^\infty d\omega\Big[\frac{\zeta_+(0,\omega)}{e^{\be(\omega-\mu)}\pm1}+\frac{\zeta_+(0,-\omega)}{e^{\be(\omega+\mu)}\pm1}\Big].
\end{equation}
It is convenient to introduce the function
\begin{equation}\label{Inu}
    I_\nu(\mu):=\int_0^\infty \frac{d\omega\zeta_+(\nu,\omega)}{e^{\be(\omega-\mu)}\pm1},\qquad\re\nu<1.
\end{equation}
On substituting \eqref{zeta+} into \eqref{Inu}, the integral $I_\nu(\mu)$ falls into two pieces corresponding to two summands in \eqref{zeta+}. Suppose $\mu$ does not lie on the real positive semiaxis $\omega$, where $\e_k(\omega)=0$ for some $k$. Then, taking into account \eqref{vac_stab}, we have for the first contribution
\begin{equation}
    \sum_k\int_0^\infty\frac{d\e\omega'_k(\e)\e^{-\nu}}{e^{\be(\omega_k(\e)-\mu)}\pm1},
\end{equation}
where $\omega_k(\e)$ is the inverse function to $\e_k(\omega)$. According to \cite{GSh}, each integral in the sum over $k$ as an analytical function of $\nu$ has singularities in the form of simple poles at the points $\nu\in\N$. It is convenient to write the second contribution as
\begin{equation}\label{second_contr}
    \int_0^\infty d\e\e^{-\nu}\int_{\omega_c(\e)}^\infty\frac{d\omega n(\e,\omega)}{e^{\be(\omega-\mu)}\pm1}=\int_0^\infty d\e\e^{-\nu}\int_0^\infty\frac{d\omega n(\e,\omega+\omega_c(\e))}{e^{\be(\omega+\omega_c(\e)-\mu)}\pm1}.
\end{equation}
Suppose the function,
\begin{equation}
    \int_{\omega_c(\e)}^\infty\frac{d\omega n(\e,\omega)}{e^{\be(\omega-\mu)}\pm1},
\end{equation}
is finite and has finite derivatives with respect to $\e$ at $\e=0$ or, put another way, recalling the dependence of $H(\omega)$ on $m^2$, we suppose that the average number of particles has finite derivatives with respect to $m^2$. This is a rather strong restriction on the class of operators under consideration. In particular, this condition fails to hold for massless particles in the Minkowski space. That is why we shall calculate the partition function of massless particles proceeding to the limit $m^2\rightarrow0$. If the condition mentioned fulfills, the contribution \eqref{second_contr} as a function of $\nu$ has simple poles at the points $\nu\in\N$. In this case, the function $I_\nu(\mu)/\Ga(1-\nu)$ is an entire function of $\nu$.

\subsection{Function $\s^{-1}_\nu(\omega)$}

It is natural to associate with $\zeta_+(\nu,\omega)$ another function $\s^{-1}_\nu(\omega)$, which appears in the high-temperature expansion. Introduce the function
\begin{equation}\label{sigma-1}
    \s^{-1}_\nu(\omega):=\reg\int_0^\infty d\omega'\frac{\zeta_+(\nu,\omega')}{\omega'-\omega},\qquad\re\nu>d/2,\qquad\nu\not\in\N.
\end{equation}
Henceforward, the symbol $\reg$ denotes the analytical regularization of the integral \cite{GSh} with respect to the parameter $\nu$. This function is analytic in the $\omega$ plane and has a branch cut discontinuity on the positive real semiaxis, where
\begin{equation}
    \s^{-1}_\nu(\omega+i0)-\s^{-1}_\nu(\omega-i0)=2\pi i\zeta_+(\nu,\omega).
\end{equation}
At large values of $\omega$, from \eqref{asympt_UV} and \eqref{sigma-1} the asymptotics follows
\begin{equation}\label{sigma-1_asymp}
    \s^{-1}_\nu(\omega)\sim \frac{\Ga(1-\nu)}{\Ga(d/2-\nu+1)}\frac{\pi(-\omega)^{d-2\nu}}{\sin\pi(2\nu-d)},\qquad (d+1)/2>\re\nu>d/2,
\end{equation}
where the exponentiation is defined as $x^\al:=|x|^\al e^{i\al\arg x}$, $\arg\al\in[0,2\pi)$, i.e., it possesses the cut along the real positive semiaxis in the $\omega$ plane. This asymptotic behavior holds true for all $\re\nu<(d+1)/2$. Indeed, for $\re\nu\in((d-1)/2,d/2)$, differentiating \eqref{sigma-1} with respect to $\omega$, we come to
\begin{equation}
    \partial_\omega\s^{-1}_\nu(\omega)\sim \frac{\Ga(1-\nu)}{\Ga(d/2-\nu+1)}\partial_\omega\frac{\pi(-\omega)^{d-2\nu}}{\sin\pi(2\nu-d)}.
\end{equation}
On integrating this expression and taking into account that $\re\nu<d/2$, we arrive at \eqref{sigma-1_asymp} for the strip $\re\nu\in((d-1)/2,d/2)$. Proceeding further, we can extend the domain of applicability of the asymptotics \eqref{sigma-1_asymp} up to $\re\nu<(d+1)/2$.

Taking into consideration the asymptotic behavior \eqref{asympt_UV}, one can see that the function $\s^{-1}_\nu(\omega)$ has the singularities for $\re\nu>d/2$ only in the form of simple poles at $\nu\in\N$ in the complex $\nu$ plane (the singularities of $\zeta_+(\nu,\omega)$) since the integral over $\omega'$ converges in that case. It also follows from the asymptotics \eqref{asympt_UV} that the integral \eqref{sigma-1} (after the substitution $\omega'\rightarrow 1/\omega'$) as the function of $\nu$ possesses the poles at $(d-2\nu+1)\in\N$ apart from the poles at $\nu\in\N$. If these new poles do not coincide with the poles at $\nu\in\N$ then they are simple. Otherwise, the second order poles emerge. The asymptotics \eqref{sigma-1_asymp} confirms this observation. If the $\omega$ expansion of the stepless part of $\zeta_+(\nu,\omega)$ in the vicinity of the infinite point has the form \eqref{zeta+expan} and $\zeta_+(\nu,0)=0$ then there are no additional singularities of the function $\s^{-1}_\nu(\omega)$ in the $\nu$ plane.

\begin{figure}[t]
\centering
\includegraphics*[width=0.4\linewidth]{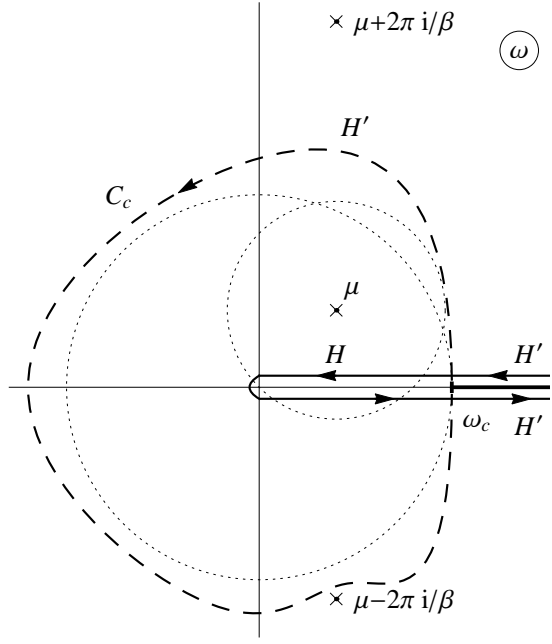}
\caption{{\footnotesize
A schematic pattern of the contours $H$ and $H'$ on the complex $\omega$ plane for the bosonic case. Here crosses denote the poles of the Bose-Einstein distribution function \eqref{Inu_Hank}. There is a cut along $\omega>\omega_c$ making the functions $\s^{-1}_\nu(\omega)$ and $\zeta_+^0(\nu,\omega)$ single-valued. As a matter of fact, the function $\s^{-1}_\nu(\omega)$ may have singularities in the interval $[0,\omega_c)$, and the function $\zeta_+^0(\nu,\omega)$ may have singularities inside the disk $|\omega|<\omega_c$. These singularities are not depicted.}}
\label{contours_HH}
\end{figure}

By the use of the function $\s^{-1}_\nu(\omega)$, the integral \eqref{Inu} can be cast into the form
\begin{equation}\label{Inu_Hank}
    I_\nu(\mu)=-\int_H\frac{d\omega}{2\pi i}\frac{\s^{-1}_\nu(\omega)}{e^{\be(\omega-\mu)}\pm1},
\end{equation}
where $H$ is the Hankel contour. If one knows the expansion of $\s^{-1}_\nu(\omega)$ in the neighborhood of the infinite point of the $\omega$ plane convergent outside of the disk with the radius $\omega_c$ \footnote{For the free fields with a homogeneous dispersion law such an expansion is presented in \cite{olopq}. As long as the expansion of $\zeta_+(\nu,\omega)$ in the vicinity of the infinite point of the $\omega$ plane differs from \eqref{zeta+expan} in this case, $\s^{-1}_\nu(\omega)$ possesses a different singularity structure in the $\nu$ plane.} then the representation \eqref{Inu_Hank} allows one to derive the high-temperature expansion of $I_\nu(\mu)$ easily. For the sake of definiteness, suppose $\im\mu\in(-\pi/\be,\pi/\be)$. Then we deform the contour $H$ into $H'$ so that $|\omega|>\omega_c$ on the contour and the contour itself lies beyond the disk of radius $|\mu|$ centered at $\mu$ (see Fig. \ref{contours_HH}). The former condition allows us to expand $\s^{-1}_\nu(\omega)$ in the vicinity of the infinite point, while the latter one permits us to expand each term of the series (see, for example, the asymptotics \eqref{sigma-1_asymp}) in terms of decreasing powers of $(\omega-\mu)$.

Upon deformation of the contour $H$ into $H'$, we get
\begin{equation}\label{Inu_Hank_2}
    I_\nu(\mu)=\left\{
                 \begin{array}{ll}
                   -\int_{H'}\frac{d\omega}{2\pi i}\frac{\s^{-1}_\nu(\omega)}{e^{\be(\omega-\mu)}+1}, & \hbox{for fermions;} \\
                   -\int_{H'}\frac{d\omega}{2\pi i}\frac{\s^{-1}_\nu(\omega)}{e^{\be(\omega-\mu)}-1}+\beta^{-1}\s^{-1}_\nu(\mu), & \hbox{for bosons.}
                 \end{array}
               \right.
\end{equation}
The last term in the expression for bosons is the contribution from the pole of the integrand \eqref{Inu_Hank} after the deformation of $H$ into $H'$ has been performed. Other poles do not contribute because of this deformation in the case when
\begin{equation}\label{mu_restr}
\begin{aligned}
    |\mu|&<\pi/\beta& \;  \text{and}&\; &|\mu\pm\pi i/\beta|>\omega_c, \qquad &\text{for fermions};\\
    |\mu|&<2\pi/\beta& \;  \text{and}&\; &|\mu\pm2\pi i/\beta|>\omega_c, \qquad &\text{for bosons}.
\end{aligned}
\end{equation}
The first inequality is the requirement that the two poles nearest to the point $\omega=\mu$ do not lie inside the disk $|\omega-\mu|<|\mu|$, while the second one is the requirement that both of them are outside of the disk $|\omega|<\omega_c$. Substituting the development of $\s^{-1}(\omega)$ in the neighborhood of the infinite point to \eqref{Inu_Hank_2}, expanding each term of the series in the inverse powers of $(\omega-\mu)$, and then integrating termwise the series (see the similar integral \eqref{first_trms} below), we arrive at the expansion in the increasing powers of $\beta$. The integrals determining the coefficients of this expansion are reduced to zeta functions. The high-temperature expansion converges provided the inequalities mentioned above hold. In case of need, the convergence radius of the high-temperature expansion can be increased by ``blowing up'' the contour $H'$ and taking into account the additional poles (the leading Matsubara frequencies) of the integrand \eqref{Inu_Hank_2}.

\subsection{High-temperature expansion}

In many cases, it is more convenient to use another representation of the function $I_\nu(\mu)$ for calculation of the high-temperature expansion. It is that representation which we shall use. Suppose that (cf. \eqref{asympt_UV})
\begin{equation}\label{zeta+expan}
    \zeta_+(\nu,\omega)=\zeta^0_+(\nu,\omega)+\zeta^q_+(\nu,\omega),\qquad\zeta^0_+(\nu,\omega)=\sum_{k=0}^\infty\zeta_k(\nu)|\omega|^{d-2\nu}\omega^{-k},\qquad |\omega|\geq\omega_c,
\end{equation}
where $\omega$ is real and $\omega_c$ is the boundary of the series convergence domain. Usually, it coincides with the boundary of the continuous spectrum, but this will be irrelevant to the derivation of a general formula for the high-temperature expansion. The function $\zeta^0_+(\nu,\omega)$ is the so-called stepless part of $\zeta_+(\nu,\omega)$  (see \cite{olopq,BalBloch,BalBlochSch,Shoenb,LandLifstat,LandLifshStat,LifshKag} and others) or the Thomas-Fermi approximation for $\zeta_+(\nu,\omega)$. It is that part of $\zeta_+(\nu,\omega)$ which is given by a naive heat kernel expansion (see, e.g., \cite{KalKaz,BalBloch,BalBlochSch}). For the Laplacian type operators, the expansion of $\zeta^0_+(\nu,\omega)$ takes the form \eqref{zeta+expan}. The discreteness of quantum numbers is completely ignored in this contribution. The second contribution to $\zeta^q_+(\nu,\omega)$ is essentially quantum one. Generally, the infinite point of the $\omega$ plane is the transcendental branch point for the function $\zeta^q_+(\nu,\omega)$ at noninteger $\nu$, i.e., the development of $\zeta^q_+(\nu,\omega)$ in the inverse powers of $\omega$ contains infinitely many terms at positive powers of $\omega$. A typical example of the function with such a singularity is $\omega^\nu\exp(-a\omega)$. Further, in Sec. \ref{NonPertCorr}, we shall see in a concrete example that exactly this type of functions arises.

At first, consider the contribution from the stepless part of $\zeta_+(\nu,\omega)$ to $I_\nu(\mu)$. Suppose that the functions $\omega^{d-2\nu-k}$ in \eqref{zeta+expan} are continued from the real axis to the complex plane as $\omega^\al=|\omega|^\al e^{i\al\arg\omega}$, $\arg\omega\in[0,2\pi)$. Then, for bosons,
\begin{multline}\label{I0nu}
    I^0_\nu(\mu)= \reg\int_0^{\omega_c} \frac{d\omega\zeta^0_+(\nu,\omega)}{e^{\be(\omega-\mu)}-1} +\int_{\omega_c}^\infty \frac{d\omega\zeta^0_+(\nu,\omega)}{e^{\be(\omega-\mu)}-1} =\reg\int_0^{\omega_c} \frac{d\omega\zeta^0_+(\nu,\omega)}{e^{\be(\omega-\mu)}-1} +\int_{H'-C_c}^\infty \frac{d\omega}{e^{-4\pi i\nu}-1} \frac{\zeta^0_+(\nu,\omega)}{e^{\be(\omega-\mu)}-1}=\\
    =\int_{H'}\frac{d\omega}{e^{-4\pi i\nu}-1} \frac{\zeta^0_+(\nu,\omega)}{e^{\be(\omega-\mu)}-1} -\int_{C_c}\frac{d\omega}{e^{-4\pi i\nu}-1} \frac{\zeta^0_+(\nu,\omega)}{e^{\be(\omega-\mu)}-1} +\reg\int_0^{\omega_c} \frac{d\omega\zeta^0_+(\nu,\omega)}{e^{\be(\omega-\mu)}-1},
\end{multline}
where the contour $C_c$ coincides with the contour $H'$ for $\re\omega\leq\omega_c$. In the second equality, it is supposed that the restrictions \eqref{mu_restr} on $\mu$ are fulfilled, i.e., we can draw the contour $H'$ as depicted in Fig. \ref{contours_HH} such that the additional contributions from the poles do not appear. Also, in this equality, using the standard trick we have passed from the integration over the ray $\omega\geq\omega_c$ to the integration over the contour $H'-C_c$ taking into account the branch cut discontinuity of the function $\omega^{d-2\nu-k}$ on this ray. Having performed the transformation \eqref{I0nu}, the high-temperature expansion of the first contribution in $I^0_\nu(\mu)$ is readily evaluated. One can develop the function $\zeta^0_+(\nu,\omega)$ on the contour $H'$ as series \eqref{zeta+expan}. Then, substituting this development to the integral, we obtain
\begin{equation}\label{first_trms}
    \int_{H'}\frac{d\omega}{e^{-4\pi i\nu}-1} \frac{\zeta^0_+(\nu,\omega)}{e^{\be(\omega-\mu)}-1}=\sum_{k,n=0}^\infty \Ga(d+1-2\nu-k)\zeta(d+1-2\nu-k-n) \frac{\zeta_k(\nu)(\be\mu)^n}{n!\be^{d+1-2\nu-k}}.
\end{equation}

As regards the second and the third contributions in $I^0_\nu(\mu)$, their high-temperature expansion can be obtained provided that $|\omega-\mu|<2\pi/\beta$ on the interval $[0,\omega_c]$ and on the contour $C_c$. Subject to this condition, the function defining the Bose-Einstein distribution can be expanded in the Laurent series in $\be$ convergent on the integration contour. Keeping in mind that $\mu$ has been already constrained by \eqref{mu_restr}, the condition stated results in one additional inequality $|\mu|+\omega_c<2\pi/\beta$, where it is implied that the contour $C_c$ can be deformed into a part of the circle $|\omega|=\omega_c$. Obviously, all of the mentioned inequalities can be satisfied since $\be\rightarrow0$. In this case,
\begin{equation}
    \reg\int_0^{\omega_c} \frac{d\omega\zeta^0_+(\nu,\omega)}{e^{\be(\omega-\mu)}-1}
    -\int_{C_c}\frac{d\omega}{e^{-4\pi i\nu}-1} \frac{\zeta^0_+(\nu,\omega)}{e^{\be(\omega-\mu)}-1} =\sum_{l=-1}^\infty\frac{(-1)^l\zeta(-l)}{\Ga(l+1)}\bar{\s}^l_\nu(\mu)\be^l,
\end{equation}
where
\begin{equation}
    \bar{\s}^{l}_\nu(\mu):=\Big[\reg\int_0^{\omega_c}d\omega-\int_{C_c}\frac{d\omega}{e^{-4\pi i\nu}-1}\Big](\omega-\mu)^l\zeta^0_+(\nu,\omega).
\end{equation}
The expression for $\bar{\s}^l_\nu(\mu)$ can be rewritten in a more simple form, when $\re\nu>(d+1+l)/2$. We deform the contour $C_c$ so that it runs from above and below the part of the real axis $[\omega_c,+\infty)$ and is closed by an arch of an infinite radius. Due to the condition on $\nu$, the contribution from the arch vanishes and the integral over the contour adjoining the semiaxis $[\omega_c,+\infty)$ can be expressed in terms of the branch cut discontinuity of the function. Then
\begin{equation}
    \bar{\s}^{l}_\nu(\mu)=\reg\int_0^\infty d\omega(\omega-\mu)^l\zeta^0_+(\nu,\omega).
\end{equation}
Similarly to how it was done for the function $\s^{-1}_\nu(\mu)$, one can prove that for $\re\nu>(d+1+l)/2$ the function $\bar{\s}^l_\nu(\mu)$ has singularities in the $\nu$ plane in the form of simple poles at $\nu\in\N$. In case when $\re\nu\leq(d+1+l)/2$, there appears the poles at $(d+l-2\nu+2)\in\N$ in addition to the poles at $\nu\in\N$. If the last condition is satisfied for some $\nu\in\N$ then there are second order poles at these points. All the rest poles of the function $\bar{\s}^{l}_\nu(\mu)$ in the $\nu$ plane are simple. The analytic properties in the $\nu$ plane of the function
\begin{equation}\label{sigma_lnu}
    \s^{l}_\nu(\mu):=\reg\int_0^\infty d\omega(\omega-\mu)^l\zeta_+(\nu,\omega)
\end{equation}
coincide with the properties of the function $\bar{\s}^l_\nu(\mu)$.

It should be mentioned that if $H(\omega)$ depends on the mass squared such that $\partial_{m^2}H(\omega)=-1$ then the following recurrence relations hold
\begin{equation}\label{recurr_rels}
    \partial_{m^2}H_+^{-\nu}=\nu H_+^{-\nu-1},\qquad\partial_{m^2}\zeta_+(\nu,\omega)=\nu\zeta_+(\nu+1,\omega),\qquad\partial_{m^2}I_\nu(\mu)=\nu I_{\nu+1}(\mu),\qquad \partial_{m^2}\s_\nu^l(\mu)=\nu \s_{\nu+1}^l(\mu).
\end{equation}
In particular, the recurrence relation (91) of \cite{olopq} can be deduced from these ones.

So, in the case of the Bose-Einstein statistics, we obtain the high-temperature expansion of $I^0_\nu(\mu)$ (cf. \cite{olopq,KalKaz})
\begin{equation}\label{Inu_bos}
    I^0_\nu(\mu)=\sum_{k,n=0}^\infty \Ga(d+1-2\nu-k)\zeta(d+1-2\nu-k-n) \frac{\zeta_k(\nu)(\be\mu)^n}{n!\be^{d+1-2\nu-k}}+\sum_{l=-1}^\infty\frac{(-1)^l\zeta(-l)}{\Ga(l+1)}\bar{\s}^l_\nu(\mu)\be^l,
\end{equation}
If the series \eqref{zeta+expan} converges for $\omega\geq\omega_c$, the convergence domain of the series \eqref{Inu_bos} is determined by the restrictions on $\mu$ specified above. The analogous expansion for the fermions reads as
\begin{multline}\label{Inu_ferm}
    I^0_\nu(\mu)=\sum_{k,n=0}^\infty (1-2^{2\nu+k+n-d})\Ga(d+1-2\nu-k)\zeta(d+1-2\nu-k-n) \frac{\zeta_k(\nu)(\be\mu)^n}{n!\be^{d+1-2\nu-k}}+\\
    +\sum_{l=0}^\infty(1-2^{1+l})\frac{(-1)^l\zeta(-l)}{\Ga(l+1)}\bar{\s}^l_\nu(\mu)\be^l,
\end{multline}
where $|\mu|+\omega_c<\pi/\beta$ and the first condition in \eqref{mu_restr} should be met.

It is instructive to compare the derived expansions with the ones presented in \cite{olopq,KalKaz}. In fact, in these papers the analytical regularization (continuation) of the integral
\begin{equation}
    \tilde{\zeta}_+(\nu,\omega):=\frac{e^{2\pi i\nu}-1}{2\pi i}\Ga(\nu)\zeta_+(\nu,\omega)=e^{i\pi\nu}\frac{\zeta_+(\nu,\omega)}{\Ga(1-\nu)}
\end{equation}
is considered instead of $\zeta_+(\nu,\omega)$ and so is the function
\begin{equation}\label{tildeI}
    \tilde{I}_\nu(\mu)=\frac{e^{i\pi\nu}I_\nu(\mu)}{\Ga(1-\nu)}
\end{equation}
instead of $I_\nu(\mu)$. This function is the entire function of $\nu$ subject to the restrictions imposed on the spectrum of $H(\omega)$. In this connection, it should be noted that an inaccuracy was made in formula (35) of \cite{KalKaz}. The revised version of this formula is
\begin{equation}\label{sigma_lnu_tild}
    \s_\nu^l(m^2):=\reg\int_0^\infty d\omega\omega^l\int_C\frac{dss^{\nu-1}}{2\pi i}\Tr_d e^{-sH(\omega)}.
\end{equation}
Nevertheless, this inaccuracy does not affect the rest formulas of the paper. The recurrence relation of the form \eqref{recurr_rels} becomes
\begin{equation}
    \partial_{m^2}\tilde{I}_\nu(\mu)=\tilde{I}_{\nu+1}(\mu),
\end{equation}
and all the rest relations can be written in a similar fashion.

For the stepless contribution from the antiparticles,
\begin{equation}
    J^0_\nu(\mu):=\int_0^\infty \frac{d\omega\zeta^0_+(\nu,-\omega)}{e^{\be(\omega+\mu)}\pm1},
\end{equation}
we take into account that
\begin{equation}\label{zeta+expan-}
    \zeta^0_+(\nu,-\omega)=\sum_{k=0}^\infty(-1)^k\zeta_k(\nu)\omega^{d-2\nu-k},\qquad \omega\geq\omega_c.
\end{equation}
It is convenient to continue the functions $\omega^{d-2\nu-k}$ from the real axis to the complex plane just as it was done for the contribution from the particles. Then, for bosons, we have
\begin{equation}
    J_\nu^0(\mu)=\sum_{k,n=0}^\infty \Ga(d+1-2\nu-k)\zeta(d+1-2\nu-k-n) \frac{(-1)^{k+n}\zeta_k(\nu)(\be\mu)^n}{n!\be^{d+1-2\nu-k}}+\sum_{l=-1}^\infty\frac{(-1)^l\zeta(-l)}{\Ga(l+1)}\bar{\tau}^l_\nu(\mu)\be^l,
\end{equation}
where
\begin{equation}
    \bar{\tau}^{l}_\nu(\mu)=\reg\int_0^\infty d\omega(\omega+\mu)^l\zeta^0_+(\nu,-\omega).
\end{equation}
If $\zeta^0_+(\nu,-\omega)=\zeta^0_+(\nu,\omega)$, which is equivalent to $\zeta_k(\nu)=0$ for $k$ odd, then
\begin{equation}
    I^0_\nu(\mu)+J^0_\nu(\mu)
\end{equation}
is an even function of $\mu$. If $\zeta_+(\nu,-\omega)=\zeta_+(\nu,\omega)$ then $I_\nu(\mu)+J_\nu(\mu)$ is an even function of $\mu$ too. In particular, the $\Omega$-potential \eqref{omeg_pot}, which includes the contributions from particles and antiparticles, is an even function of $\mu$ in this case.

Now we are going to show how to take into account the essentially quantum corrections $\zeta_+^q(\nu,\omega)$ to the high-temperature expansion of $I_\nu(\mu)$. Let us assume that, for  $\omega\rightarrow+\infty$, the function $\zeta_+^q(\nu,\omega)$ is developed as a series with a typical term
\begin{equation}\label{oscil_corr}
    \omega^{\al(\nu)}e^{-a\omega},\qquad\re a\geq0.
\end{equation}
Furthermore, $\al(\nu)$ decreases as the term number increases and $|a|>a_0>0$. In this case, we have for bosons
\begin{equation}\label{Iq}
\begin{split}
    I^q_\nu(\mu):=&\int_0^\infty \frac{d\omega\zeta^q_+(\nu,\omega)}{e^{\be(\omega-\mu)}-1}=\Big[\int_0^{\omega_0}+\int_{\omega_0}^\infty\Big] \frac{d\omega\zeta^q_+(\nu,\omega)}{e^{\be(\omega-\mu)}-1}\\
    =&\sum_{l=-1}^\infty\frac{(-1)^l\zeta(-l)}{\Ga(l+1)}\be^l\int_0^{\omega_0} d\omega(\omega-\mu)^l\zeta^q_+(\nu,\omega)+\int_{\omega_0}^\infty \frac{d\omega\zeta^q_+(\nu,\omega)}{e^{\be(\omega-\mu)}-1}\\
    =&\sum_{l=-1}^\infty\frac{(-1)^l\zeta(-l)}{\Ga(l+1)}\be^l\int_0^\infty d\omega(\omega-\mu)^l\zeta^q_+(\nu,\omega) -\sum_{l=-1}^\infty\frac{(-1)^l\zeta(-l)}{\Ga(l+1)}\be^l\int_{\omega_0}^\infty d\omega(\omega-\mu)^l\zeta^q_+(\nu,\omega)+\\
    +&\int_{\omega_0}^\infty \frac{d\omega\zeta^q_+(\nu,\omega)}{e^{\be(\omega-\mu)}-1},
\end{split}
\end{equation}
where one can put $\omega_0$ to be equal to $\mu+2\pi/\be$ for real $a$ and $\mu$, $|\mu|<2\pi/\be$. If $a$ is a complex number then the integration contour has to be rotated so that it goes along the line of the steepest descent of the function $e^{-a\omega}$.  Then the last two terms in \eqref{Iq} are suppressed by the exponent $e^{-a\omega_0}$ for $\be\rightarrow0$. In what follows, such terms will be neglected. However, it should be noted that the expansion \eqref{Iq} becomes asymptotic after these exponentially suppressed terms have been cast out. The function $I^q_\nu(\mu)$ possesses the same singularities in the $\nu$ plane as $\zeta^q_+(\nu,\omega)$ does and may have only simple poles at $\nu\in\N$. For fermions, the expansion can be obtained in a similar way, whereas, in this case, $|\mu|<\pi/\be$ and $\omega_0=\mu+\pi/\beta$ for $a$ real.

To sum up, if we neglect the exponentially suppressed terms at $\be\rightarrow0$, the high-temperature expansion of $I_\nu(\mu)$ reads as \eqref{Inu_bos}, \eqref{Inu_ferm}, where $\bar{\s}_\nu^l(\mu)$ have to be replaced by $\s_\nu^l(\mu)$. This holds true for the contribution from the antiparticles too.

\section{Descent formulas}\label{DescForm}

The main tool to derive the high-temperature expansions is the heat kernel expansion technique \cite{DowKen} and its various resummations \cite{GusZeln,Page}. As a rule, the coefficients of the high-temperature expansion depend explicitly on the Killing vector, which determines the stationarity of the space-time, singles out the privileged set of mode functions of the quantum fields, and, consequently, determines the Fock vacuum state. However, some coefficients turn out to be independent of the Killing vector \cite{DowKen,DowSch,DowSch1,Fursaev1,Fursaev2,KalKaz}. One can convince oneself in this fact by a direct calculation, but it is easier to use the descent formulas \cite{Fursaev2} that relate the expansion coefficients of the heat kernel in $(d+1)$ dimensional space to the expansion coefficients of the same heat kernel in $d$ dimensional space. In deriving these formulas, it is supposed that the coefficients of the Laplacian type operator (the background fields) determining the heat kernel are independent of the $x^0$ variable. The method connecting the Green functions in $(d+1)$ dimensional space with the Green functions in $d$ dimensional space is known in mathematical physics as the descent method (see, e.g., \cite{Cour}). Thus, we shall call the formulas relating the heat kernel expansion coefficients in $(d+1)$ dimensional space with the corresponding coefficients in $d$ dimensional space as the descent formulas. Similar formulas can be found in \cite{Gilkey1,Gilkey2,VasilHeatKer}.

Consider the operator of a Laplacian type,
\begin{equation}
    H=:\bar{H}-m^2,
\end{equation}
acting on the space of square-integrable functions depending on $D:=d+1$ arguments. Assume that the coefficients of this operator are independent of the variable $x^0$, i.e., in particular, the Riemannian metric $g_{\mu\nu}$ associated with this operator possesses the Killing vector $\xi^\mu\partial_\mu=\partial_{x^0}$,
\begin{equation}
    \mathcal{L}_\xi g_{\mu\nu}=0.
\end{equation}
Performing the Fourier transform over the variable $x^0$ and rewriting the resulting operator $H(\omega)$ in terms of the Killing vector, as done in (8) of \cite{prop} and (15) of \cite{KalKaz}, we arrive at
\begin{equation}\label{traceD}
    \Tr_D(e^{-\tau H}f(\spx))=\int\frac{dx^0d\omega}{2\pi}\Tr_d(e^{-\tau H(\omega)}f(\spx)),
\end{equation}
where $f(\spx)$ is some function independent of $x^0$ . Expanding the left- and the right-hand sides of this equation in $\tau$ and canceling the arbitrary function $f(\spx)$, we come to
\begin{equation}\label{traceD_rel}
    \sqrt{g}\sum_{k=0}^\infty a_k(x)\frac{\tau^{k-D/2}}{(4\pi)^{D/2}}=\sqrt{\bar{g}}\sum_{k=0}^\infty\int\frac{d\omega}{2\pi} e^{-\tau(\frac{\omega^2}{\xi^2}-\tilde{m}^2)}\tilde{a}_k(\omega,x)\frac{\tau^{k-d/2}}{(4\pi)^{d/2}}.
\end{equation}
We have resummed some terms of the heat kernel expansion into the exponent on the right-hand side as made in (18) of \cite{KalKaz}. Writing $\tilde{a}_k(\omega,x)$ in the form (see (38) of \cite{KalKaz})
\begin{equation}\label{akj}
    \tilde{a}_k(\omega,x)=\sum_{j=0}^{[4k/3]}\tilde{a}^{(j)}_k(x)(g^2\omega^2)^{j/2},
\end{equation}
and integrating over $\omega$ in \eqref{traceD_rel}, we obtain the equality
\begin{equation}
    \sum_{k=0}^\infty a_k(x)\tau^k=\sum_{k=0}^\infty\sum_{n=0}^{[2k/3]}\tilde{a}^{(2n)}_k(x)\tau^ke^{\tau\tilde{m}^2}\frac{\Ga(n+1/2)}{\sqrt{\pi}}.
\end{equation}
Whence, equating the coefficients at the same power of $\tau$, we deduce the descent formula
\begin{equation}\label{desc_f0}
    a_s=\sum_{k=0}^{3s}\sum_{n=0}^{[2k/3]}\frac{\Ga(n+1/2)}{\sqrt{\pi}}\frac{\tilde{m}^{2s+2n-2k}}{\Ga(s+n-k+1)}\tilde{a}^{(2n)}_k =\sum_{k=0}^s\frac{\tilde{m}^{2s-2k}}{(s-k)!}\sum_{n=0}^{2k}\frac{\Ga(n+1/2)}{\sqrt{\pi}}\tilde{a}^{(2n)}_{k+n}.
\end{equation}
The left-hand side of this expression is independent of the Killing vector field and, consequently, the explicit dependence on the Killing vector field on the right-hand side cancels out. Such a cancelation happens for the metric of an arbitrary signature since it has a pure algebraic origin. The explicit form of the first three formulas read as
\begin{equation}\label{desc_f1}
\begin{gathered}
    a_0=\tilde{a}_0^{(0)},\qquad a_1=\tilde{m}^2\tilde{a}_0^{(0)}+\tilde{a}^{(0)}_1+\frac12\tilde{a}^{(2)}_2+\frac34\tilde{a}^{(4)}_3,\\ a_2=\frac{\tilde{m}^4}2\tilde{a}_0^{(0)}+\tilde{m}^2\Big(\tilde{a}^{(0)}_1+\frac12\tilde{a}^{(2)}_2+\frac34\tilde{a}^{(4)}_3\Big)+\tilde{a}^{(0)}_2+\frac12\tilde{a}^{(2)}_3+\frac34\tilde{a}^{(4)}_4 +\frac{15}8\tilde{a}^{(6)}_5+\frac{105}{16}\tilde{a}^{(8)}_6.
\end{gathered}
\end{equation}
Notice that the coefficient at the logarithm in formulas (44), (45) of \cite{KalKaz} (see also \eqref{F_f} below) coincides with the coefficient $a_2$ \cite{DowKen,DowSch,DowSch1,Fursaev1,Fursaev2}.

Now we prove another descent formula \cite{Fursaev2}. Let us given
\begin{equation}\label{ak_eucl}
\begin{split}
    \Tr_D(e^{-\tau H}f(\spx)) \approx &\sum_{k=0}^\infty\frac{\tau^{k-D/2}}{(4\pi)^{D/2}}e^{\tau m^2}\int dx\sqrt{g} f(\spx) b_k(x),\\
    \Tr_d(e^{-\tau H(\omega)}f(\spx)) \approx &\sum_{k=0}^\infty\frac{\tau^{k-d/2}}{(4\pi)^{d/2}}e^{\tau m^2}\int d\spx\sqrt{\bar{g}}e^{-\tau\omega^2/\xi^2} f(\spx) \bar{a}_k(\omega,x).
\end{split}
\end{equation}
Then, substituting these asymptotic expansions to \eqref{traceD} and equating the terms at the same powers of $\tau$, we find \cite{Fursaev2}
\begin{equation}\label{desc_from}
    b_k=\sum_{n=0}^{2k}\frac{\Ga(n+1/2)}{\sqrt{\pi}}\bar{a}^{(2n)}_{k+n},
\end{equation}
where $\bar{a}_k^{(j)}$ are determined in the same way as $\tilde{a}_k^{(j)}$ in formula \eqref{akj}. The coefficients $\bar{a}_k(\omega,x)$ can be expressed in terms of $\tilde{a}_k(\omega,x)$ with the aid of the equality
\begin{equation}\label{ak_tild_ak_bar}
    \sum_{k=0}^\infty e^{\tau(\tilde{m}^2-m^2)}\tilde{a}_k(\omega,x)\tau^{k}=\sum_{k=0}^\infty \bar{a}_k(\omega,x)\tau^{k},
\end{equation}
where the left-hand side has to be expanded in a series in $\tau$. The descent formula \eqref{desc_from} follows from \eqref{desc_f0} at $\tilde{m}=0$.

The descent formulas \eqref{desc_from} enable us to prove an interesting property of the high-temperature expansion. If we expand the terms in the free energy standing at the zeroth power of temperature and at the negative powers of the effective mass squared $\tilde{m}^2$ in the inverse powers of $m^2$ then the coefficients of the expansion will be independent of the Killing vector field. The resulting asymptotic series in $m^{-2}$ is the standard large mass expansion of the effective action at zero temperature. This may serve as an indirect verification of formulas (41), (42) of \cite{KalKaz} for the contribution of the stepless part of $\zeta_+(\nu,\omega)$ and, in general, of the correctness of the energy cutoff regularization scheme.

Indeed, according to formulas (41), (42) of \cite{KalKaz}, the terms described above are of the form
\begin{equation}\label{fin_part_m}
    F_b=\int\frac{d\spx\sqrt{|g|}}{(4\pi)^{d/2}}\sum_{k,j=0}^\infty(-1)^k\tilde{a}_k^{(j)}\tilde{m}^{D+j-2k}\frac{\Ga(k-(D+j)/2)}{4\Ga\big((1-j)/2\big)} +\cdots,
\end{equation}
where the ellipses denote the remaining terms. The same formula, but with an opposite sign, holds for fermions too. Being rewritten through $m^2$, this expression reads
\begin{equation}
    F_b=\int\frac{d\spx\sqrt{|g|}}{(4\pi)^{d/2}}\sum_{k,j=0}^\infty(-1)^k\bar{a}_k^{(j)}m^{D+j-2k}\frac{\Ga(k-(D+j)/2)}{4\Ga\big((1-j)/2\big)} +\cdots.
\end{equation}
Now, taking into account that $\bar{a}_k^{(j)}=0$ for $j$ odd, we can write this formula as
\begin{equation}
\begin{split}
    F_b = &\int\frac{d\spx\sqrt{|g|}}{(4\pi)^{d/2}}\sum_{k=0}^\infty\sum_{n=0}^{[2k/3]}(-1)^k\bar{a}_k^{(2n)}m^{D+2n-2k}\frac{\Ga(k-n-D/2)}{4\Ga(1/2-n)} +\cdots=\\
    = &\int\frac{d\spx\sqrt{|g|}}{2(4\pi)^{D/2}}\sum_{k=0}^\infty (-1)^km^{D-2k}\Ga(k-D/2)\sum_{n=0}^{2k} \frac{\Ga(n+1/2)}{\sqrt{\pi}}\bar{a}^{(2n)}_{k+n}+\cdots=\\
    = &\int\frac{d\spx\sqrt{|g|}}{2(4\pi)^{D/2}}\sum_{k=0}^\infty (-1)^km^{D-2k}\Ga(k-D/2)b_{k}(x)+\cdots,
\end{split}
\end{equation}
where, in the last equality, we have exploited the descent formula \eqref{desc_from}. The last formula is the standard form of the expansion of the effective action at zero temperature in the inverse powers of a large mass. This expansion can be found, for example, in (6.42) of \cite{BirDav}. The difference in sign $(-1)^k$ results from the different definition of $\bar{a}_k$ (compare \eqref{ak_eucl} with (6.39) of \cite{BirDav}). Note that formula \eqref{fin_part_m} is correct for $D$ odd. As far as even $D$ is concerned, one ought to get rid of all the terms, which are singular due to the gamma function in the numerator, i.e., all the terms at the nonnegative powers of the mass. So, we see that if $D$ is odd, the finite part of the effective action at zero temperature (without the non-perturbative corrections) being expanded in asymptotic series in the inverse powers of a large mass $m$ does not depend on the Killing vector. If $D$ is even, this is valid for the terms at the negative powers of $m^2$ and the logarithmic term only. Later on, when we derive the explicit expressions for the non-perturbative contributions to the high-temperature expansion, we shall turn back to the interpretation of this result.

\section{Non-perturbative corrections induced by a scalar field}\label{NonpertContrScalField}

\subsection{Perturbation theory}\label{PertTheor_sec}

Now we apply the general formulas obtained above to the concrete model. Consider a massive scalar field on a stationary gravitational background at a finite reciprocal temperature $\be$. For simplicity, we restrict our considerations to the case of a vanishing chemical potential. In the adapted coordinates, where $\xi^\mu=(1,0,0,0)$, the free energy $F$ takes the standard form
\begin{equation}\label{free_energy}
    e^{-\beta F}:=\Tr e^{-\be \mathcal{H}},\qquad \mathcal{H}=\int_\Si d\Si_\mu T^\mu_\nu \xi^\nu=\int d\spx\sqrt{|g|}T^0_0,
\end{equation}
where $T^\mu_\nu$ is the energy-momentum tensor, $\Si$ is the Cauchy surface, which we take to be $x^0=const$. The operator $\mathcal{H}$ is the Hamiltonian of the scalar field expressed in terms of the creation-annihilation operators associated with the stationary mode functions $(u_\al,\bar{u}_\al)$. These mode functions are the eigenvectors of the Lie derivative with respect to the Killing vector (see, e.g., \cite{DeWGAQFT,GriMaMos,DeWQFTcspt})
\begin{equation}
    i\mathcal{L}_\xi u_\al=\omega_\al u_\al,
\end{equation}
viz., they depend on time as $e^{-i\omega_at}$ in the adapted system of coordinates. The mode functions corresponding to the energy $\omega$ span the kernel of the Klein-Gordon operator,
\begin{multline}\label{KG_eq}
    H(x,y)=(-\nabla^2_x-m^2)\frac{\de(x-y)}{|g|^{1/4}(x)|g|^{1/4}(y)}=\\
    =|g|^{-1/4}(x)\biggl[-|g|^{-1/4}(x)\partial_\mu\sqrt{|g|}g^{\mu\nu}\partial_\nu|g|^{-1/4}(x)-m^2\biggr]\frac{\de(x-y)}{|g|^{1/4}(y)},
\end{multline}
where all the time derivatives should be replaced by $-i\omega$. This operator, which we denote as $H(\omega)$, must be supplemented by the appropriate boundary conditions. To simplify further calculations, we assume that the system considered is large enough to neglect the boundary effects or the space represents a compact manifold without boundary. The operator $H(\omega)$ is of a Laplacian type, it is Hermitian with respect to the measure $\sqrt{|g|}$ on the square-integrable functions depending on $\spx$, and possesses the spectrum bounded from above at fixed $\omega$.

The one-loop correction to the free energy can be cast into the form \eqref{omega_pot_1} with $\mu=0$. As we derived in the previous section, the high-temperature expansion is written as \eqref{Inu_bos}, \eqref{Inu_ferm} with the replacement $\bar{\s}^l_\nu\rightarrow \s^l_\nu$, when the exponentially suppressed contributions are neglected. The first terms in \eqref{Inu_bos} and \eqref{Inu_ferm} (i.e., the terms in these formulas that are proportional to the product of the gamma and zeta functions) were found in \cite{BalBloch,DowKen,DowSch,DowSch1,Fursaev1,Fursaev2,NakFuk,KalKaz}. They are determined by a stepless part of $\zeta_+(\nu,\omega)$ and can be obtained with the help of the asymptotic heat kernel expansion in the large mass $m$. In order to find the second terms of the high-temperature expansions \eqref{Inu_bos} and \eqref{Inu_ferm}, one needs a non-perturbative (that is not in the form of an asymptotic series in $\tau$) expression for the heat kernel $\exp(-\tau H(\omega))$ taken on the diagonal.

\paragraph{Hamiltonian.}

The heat kernel is a mere evolution operator with the Hamiltonian $H(\omega)$ taken at the imaginary time $-i\tau$. Therefore, in order to find the approximate, but non-perturbative (in the sense mentioned above), expression for it, we can employ the standard perturbation theory in quantum mechanics assuming that the coefficients of $H(\omega)$ are nearly constant. For the reader convenience and for the conformity of notation we describe such a perturbation theory in Appendix \ref{Pert_Theor} in some detail. In order to use the ordinary formulas of quantum mechanics and to take completely into account the dependence of the heat kernel on the metric, we shall work with the operators self-adjoined with respect to the standard scalar product with a trivial measure (not $\sqrt{|g|}$). To this end, one has to perform a similarity transform changing the measure and making it trivial. As a result, the operator $H(\omega)$ passes to $\tilde{H}(\omega)$ (see \cite{prop} for details, the Euclidean version see in \cite{HuOCon}),
\begin{multline}\label{Hamiltonian2}
    -\tilde{H}=\bar{g}^{-1/4}(p_i+\omega g_i)\sqrt{\bar{g}}\bar{g}^{ij}(p_j+\omega g_j)\bar{g}^{-1/4}+\frac12\bnabla^ih_i+\frac14h_ih^i-\frac{\omega^2}{\xi^2}+m^2=\\
    =(p_i+\omega g_i)\bar{g}^{ij}(p_j+\omega g_j) +\frac12\partial_i(\bar{g}^{ij}\partial_j\ln\sqrt{\bar{g}}) +\frac14\partial_i\ln\sqrt{\bar{g}}\bar{g}^{ij}\partial_j\ln\sqrt{\bar{g}} +\frac12\bnabla^ih_i +\frac14h_ih^i -\frac{\omega^2}{\xi^2}+m^2,
\end{multline}
where $p_i:=-i\partial_i$, $\xi^2=g_{\mu\nu}\xi^\mu\xi^\nu$, $g_\mu:=\xi_\mu/\xi^2$, $h_\mu:=\partial_\mu\ln\sqrt{\xi^2}$, and the relations \eqref{gbar_rels} hold. The connection  $\bnabla_i$ is constructed by the use of the positive-definite metric $\bar{g}_{ij}$. Then the heat kernel and its trace become
\begin{equation}\label{heat_kern}
  G(\omega,\tau;\spx,\spy):=\lan\spx|e^{-\tau\tilde{H}}|\spy\ran,\qquad \Tr e^{-\tau\tilde{H}}=\int d\spx\lan\spx|e^{-\tau\tilde{H}}|\spx\ran,
\end{equation}
respectively. Henceforward, it will be convenient to change the sign of the Hamiltonian and regard $-\tilde{H}$ as the generator of evolution instead of $\tilde{H}$. In that case, the quantum mechanical evolution operator is $\exp(-is[-\tilde{H}])$. Putting $s=i\tau$, we obtain the heat kernel from the latter operator.

\paragraph{System of coordinates.}

According to the general procedure expounded in Appendix \ref{Pert_Theor}, it is necessary to split the Hamiltonian \eqref{Hamiltonian2} into the free part $H_0$, quadratic in the variables $x^i(\tau)$ and $p_i(\tau)$, and the perturbation $V$. Besides, we demand that, for an every finite order of the perturbation series, the approximate evolution operator possesses all the symmetries of the exact evolution operator. In our case this means that the contributions of the perturbation theory are to be written in a general covariant form in terms of the metric  $g_{\mu\nu}$ and the Killing vector $\xi^\mu$, which defines the vacuum state as described above. Since the perturbation theory will represent a certain expansion in derivatives of the coefficients of the operator $\tilde{H}$, we need to define a covariant gradient expansion. With this end in view, we introduce the Riemann normal frame of the metric $\bar{g}_{ij}$ with the origin at the middle of the geodesic of the metric $\bar{g}_{ij}$ connecting the points $\spx$ and $\spy$ (the so-called midpoint prescription). This system of coordinates is not uniquely defined: apart from the global Euclidean rotations in space around the origin, one can change the time variable as $t\rightarrow t+\vf(\spx)$, where $\vf(\spx)$ is some smooth function of $\spx$ (see \cite{LandLifshCTF}, Sec. 88, for details). Under the latter transform, the fields $g_i$ are changed by a gradient of the function $\vf(\spx)$ similarly to the electromagnetic potentials. Let us seize this opportunity to redefine $t$ and impose the Fock gauge on $g_i$:
\begin{equation}\label{gi}
    g_i(x)x^i=0\quad\Leftrightarrow\quad g_i(x)=\sum_{n=1}^\infty\frac{n}{(n+1)!}x^{j_1}\cdots x^{j_n}\partial_{j_1}\ldots\partial_{j_{n-1}}f_{j_ni},
\end{equation}
in the Riemann normal frame specified above. The tensor $f_{\mu\nu}$ is defined in \eqref{fmunu}. These conditions fix unambiguously (up to global space rotations) the system of coordinates in the spacetime. This allows us to restore in a unique way the general covariant expressions from the derivatives of $\bar{g}_{ij}$ and $g_{i}$ taken at the origin of the frame \cite{Petrov}. For instance,
\begin{equation}\label{grav_pot}
\begin{split}
    \bar{g}^{ij}&=\de_{ij}+\frac13\bar{R}_{ikjl}x^kx^l+\cdots,\\
    \frac12&\partial_i(\bar{g}^{ij}\partial_j\ln\sqrt{\bar{g}})+\frac14\partial_i\ln\sqrt{\bar{g}}\bar{g}^{ij}\partial_j\ln\sqrt{\bar{g}}=-\frac16\bar{R}-\frac16\bnabla_i\bar{R}x^i+ \frac12\bar{r}_{ij}x^ix^j+\cdots,\\
    \bar{r}_{ij}&:=
    \frac1{5}\bigl(\frac13\bar{R}_{ik}\bar{R}^k_{\ j}-\frac16\bar{R}^{kl}\bar{R}_{kilj}-\frac16\bar{R}_i^{\ mnk}\bar{R}_{jmnk}-\frac14\bnabla^2\bar{R}_{ij}-\frac34\bnabla_{ij}\bar{R}\bigr),
\end{split}
\end{equation}
where we have employed the formulas for the developments of $\bar{g}_{ij}$ and $\bar{g}$ in the Riemann normal coordinates (see, e.g., \cite{prop,HuOCon,BekPark,Petrov}). The tensor $\bar{R}_{ijkl}$, its covariant derivatives, and contractions constructed with the help of the metric $\bar{g}_{ij}$ are expressed through the curvature tensor $R_{\mu\nu\rho\s}$ of the metric $g_{\mu\nu}$, the Killing vector $\xi^\mu$, and their covariant derivatives and contractions (see, e.g., Appendices in \cite{prop,KalKaz,FrZel}).

\paragraph{Free Hamiltonian.}

Now we need to single out unambiguously the quadratic part $H_0$ of the Hamiltonian and define the power counting scheme for the diagrams. The quadratic part $H_0$ determines the base of the perturbation theory and its propagators, while the power counting scheme allows us to order the infinite set of diagrams and distinguish the most relevant ones under the assumption of smallness of the field derivatives. In many respects this procedure is analogous to the effective field theory approach \cite{Weinb,DonGolHol}, but slightly simpler as long as we consider quantum mechanics with a finite number of degrees of freedom. In order to introduce such a grading, we shall make, at first, some estimations of the typical magnitudes of the structures appearing in the expansion of the coefficients of the operator $\tilde{H}$.

Let $L$ be a characteristic scale of variations of the gravitational field. For example, for a spherically symmetric metric, this quantity is of the order of the distance from the center of a gravitating object to the point where the derivative expansion is sought. In the weak field limit, at a large distance from the gravitating object, where
\begin{equation}
    \e:=1-\xi^2\sim r_g/L\ll1,
\end{equation}
we can use the expressions given in \cite{LandLifshCTF}, Sec. 105. As a result, we have
\begin{equation}\label{estim_weak}
    \partial_i\sim r^{-1}\sim L^{-1},\qquad h_i\sim\e/L,\qquad\bar{R}_{ij}\sim\e/L^2,\qquad f_{ij}\sim\e^2/L,\qquad \text{etc.}
\end{equation}
for solutions to the Einstein equations. The estimation of $f_{ij}$ has been obtained for the maximal (critical) value of the angular momentum of a gravitating object $J=Mr_g/2$, where $M$ is a mass of a body and $r_g$ is the Schwarzschild radius. In the strong field limit, where $\e\approx1$, i.e., near the ergosphere, we can use the Kerr solution to find the estimations. In this case, the most relevant contributions come from the terms containing the negative powers of $\xi^2$ and the derivatives acting on $\xi^{-2}$, viz.,
\begin{equation}\label{estim_strong}
    \partial_i\sim h_i\sim \xi^{-2}/L,\qquad f_{ij}\sim \xi^{-4}/L,\qquad\bnabla_ih_j\sim\bar{R}_{ij}\sim \xi^{-4}/L^2,\qquad \text{etc.}
\end{equation}
In the weak field limit, the condition of slow variation of the fields $g_{\mu\nu}$ and $\xi^\mu$ turns into
\begin{equation}
    |\omega| L\gg1.
\end{equation}
As follows from \eqref{estim_strong}, in the strong field limit, $\e\approx1$, this condition is substituted for
\begin{equation}\label{UV_strong}
    \xi^2|\omega| L\gg1,
\end{equation}
i.e., the derivatives of $\bar{g}_{ij}$ and $g_i$ are made dimensionless with the aid of the appropriate power of $\omega$. Further, we shall see that this is indeed the case.

We specify the free Hamiltonian $H_0$ determining the base of a perturbation theory by imposing the following two requirements:
\begin{enumerate}
  \item $H_0$ is no more than quadratic in $x^i(\tau)$ and $p_i(\tau)$;
  \item $H_0$ is quadratic in $\omega$ and $p_i$, and does not include the terms at lower powers of $\omega$ and $p_i$ (apart from the constant term $m^2$).
\end{enumerate}
The first condition is necessary to construct the perturbation theory stated in Appendix \ref{Pert_Theor}. It is concerned with the fact that, for the systems that do not possess some special symmetries, i.e., for a general background, a general solution to the Heisenberg equations can be constructed only for the quadratic Hamiltonians. The second condition is related to the requirement that the free Hamiltonian must include the most relevant terms in the short-wave approximation \eqref{UV_strong}. As we shall see below \eqref{aver_props}, $p_i\sim\omega$ for the solutions to the Heisenberg equations. The term $m^2$ is taken into account non-perturbatively since usually $m^2L^2\gg1$. On the other hand, the inclusion of the terms at lower powers of $\omega$ and $p_i$ into $H_0$ is unjustified since these small corrections are overlapped by the succedent terms of the perturbation series\setcounter{local}{\thepage}\footnote{Note in this connection that such terms were taken into account non-perturbatively in the papers \cite{prop,KalKaz} without a rigorous evaluation of the subsequent terms of the perturbation series. As a result, the wrong conclusion was made on instability of a massless scalar field on stationary gravitational backgrounds. For the case of a spherically symmetric metric \cite{Comm}, this formally appeared as that the quasiclassical estimations were applied to a non quasiclassical potential $\hbar^2 r''(x)/r(x)$ vanishing in the classical limit. The general proof of stability of a massless scalar field on a stationary gravitational background can be found, for example, in \cite{DeWGAQFT}, Sec. 17. Also notice that the well-known corrections to the mass squared like $R/6$ \cite{BekPark,Parker,Prokhor} or $R/4$ \cite{DeWGAQFT} are proportional to $\hbar^2$ and must be taken into account perturbatively too.}. The Hamiltonian $H_0$ containing all the terms satisfying the above two conditions is uniquely defined
\begin{equation}\label{Ham_func}
    H_0=\big(p_i+\frac12x^j\omega f_{ji}\big)^2-\omega^2\big(\xi^{-2}+b_ix^i+\frac{1}{2}E_{ij}x^ix^j\big)+m^2,
\end{equation}
where
\begin{equation}
    b_i:=\partial_i\xi^{-2}\Big|_{x=0}=-2\xi^{-2}h_i,\qquad E_{ij}:=\partial_{ij}\xi^{-2}\Big|_{x=0}=\xi^{-2}(4h_ih_j-2\bnabla_ih_j).
\end{equation}
Recall that we changed the overall sign of the Hamiltonian.

The first condition imposed on the Hamiltonian $H_0$ looks rather technical and requires some extra substantiations. First of all, recall that our chief goal is to obtain the generating functional of one-particle irreducible Green functions (the effective action) at a finite temperature and small external momenta. In constructing the effective action with the aid of the background field method, the background metric has not to be a solution of the classical equations of motion (the Einstein equations), in general. So as to find
\begin{equation}
    \frac{\de^n\Ga[g_{\mu\nu}]}{\de g_{\mu_1\nu_1}\cdots\de g_{\mu_n\nu_n}}\Big|_{g_{\mu\nu}=g^{(0)}_{\mu\nu}},\qquad g^{(0)}_{\mu\nu}\approx\eta_{\mu\nu},
\end{equation}
at small external momenta we only need a sufficiently wide class of metrics slowly varying in space. It is clear that the quadratic approximation described above works well for stationary metrics closely approximated by, for example,
\begin{equation}
    \bar{g}_{ij}=\de_{ij},\quad g_i=\frac12 x^jf_{ji},\quad g_{00}=1;\qquad \bar{g}_{ij}=\de_{ij},\quad g_i=0,\quad g^{-1}_{00}=c+b_ix^i+\frac12 E_{ij}x^ix^j,
\end{equation}
in the adapted system of coordinates. In the first case, the quadratic approximation gives the exact answer, whereas in the second case the fulfillment of the condition \eqref{UV_strong} is implied. It is also reasonable to expect that the quadratic approximation is good enough for the solutions to the Einstein equations, when the metric varies slowly (see the approximation \cite{Page} and its numerical verifications in \cite{AndHisSam,Howard} and others). In terms of the contributions of classical trajectories to the evolution operator (see \cite{BalBlochSch} for details), the quadratic approximation accurately describes the two leading contributions: the contribution from the shortest geodesic of a given energy connecting the points $\spx$ and $\spy$ (this provides the Thomas-Fermi contribution to $\zeta_+(\nu,\omega)$) and the contribution from the geodesic of a fixed energy connecting the points $\spx$ and $\spy$, reflected once from the turning point (this provides an oscillating contribution to $\zeta_+(\nu,\omega)$). For example, the approximation made should work well for the potential of two spherically symmetric gravitating bodies in the vicinity of the point where the attractive force is approximately zero. In the neighbourhood of this point, the gravitational potential along the line connecting the centers of the gravitating bodies has the form of an inverted parabola. Consequently, there are complex classical trajectories under the potential barrier, which are ``wound'' around this parabola. According to the general results of (\cite{BalBlochSch}, see also \cite{LandLifshQM}), the contributions of such trajectories are represented by oscillating exponentially suppressed terms in $\zeta_+(\nu,\omega)$.

Usually, in QFT, the particle creation process (in our case, the creation of particles by a gravitational field \cite{GriMaMos,Hawk,BrilWhil,MamMostStar}) is related to such trajectories in the sense that the module of the matrix element between the vacuum state of a scalar field in a flat spacetime and the vacuum state defined with respect to the creation-annihilation operators associated with the stationary mode functions, which take into account the interaction with gravity, is less than unity. This fact can be revealed by the presence of imaginary terms in the effective action constructed in an appropriate way. Note that in the framework we develop, which is based on the representation of a free energy \eqref{zeta+_nu}, \eqref{omeg_pot}, the imaginary contributions to the free energy are absent by virtue of the fact that \eqref{omeg_pot} is real. However, if one uses the Schwinger representation
\begin{equation}
  \zeta(\nu,\omega):=\int_0^\infty\frac{id\tau}{\Ga(\nu)}(i\tau)^{\nu-1}\Tr e^{-i\tau [\tilde{H}(\omega)-i0]},
\end{equation}
where that integral is understood in the sense of analytical regularization over $\nu$ by analogy with the generalized function $(x-i0)^{-\nu}$ \cite{GSh}, then the effective action at a finite temperature, formally constructed as \eqref{omeg_pot} with the replacement of $\zeta_+(\nu,\omega)$ by $\zeta(\nu,\omega)$, will possess imaginary terms. A more detailed investigation of this question will be given elsewhere.

\paragraph{Ingredients of the perturbation theory.}

The Hamiltonian  $H_0$ determines the averages and propagators \eqref{phi_prop} in the interaction picture and the matrix element $\lan \overline{out}|\overline{in}\ran$, where the states $|\overline{in}\ran$ and $|\overline{out}\ran$ are specified in \eqref{xy_inout} and \eqref{vacua}. The explicit expressions for these ingredients of the perturbation theory can be obtained for an arbitrary quadratic Hamiltonian. Nevertheless, to simplify the subsequent formulas we consider the case when
\begin{equation}\label{E_f}
    [E,f]=0\;\Rightarrow\; f=-iH\ups^1_{[i}\bar{\ups}^1_{j]},\qquad E=\la_1\ups^1_{(i}\bar{\ups}^1_{j)}+\la_2\ups^2_i\ups^2_j.
\end{equation}
The vectors $\ups^1_i$, $\bar{\ups}^1_i$, and $\ups^2_i$ are orthonormal with respect to the standard Hermitian scalar product, the overbar denotes complex conjugation, the vector $\ups^2_i$ having real components. Also, for definiteness, we assume that $\la_1<0$ and $\la_2>0$. In a weak gravitational field, this relations hold for a vacuum solutions to the Einstein equations (see (48) of \cite{prop}). In that case, we obtain \cite{prop} (for the Euclidean version see \cite{HuOCon})
\begin{equation}\label{HK_resum}
\begin{split}
    \lan \overline{out}|\overline{in}\ran&=(4\pi i s)^{-d/2}\det\Bigl(\frac{\sin(s\omega\ka)}{s\omega\ka}\Bigr)^{-1/2}e^{iS(s,\bs_i)-ism^2},\\
    S:&=\frac14\bs\omega\ka\ctg(s\omega\ka)\bs-\frac12(\frac{\bs}2-bE^{-1})\omega\ka\bigl(\ctg(s\omega\ka) -\frac{e^{s\omega f}}{\sin(s\omega\ka)}\bigr)(\frac{\bs}2+E^{-1}b) -\frac{s}2\omega^2bE^{-1}b+\\
    &+\frac{\omega}2bE^{-1}f\bs+s\frac{\omega^2}{\xi^2},
\end{split}
\end{equation}
where $\bs^i=x^i-y^i=2x^i=-2y^i$ and $\varkappa:=\sqrt{-f^2-2E}$. By construction, the above expression for the matrix element $\lan \overline{out}|\overline{in}\ran$ is a bi-density with respect to the coordinates $\spx$ and $\spy$. So as to obtain the bi-scalar, one has to multiply this expression by
\begin{equation}
    \bar{\De}^{1/2}(\spx,\spy)=\bar{g}^{-1/4}(\spx)\bar{g}^{-1/4}(\spy),
\end{equation}
where $\bar{\De}(\spx,\spy)$ is the covariant van Vleck determinant for the metric $\bar{g}_{ij}$. Notice that the formula (30) of \cite{prop} contains a misprint in the sign of the expression standing in the exponent: $\exp(-i\omega^\pm\tau)$ should be substituted for $\exp(i\omega^\pm\tau)$.

The averages and the propagators \eqref{phi_prop} are written as (see Appendix \ref{Pert_Theor})
\begin{equation}\label{aver_props}
\begin{split}
    \bar{x}(\tau)=&\frac{\sin(\tau\omega\ka)}{\sin(s\omega\ka)}e^{(s-\tau)\omega f}(x_0+x)+\frac{\sin((s-\tau)\omega\ka)}{\sin(s\omega\ka)}e^{-\tau\omega f}(x_0+y)-x_0,\\
    \bar{p}(\tau)=&\frac{\omega\ka}2\frac{\cos(\tau\omega\ka)}{\sin(s\omega\ka)}e^{(s-\tau)\omega f}(x_0+x)-\frac{\omega\ka}2\frac{\cos((s-\tau)\omega\ka)}{\sin(s\omega\ka)}e^{-\tau\omega f}(x_0+y)-\frac12\omega fx_0,\\
    D_{xx}(\tau_1,\tau_2)=&\frac{2}{\omega\ka}\Big[\theta(\tau_1-\tau_2)\frac{\sin((s-\tau_1)\omega\ka)\sin(\tau_2\omega\ka)}{\sin(s\omega\ka)} +\theta(\tau_2-\tau_1)\frac{\sin(\tau_1\omega\ka)\sin((s-\tau_2)\omega\ka)}{\sin(s\omega\ka)}\Big]e^{\omega f(\tau_2-\tau_1)},\\
    D_{xp}(\tau_1,\tau_2)=&\Big[\theta(\tau_1-\tau_2)\frac{\sin((s-\tau_1)\omega\ka)\cos(\tau_2\omega\ka)}{\sin(s\omega\ka)} -\theta(\tau_2-\tau_1)\frac{\sin(\tau_1\omega\ka)\cos((s-\tau_2)\omega\ka)}{\sin(s\omega\ka)}\Big]e^{\omega f(\tau_2-\tau_1)},\\
    D_{pp}(\tau_1,\tau_2)=&-\frac{\omega\ka}{2}\Big[\theta(\tau_1-\tau_2)\frac{\cos((s-\tau_1)\omega\ka)\cos(\tau_2\omega\ka)}{\sin(s\omega\ka)} +\theta(\tau_2-\tau_1)\frac{\cos(\tau_1\omega\ka)\cos((s-\tau_2)\omega\ka)}{\sin(s\omega\ka)}\Big]e^{\omega f(\tau_2-\tau_1)},
\end{split}
\end{equation}
where $x_0=E^{-1}b$ and $x=-y$. We shall depict the propagators $D_{xx}$ and the averages $\bar{x}$ by solid lines on the Feynman graphs. The propagators $D_{pp}$ and the averages $\bar{p}$ will be denoted by dashed lines, while for the propagators $D_{xp}$ the half solid half dashed lines will be used. On developing the Hamiltonian \eqref{Hamiltonian2} as a covariant Taylor series and casting out the quadratic part $H_0$, we can distinguish four types of vertices:
\begin{equation}\label{vertices_type}
    V_{2p}:\udg{0.8}{\diagViip}\sim\al^{n},\quad V_{1p}:\udg{0.8}{\diagVip}\sim\al^{n}, \quad V_{x}:\udg{0.8}{\diagVx}\sim\al^{n},\quad V_{0}:\udg{0.8}{\diagVo}\sim\al^{n+2},
\end{equation}
where $n$ is the number of $x$ lines joined to the vertex and the dots denote possible additional solid lines. The vertices of the types $V_{2p}$ and $V_{1p}$ have no less than two $x$ lines joining to the vertex, while the vertices $V_x$ possess no less than three such lines. The ordering of operators in the vertex is taken into account by the infinitesimal shifts of the time arguments of operators. The time arguments are shifted in such a way that the $T$-ordering places them in the proper order as they stand in the Hamiltonian (see, for instance, \eqref{V2}).

Let us introduce a grading on the set of diagrams. We attribute, formally, every vertex by the coupling constant $\al^n$, where $n$ is the number of derivatives of the fields $\bar{g}_{ij}$ and $g_i$ in the vertex. As an example, see \eqref{vertices_type} and the vertices of the order $\al^2$ in \eqref{V2}. For the vertices $V_{2p}$, $V_{1p}$ and $V_x$, the power of $\al$ is equal to the number of the lines $D_{xx}$ and $D_{xp}$ joining by the $x$ leg to the vertex. As for the $V_0$ type vertices, one should keep in mind the fact that the vertex without external lines has already the order $\al^2$. The order of the whole diagram in $\al$ is defined as the product of the ``coupling constants'' $\al^n$ of all the vertices of the diagram. The order of the diagram in $\omega$ is defined as
\begin{equation}\label{omega_count}
    E_p+I_{pp}-I_{xx}+V_x-V_{2p}-V_0,
\end{equation}
where $E_p$ is the number of the external $p$ lines, while $I_{pp}$ and $I_{xx}$ are the numbers of the internal lines $D_{pp}$ and $D_{xx}$, respectively. The numbers of the corresponding vertices are denoted by $V_x$, $V_{2p}$, and $V_0$. Formula \eqref{omega_count} easily follows from the explicit expressions for the averages and propagators \eqref{aver_props}. This formula also allows for the fact that the integral over $\tau$ in the vertex produces extra $\omega^{-1}$. It can be seen from that the proper-time $\tau$ enters the arguments of the trigonometric functions and exponents in the combination $\tau\omega$. Having redefined $\tau\rightarrow\omega^{-1}\tau$, the arguments of the functions mentioned cease to depend on $\omega$, and every integration over $\tau$ results in $\omega^{-1}$ after that. It follows from \eqref{omega_count} that every closure of the external lines into a loop diminishes the order of a diagram in $\omega$ by one. Thus, in order to find the contributions to the connected part of the evolution operator matrix element \eqref{con_part}, one ought to draw all the connected tree diagrams of a given order in $\al$, which determines the order of the diagrams in derivatives. The tree diagrams give the leading contribution in $\omega$. Then, the external lines are closed into loops, which results in the corrections to the tree contribution of the order of $\omega^{-L}$, i.e., the expressions will look like
\begin{equation}
    \omega^{1-2V_0-L}f(s\omega),
\end{equation}
where $f(x)$ is some function independent of $s$ and $\omega$ specified by the Feynman rules, while $L$ is the number of loops. The loop expansion is equivalent to the quasiclassical procedure for expansion in $\omega^{-1}$ or $\hbar$ elaborated in \cite{BuldNomof,BBTY}.

In the paper \cite{prop}, it was verified by the explicit calculation that the development in $s$ of the evolution operator ensuing from the perturbation theory described above reproduces the standard asymptotic expansion of the heat kernel (see, e.g., \cite{VasilHeatKer}). As a matter of fact, in order to reproduce all the terms of the asymptotic expansion of a given order $n$ in derivatives, one has to calculate all the contributions of the perturbation theory up to the order $\al^n$. The resultant perturbation theory is rather cumbersome even for the diagonal matrix elements of the evolution operator. In Appendix \ref{Pert_Theor}, we provide all the terms of the perturbation series of the order $\al^2$ for the logarithm of the heat kernel diagonal (see \eqref{a_1_ap}-\eqref{c_ap}). Nevertheless, such a perturbation theory, as it is formulated in Appendix \ref{Pert_Theor}, admits a simple realization in a computer program. The only technical issue, which could arise, is the evaluation of the integrals in vertices. However, as seen from \eqref{aver_props}, in our case the integrals of the perturbation theory are reduced to the integrals of a product of exponents, which are easily calculated analytically (see formulas \eqref{t_tens_ap} and \eqref{l_tens_ap}).

\subsection{Non-perturbative corrections}\label{NonPertCorr}

The purpose of our investigation is to derive the explicit expression for the divergent and finite parts of the high-temperature expansion of a free energy. For a four dimensional spacetime, it implies we need to find all the contributions of the perturbation theory for the heat kernel with fourth derivatives of the fields, i.e., up to the order $\al^4$. The prospects are rather ominous having in view the explicit expressions for the terms of the order of $\al^2$ (see \eqref{a_1_ap}-\eqref{c_ap}). Fortunately, as we shall see below, the higher contributions of the perturbation theory are not necessary for our aim. In fact, a good approximation for the high-temperature expansion to the order we consider can be obtained by a mere combination of the results of the papers \cite{KalKaz} and \cite{prop}. It is important at this point that the higher orders of perturbation theory do not change the arrangement of singularities of the evolution operator in the $s$ plane (see \eqref{aver_props}).

\begin{figure}[t]
\centering
\begin{tabular}{cr}
    \multirow{2}{0.5\linewidth}[0.25\linewidth]{\raisebox{0.5\linewidth}{a)}\,\hbox{\includegraphics*[width=\linewidth]{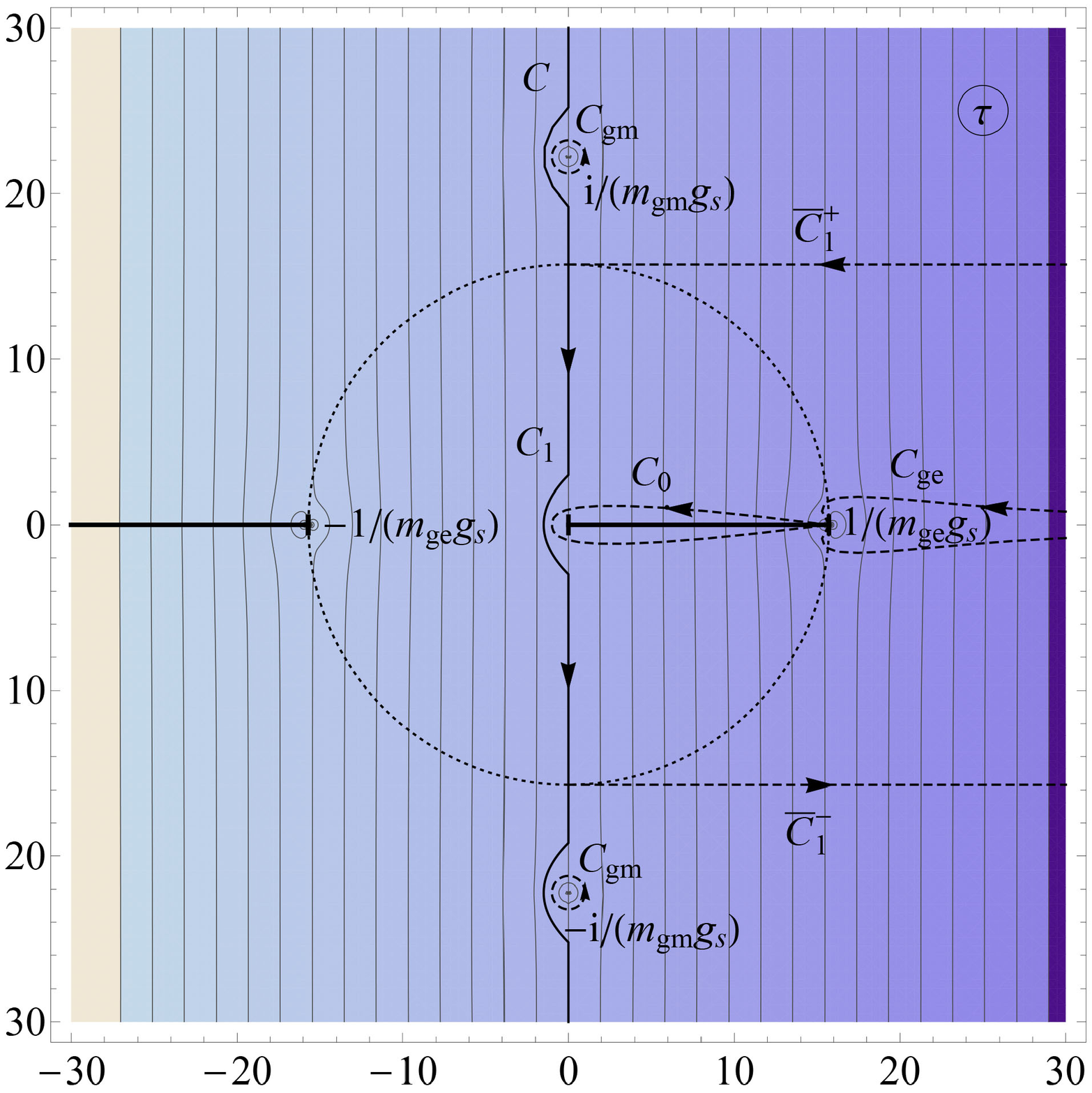}}} &
    \quad\raisebox{0.15\linewidth}{b)}\,\hbox{\includegraphics*[height=0.31\linewidth]{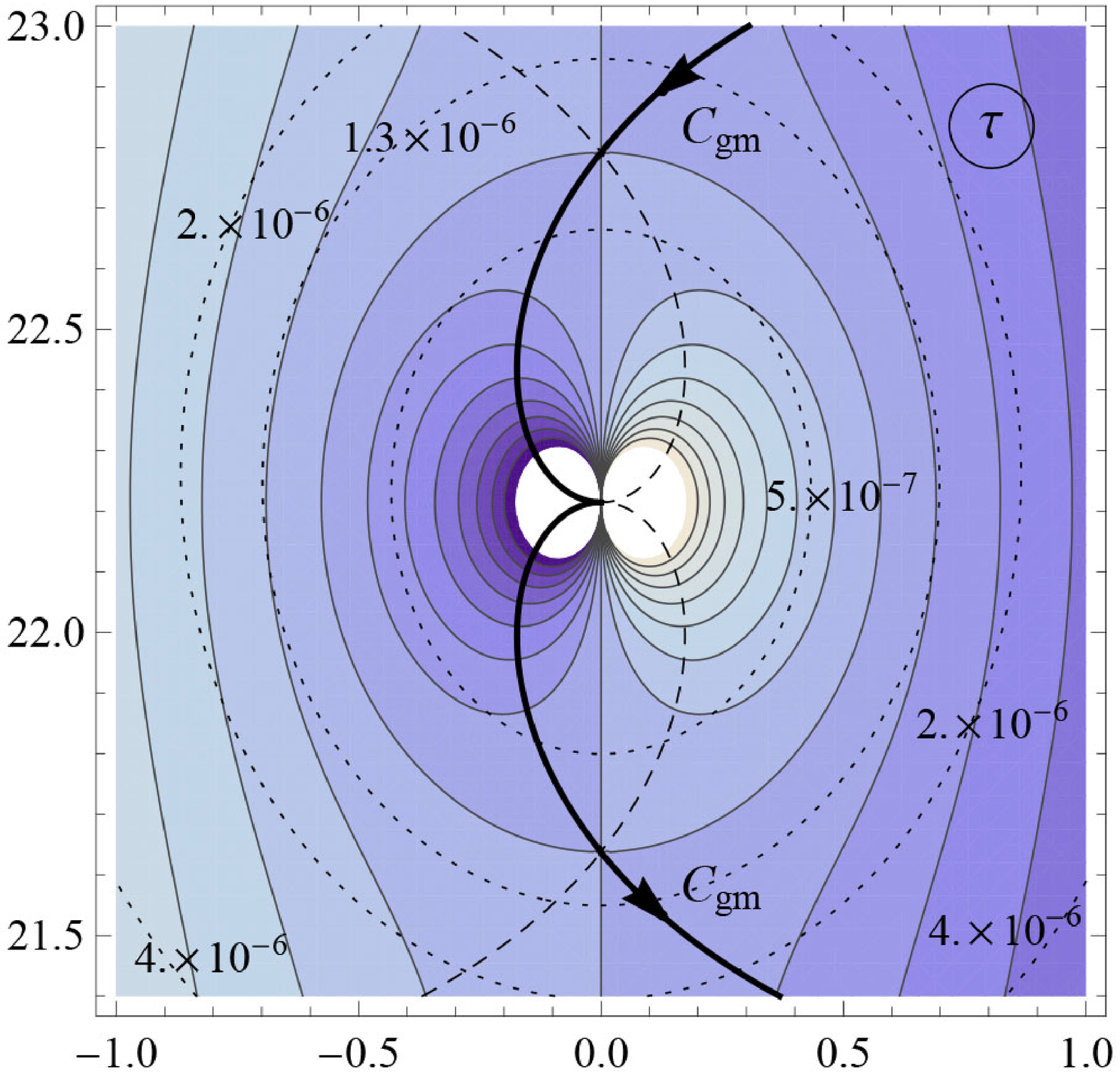}} \\
     & \quad\raisebox{0.155\linewidth}{c)}\,\hbox{\includegraphics*[height=0.31\linewidth]{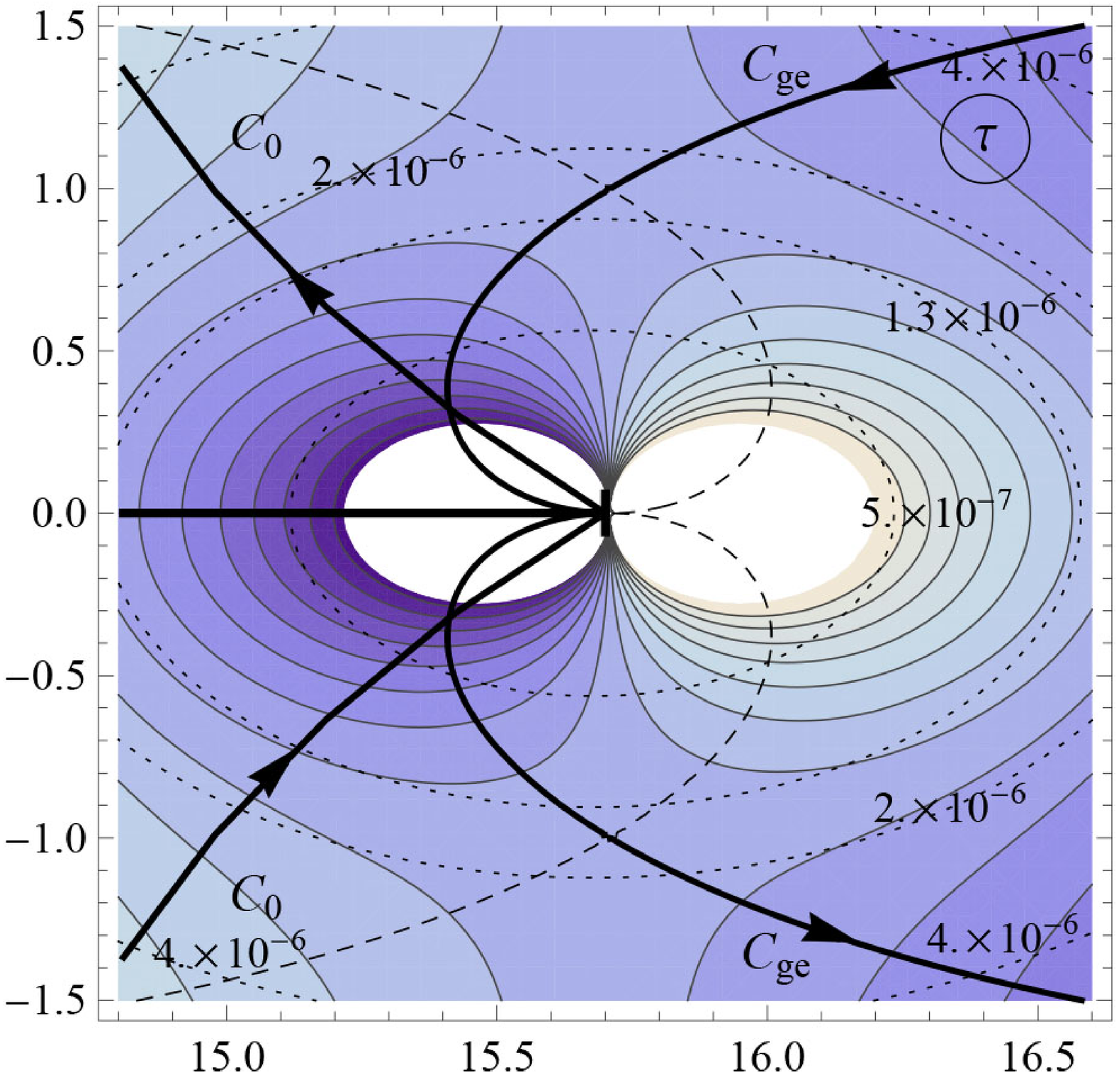}}\\
\end{tabular}
\caption{{\footnotesize The structure of singularities of the function $G_0(\omega,\tau;\spx,\spx)$ and the integration contours in the $\tau$ plane after the variable $\tau$ has been stretched $\tau\rightarrow\omega^{-1}\tau$. Fig. a): The level lines of the real part of the function $S_0+\tau m^2/\omega$ are depicted for the case when the inequality \eqref{omega_restr} holds. The concrete values are taken to be $L=1$, $\e=10^{-2}$, $H=\e^2/L$, $\la_1=-\e/L^2$, $\la_2=2\e/L^2$, $g_s^2=1+\e$, $b_\parallel=2\e/L$, $b_\perp=\e/(2L)$, $\omega=4$, and $m=2$. In the massless case $m=0$, the level lines look similarly. The dotted circle of the radius $(m_{ge}g_s)^{-1}$ designates the convergence domain of the series \eqref{HKE_converg}. Note that the form of the contour $C_{ge}$ is ``duplicated'' with a period of two singular points on the real axis, i.e., it approaches the singular points as drawn in Fig. c) at $\tau=(2n-1)/(m_{ge}g_s)$, $n\in\N$. Fig. b): The enlarged part of Fig. a) in the vicinity of the singular point on the imaginary axis. The lines of the steepest descent are depicted by dashed lines. They intersect at the saddle points. The contour $C_{gm}$ is deformed and passes along the line of the steepest descent through the saddle points. The dotted lines are the level lines of the relative error arising due to the replacement of $S_0+\tau m^2/\omega$ by the three leading terms of the Laurent series in the neighbourhood of the singular point. Fig. c): The same as in Fig. b), but for the vicinity of the singular point on the real axis in the massless limit $m=0$.}}
\label{cuts}
\end{figure}

According to the general results of the previous sections \eqref{Inu_bos}, \eqref{Inu_ferm}, and \eqref{Iq}, so as to find the high-temperature expansion, it is sufficient to obtain the explicit expressions for the functions $\s^l_\nu$ defined in \eqref{sigma_lnu} provided we neglect exponentially suppressed terms at $\be\rightarrow0$. The first terms in the expansion \eqref{Inu_bos}, \eqref{Inu_ferm} were found in \cite{KalKaz} up to the required order in $\beta$. To simplify the calculations and to adjust the notation to \cite{KalKaz}, we shall use the normalization \eqref{tildeI}, \eqref{sigma_lnu_tild}. At coinciding arguments, the Gaussian contribution \eqref{HK_resum} to the diagonal of the heat kernel can be cast into the form
\begin{equation}\label{HK_diag_Gauss}
    \lan\spx|e^{-\tau\tilde{H}(\omega)}|\spx\ran\approx G_0(\omega,\tau;\spx,\spx)=\frac{e^{3\pi i/2}}{(4\pi)^{3/2}}\frac{\omega^{3/2}\sqrt{H^2-2\la_1}}{\sh (\tau\omega\sqrt{H^2-2\la_1})}\Big(\frac{\sin (\tau\omega\sqrt{2\la_2})}{\sqrt{2\la_2}}\Big)^{-1/2}e^{S_0+\tau m^2},
\end{equation}
where
\begin{equation}\label{S_0}
\begin{split}
    S_0& = \omega\bigg[\frac{b^2_\perp}{2\la_1^2}\sqrt{H^2-2\la_1}\Bigl(\cth (\tau\omega\sqrt{H^2-2\la_1}) -\frac{\ch( \tau\omega H)}{\sh( \tau\omega\sqrt{H^2-2\la_1})}\Bigr) -\frac{b_\parallel^2}{2\la_2^2}\sqrt{2\la_2}\tg(\frac{\tau\omega}{2}\sqrt{2\la_2})-\\
    & -\tau\omega\Big(\frac{1}{\xi^2}-\frac{b_\perp^2}{2\la_1} -\frac{b_\parallel^2}{2\la_2}\Big)\bigg],
\end{split}
\end{equation}
and $b_\parallel=\ups^2_ib_i$, $b_\perp^2=b_i\ups^1_{(i}\bar{\ups}^1_{j)}b_j$. Notice that the misprint was made in formula (56) of \cite{prop}: one has to exchange $\tnh\leftrightarrow\cth$ in this formula. The branch of the multivalued function \eqref{HK_diag_Gauss} in the complex $\tau$ plane is specified by the conditions that the function $G_0(\tau)$ should be holomorphic in the strip $\re\tau\in(-\pi/\omega\sqrt{2\la_2},0)$, it should be real-valued on the segment of the real axis belonging to this strip, and the system of cuts is to be symmetric with respect to the reflection in the real axis. Consequently, the structure of singularities of the function $G_0(\tau)$ looks as depicted in Fig. \ref{cuts} and the square root of the sine in \eqref{HK_diag_Gauss} is defined as $x^{-1/2}=|x|^{-1/2}e^{-\arg(x)/2}$, where $\arg x\in[0,2\pi)$. Later on, it will be convenient to continue \eqref{HK_diag_Gauss} analytically to the complex $\omega$ plane. Then we shall define $\omega^{3/2}$ as an analytic function with the cut along $\omega<0$, i.e., the principal branch of the function $\omega^{3/2}$ will be taken.

\subsubsection{Massive case}

\paragraph{Partition of the contour.}

Let us consider, at first, the massive case $m^2>0$, $mL\gg1$. We shall assume that the singularities of the functions $G(\tau)$ and $G_0(\tau)$ are located at the same points of the $\tau$ plane (this is the case at any finite order of the perturbation theory) and have the same ``structure''. Also we shall suppose that the nearest to the origin singularity lies on the real axis of the $\tau$ plane and rather than on the imaginary axis, that is we assume $\sqrt{H^2-2\la_1}<\sqrt{2\la_2}$. Usually, this inequality is satisfied (see (48) of \cite{prop}) in the weak field limit. Then the integral determining  $\tilde{\zeta}_+(\nu,\omega)$ is rewritten as
\begin{equation}\label{contour_partit}
    \int_C\frac{d\tau\tau^{\nu-1}}{2\pi i}\Tr e^{-\tau\tilde{H}(\omega)}=\int_{C_1}\frac{d\tau\tau^{\nu-1}}{2\pi i}\Tr e^{-\tau\tilde{H}(\omega)} +\int_{\bar{C}_1}\frac{d\tau\tau^{\nu-1}}{2\pi i}\Tr e^{-\tau\tilde{H}(\omega)},
\end{equation}
where $C=C_1\cup\bar{C}_1$ and the contour $C_1$ goes from the point $i/(m_{ge}g_s\omega)$ to the point $-i/(m_{ge}g_s\omega)$ passing the origin of the $\tau$ plane from the left. Henceforth, we use the convenient notation
\begin{equation}
    m_{ge}:=\frac{\sqrt{2\la_2}}{\pi g_s},\qquad m_{gm}:=\frac{\sqrt{H^2-2\la_1}}{\pi g_s},
\end{equation}
and also $g_s:=(g_\mu g^\mu)^{1/2}$. Notice that $m^2_{ge}$ and $m^2_{gm}$ are not expressible in a covariant way in terms of the metric $g_{\mu\nu}$, its curvature, and their contractions. This can be easily checked by considering $m^2_{ge}$ and $m^2_{gm}$ at the origin of the system of coordinates we work. The expressions for $m^2_{ge}$ and $m^2_{gm}$ do not depend on $\bar{g}_{ij}$, whereas any scalar possessing the dimension $m^2$ constructed solely in terms of the metric does depend on $\bar{g}_{ij}$ and its derivatives of the second and higher orders taken at the origin of this frame.

In virtue of the assumption on the structure of singularities of $G(\tau)$, the heat kernel can be developed as a convergent series in $\tau$ on the contour $C_1$:
\begin{equation}\label{HKE_converg}
    \Tr e^{-\tau \tilde{H}(\omega)}=e^{i\pi d/2}\sum_{k=0}^\infty\int \frac{d\spx\sqrt{\bar{g}}}{(4\pi)^{d/2}} \bar{a}_k(\omega,x)\tau^{k-d/2}e^{-\tau(g^2\omega^2-m^2)},
\end{equation}
where we have saved out the exponential factor from the series. The expression standing in the exponent gives the major contribution in the short-wave approximation and is determined by this condition unambiguously. Also if we redefine the variable $\omega\rightarrow m\omega$ and take into account that $mL$ is a very large quantity then, at the leading order, we come to the expression standing in the exponent again. So it is that form which is suitable for the large mass or the short-wave expansions. As a result, the contribution from the contour $C_1$ becomes
\begin{multline}\label{C1_contrib}
    \int_{C_1}\frac{d\tau\tau^{\nu-1}}{2\pi i}\Tr e^{-\tau\tilde{H}(\omega)}=e^{i\pi d/2}\sum_{k=0}^\infty\int \frac{d\spx\sqrt{\bar{g}}}{(4\pi)^{d/2}} \bar{a}_k(\omega,x)\times\\
    \times\Big[\int_C\frac{d\tau}{2\pi i}\tau^{k+\nu-1-d/2}e^{-\tau(g^2\omega^2-m^2)} -\int_{\bar{C}_1}\frac{d\tau}{2\pi i}\tau^{k+\nu-1-d/2}e^{-\tau(g^2\omega^2-m^2)}\Big].
\end{multline}
The contribution of the first term in the square brackets provides the stepless part of $\tilde{\zeta}_+(\nu,\omega)$. It is this contribution which was found in \cite{KalKaz}. The second term in the square brackets leads to an appearance of the essentially quantum oscillating contributions \eqref{oscil_corr} to $\tilde{\zeta}_+(\nu,\omega)$. These contributions and the contributions coming from the second term in \eqref{contour_partit} were not taken into account in \cite{KalKaz}. Notice that the series in \eqref{C1_contrib} consisting of the contributions from the first term in the square brackets does not converge. It is an asymptotic series. Only does the inclusion of the exponentially suppressed at $m^2\rightarrow\infty$ terms (the contributions of the the second term in the square brackets) make the series \eqref{C1_contrib} convergent. A wrong impression may also arise that the contribution of the second term in the square brackets in \eqref{C1_contrib} cancels out the second term in \eqref{contour_partit}. This is not the case as the order of the summation over $k$ and integration over $\tau$ are not interchangeable in \eqref{C1_contrib}. Often, carrying out considerations on the so-called physical level of rigor, one disregards such mathematical ``subtleties''. However, they are relevant in our case. Further, we shall ascertain that these two contributions to $\tilde{\zeta}_+(\nu,\omega)$ and, correspondingly, to the free energy are represented by different expressions.

\paragraph{Contour $C_1$.}

Thus, we can distinguish three types of terms in $\tilde{\s}^l_\nu$: the two contributions \eqref{C1_contrib} from the contour $C_1$ and the contribution \eqref{contour_partit} from the contour $\bar{C}_1$. The first contribution coming from the integration along the contour $C$ was found in \cite{KalKaz}. Denoting by $\tilde{\tau}^l_\nu$ the contribution of the stepless part of $\zeta_+^0(\nu,\omega)$ to $\tilde{\s}^l_\nu$, we have (see (33) of \cite{KalKaz})
\begin{equation}\label{tau_tild}
\begin{split}
    \tilde{\tau}^l_\nu & =e^{i\pi\nu}\sum_{k,j=0}^\infty(-1)^k\int d\spx\frac{\sqrt{|g|}\tilde{a}_k^{(j)}}{(4\pi)^{d/2}}\frac{\Ga(k+\nu-(D+j+l)/2)}{2\Ga\big((1-j-l)/2\big)g_s^{l}}\tilde{m}^{D+j-2\nu-2k+l}=\\
    & = e^{i\pi\nu}\sum_{k,j=0}^\infty(-1)^k\int d\spx\frac{\sqrt{|g|}\bar{a}_k^{(j)}}{(4\pi)^{d/2}}\frac{\Ga(k+\nu-(D+j+l)/2)}{2\Ga\big((1-j-l)/2\big)g_s^{l}}m^{D+j-2\nu-2k+l},
\end{split}
\end{equation}
where $j\leq[4k/3]$. In the last equality, we have expanded the expression in $m^2$ instead of $\tilde{m}^2$. The introduction of the effective mass allows us to simplify drastically the calculation of higher terms of the heat kernel expansion, but the assignment of a physical sense to it is not always warrantable (see the footnote on p. \arabic{local}).

So, the two other contributions to $\tilde{\s}^l_\nu$ are left to calculate. Let us start with the second term in \eqref{C1_contrib}. Recovering the explicit dependence of the coefficients of the heat kernel expansion on $\omega$ as in \eqref{akj}, we see that we need to take the integral,
\begin{equation}
    \frac{e^{i\pi d/2}g_s^j}{(4\pi)^{d/2}}\int_0^\infty d\omega\omega^{l+j}\Big[\int_{\bar{C}^+_1}\frac{d\tau}{2\pi i} +\int_{\bar{C}^-_1}\frac{d\tau}{2\pi i}\Big] \tau^{k+\nu-1-d/2}e^{-\tau(g^2\omega^2-m^2)},
\end{equation}
in order to find the contribution to $\tilde{\s}^l_\nu$. The contour $\bar{C}^+_1$ goes to the point $i/(m_{ge}g_s\omega)$ and the contour $\bar{C}^-_1$ goes from the point $-i/(m_{ge}g_s\omega)$ so that the integrals over $\tau$ converge.

Consider the first integral. Rotate the integration contours in the $\omega$ and $\tau$ planes in such a way that during this rotation the integral remains convergent and its value is left unchanged. In the $\omega$ plane we rotate the integration contour clockwise by the angle $\pi/2$, while in the $\tau$ plane we rotate the contour counterclockwise such that it runs along the real axis from $-\infty$ to $-1/(m_{ge}g_s|\omega|)$ after the rotation. Recall that the function $\tau^{k+\nu-1-d/2}$ has a branch cut discontinuity for $\tau>0$. The integration contour in the $\tau$ plane does not intersect this branch cut during the rotation. Thus we come to
\begin{multline}
    \frac{e^{i\pi d/2}g_s^j}{(4\pi)^{d/2}}\int_0^\infty d\omega\omega^{l+j}\int_{\bar{C}^+_1}\frac{d\tau}{2\pi i} \tau^{k+\nu-1-d/2}e^{-\tau(g^2\omega^2-m^2)}=\\
    =(-i)^{l+j+1}\frac{(-1)^{k+1}}{2\pi i}\frac{e^{i\pi \nu}g_s^j}{(4\pi)^{d/2}}\int_0^\infty d\omega\omega^{l+j}(g^2\omega^2+m^2)^{d/2-\nu-k}\Gamma\Big(k+\nu-d/2,\frac{g^2\omega^2+m^2}{m_{ge}g_s\omega}\Big),
\end{multline}
where $\Ga(\al,x)$ is the incomplete gamma function. The last integral can be also obtained by a straightforward calculation. At that, it is convenient to add $-i0$ to the terms standing in the round brackets in the exponent. Analogously, for the second integral, we rotate counterclockwise the integration contour in the $\omega$ plane by the angle $\pi/2$, while in the $\tau$ plane we rotate the contour clockwise such that it goes along the real axis from $-\infty$ to $-1/(m_{ge}g_s|\omega|)$ after the rotation. Then we have
\begin{multline}
    \frac{e^{i\pi d/2}g_s^j}{(4\pi)^{d/2}}\int_0^\infty d\omega\omega^{l+j}\int_{\bar{C}^-_1}\frac{d\tau}{2\pi i} \tau^{k+\nu-1-d/2}e^{-\tau(g^2\omega^2-m^2)}=\\
    =i^{l+j+1}\frac{(-1)^k}{2\pi i}\frac{e^{i\pi \nu}g_s^j}{(4\pi)^{d/2}}\int_0^\infty d\omega\omega^{l+j}(g^2\omega^2+m^2)^{d/2-\nu-k}\Gamma\Big(k+\nu-d/2,\frac{g^2\omega^2+m^2}{m_{ge}g_s\omega}\Big).
\end{multline}
Adding the above two expressions and performing a change of the variable, we arrive at
\begin{equation}
    (-1)^ke^{i\pi \nu}\frac{\sin[\pi(l+j+1)/2]}{\pi(4\pi)^{d/2}g_s^{l+1}}m^{D-2\nu-2k+l+j}\int_0^\infty d\omega\omega^{l+j}(\omega^2+1)^{d/2-\nu-k}\Ga\Big(k+\nu-d/2,\frac{m}{m_{ge}}(\omega+\omega^{-1})\Big).
\end{equation}
Remark that this expression is equal to zero for $(l+j)$ odd. Moreover, since $\bar{a}_k^{(j)}=0$ for $j$ odd, this contribution vanishes for $l$ odd. Also we see from this formula that the contribution considered is essentially quantum one, because the asymptotics of the expression standing under the integral over $\omega$ (in fact, the contribution to $\tilde{\zeta}_+(\nu,i\omega)$) has the form \eqref{oscil_corr}. Now we employ the asymptotic expansion for the incomplete gamma function at large arguments (see, e.g., \cite{GrRy}),
\begin{equation}
    \Ga(\al,x)\approx x^{\al-1}e^{-x},\quad x\rightarrow+\infty,
\end{equation}
and make the substitution $\omega=e^t$. Then the integral is easily evaluated by the WKB method in the vicinity of the saddle point $t=0$. As a result, we obtain for the leading contribution
\begin{equation}
    (-1)^ke^{i\pi \nu}\frac{\sin[\pi(l+1)/2]}{(4\pi)^{D/2}g_s^{l+1}}m^{D-2\nu-2k+l+j}\sqrt{\frac{m_{ge}}{m}}e^{-2m/m_{ge}}.
\end{equation}
Collecting the factors remaining in \eqref{C1_contrib}, we write the contribution of the second term in \eqref{C1_contrib} to $\tilde{\s}^l_\nu$ as
\begin{equation}\label{sigma-2}
    \tilde{\s}^l_\nu\Big|_{II}\approx-e^{i\pi\nu}\sin\frac{\pi(l+1)}2\sum_{k,j=0}^\infty(-1)^k\int \frac{d\spx\sqrt{|g|}}{(4\pi)^{D/2}}g_s^{-l}\bar{a}^{(j)}_k(x)m^{D-2\nu-2k+l+j}\sqrt{\frac{m_{ge}}{m}}e^{-2m/m_{ge}},
\end{equation}
where $j$ is an even number and $j\leq[4k/3]$. The expression \eqref{sigma-2} is regular at $\nu=0$.

\paragraph{Contour $\bar{C}_1$ and the singular points.}

Now we have to evaluate the contribution to $\tilde{\s}^l_\nu$ from the second term in \eqref{contour_partit}. In calculating this contribution, we shall use the Gaussian approximation \eqref{HK_diag_Gauss} to $G(\tau)$. Usually, the contributions we are about to calculate are small in comparison with the main contribution to $\tilde{\s}^l_\nu$ from the stepless part of $\tilde{\zeta}_+(\nu,\omega)$. So, the use of $G_0(\tau)$ for the evaluation of such contributions is justified. The higher corrections to $G_0(\tau)$ like those found in \eqref{a_1_ap}-\eqref{c_ap} give even smaller contributions. The analysis of the asymptotics of $G_0(\tau)$ at large $\tau$ shows up that for
\begin{equation}
    \omega^2\tilde{g}^2+\omega\sqrt{H^2-2\la_1}-m^2>0,\qquad\tilde{g}^2:=g^2-\frac12bE^{-1}b,
\end{equation}
the integration contour $\bar{C}_1$ in the $\tau$ plane can be rotated to the right, whereas for
\begin{equation}
    \omega^2\tilde{g}^2-\omega\sqrt{H^2-2\la_1}-m^2<0,
\end{equation}
it can be rotated to the left. Inasmuch as we are interested in the high-temperature expansion at a large mass $m$ and $\omega\sim m$ after the stretching of the integration variable, it is convenient to rotate the contour to the right, when
\begin{equation}\label{omega_restr}
    \omega^2\tilde{g}^2-m^2>0,
\end{equation}
and to the left, otherwise. This condition is independent of $m^2$ after the integration variable has been stretched $\omega\rightarrow m\omega$. It is clear that the final answer, $\tilde{\zeta}_+(\nu,\omega)$ and $\tilde{\s}^l_\nu$, does not depend on a choice of the concrete condition on $\omega$, when the contour $\bar{C}_1$ is to be rotated to the right or left. It is only important that the integral over $\tau$ will be convergent.

When we rotate the integration contour $\bar{C}_1$ to the right, the singular points $\tau=in/(m_{gm}g_s\omega)$, where $n\in\mathbb{Z}$, also contribute to the integral over $\tau$. We shall consider the contribution of these points subsequently, but now we find the leading contribution to $\tilde{\s}^l_\nu$ from the integral along the contour $\bar{C}_1$ after the rotation to the left or to the right. As seen in Fig. \ref{cuts}, this integral is saturated in the neighbourhood of the boundary points $\pm i/(m_{ge}g_s\omega)$. At these points, we have
\begin{multline*}
    S\equiv\pm i\Big(\frac{m^2}{m_{ge}g_s\omega}-\frac{\omega\hat{g}^2}{m_{ge}g_s}\Big):=\pm i\Big[\frac{m^2}{m_{ge}g_s\omega}-\omega\Big(\frac{\tilde{g}^2}{m_{ge}g_s}+\\
    +\pi m_{ge}g_s\frac{b^2_{\parallel}}{2\la_2^2}\tnh\frac{\pi}2 +\pi m_{gm}g_s\frac{b^2_\perp}{2\la_1^2}\frac{\cos(\pi m_{gm}/m_{ge})-\cos(H/(m_{ge}g_s))}{\sin(\pi m_{gm}/m_{ge})} \Big) \Big],
\end{multline*}\\[-3em]
\begin{multline}\label{SSprime}
    S'\equiv m^2-\omega^2\bar{g}^2:=m^2-\omega^2\Big[\tilde{g}^2 +\frac{b^2_{\parallel}}{4\la_2^2}\frac{\pi^2m_{ge}^2g^2}{\ch^2(\pi/2)}+\\
    +\frac{b^2_\perp}{2\la_1^2}\pi^2m_{gm}^2g^2\frac{1-\cos(H/(m_{gm}g_s))\cos(\pi m_{gm}/m_{ge})-H/(\pi m_{gm}g_s)\sin(H/m_{ge}g_s)\sin(\pi m_{gm}/m_{ge})}{\sin^2(\pi m_{gm}/m_{ge})} \Big].
\end{multline}
Here $S=S_0+\tau m^2$ is the expression standing in the exponent in \eqref{HK_diag_Gauss}. The preexponential factor becomes at these points
\begin{equation}
    e^{\pm3\pi i/4}\Big(\frac{\pm i}{m_{ge}g_s\omega}\Big)^{\nu-1}\frac{\omega^{3/2}g_s^{3/2}m_{gm}m_{ge}^{1/2}}{8\sh^{1/2}\pi\sin(\pi m_{gm}/m_{ge})}.
\end{equation}
Now we ascertain that, in the weak field limit, the corrections of order $\al^2$ found in \eqref{a_1_ap}-\eqref{c_ap} give smaller contributions to $S$ than the terms written above. For the reader convenience, we provide here the orders in $\e$ for the structures appearing in the development of $S$. In the weak field limit, it follows from \eqref{estim_weak} that
\begin{equation}
    g_s\sim1,\qquad m_{ge}\sim m_{gm}\sim\e^{1/2},\qquad\la_1\sim\la_2\sim b_\perp\sim b_\parallel\sim\bar{R}\sim\e,\qquad H\sim\e^2.
\end{equation}
At the boundary points $\pm i/(m_{ge}g_s\omega)$, we have $t\sim\ell\sim1$ and so
\begin{equation}\label{correc_pert}
    \udg{0.7}{\diagai}\sim\e^{3/2}\omega,\qquad \udg{0.7}{\diagaii}\sim\e,\qquad \udg{0.7}{\diagav}\sim\e^{1/2}\omega^{-1},\qquad \udg{0.7}{\diagc}\sim\e^{1/2}\omega^{-1},
\end{equation}
where we have depicted the types of the diagrams only. The contributions of the terms proportional to $\partial_bf_{ca}$ are even smaller. The corrections \eqref{correc_pert} should be compared with the terms in \eqref{SSprime}. It is evident the contributions \eqref{correc_pert} can be neglected.

Further, in order to find the contribution to $\tilde{\s}^l_\nu$, we act in the same way as we did in calculating the contribution \eqref{sigma-2}. For the part of the integration contour $\bar{C}_1$ lying in the upper half-plane of the complex $\tau$ plane ($\bar{C}_1^+$), we make a transform $\omega\rightarrow e^{-i\vf/2}\omega$, putting eventually $\vf=\pi$, and the contour in the $\tau$ plane is deformed to part of the negative real axis $(-\infty,-1/(m_{ge}g_s|\omega|)]$. For the part of the integration contour $\bar{C}_1$ lying in the lower half-plane of the complex $\tau$ plane ($\bar{C}_1^-$), we perform the analogous rotation, but to the opposite direction. The integral over $\tau$ is evaluated by the WKB method for the boundary point. Therefore, we can safely forget about the singular points on the imaginary axis when evaluating this integral. Summing up the two expressions corresponding to the lower and upper parts of the contour $\bar{C}_1$, we obtain in the leading order
\begin{equation}
    \tilde{\s}^l_\nu\Big|_{III}=e^{i\pi\nu}\sin\frac{\pi(l+1)}{2}\int \frac{d\spx\sqrt{\bar{g}}}{8\pi}\frac{g_s^{5/2-\nu}m_{gm}m_{ge}^{3/2-\nu}}{\sh^{1/2}\pi\sin(\pi m_{gm}/m_{ge})} \int_0^\infty d\omega\omega^{5/2+l-\nu}\frac{e^{-(m^2/\omega+\omega\hat{g}^2)/(m_{ge}g_s)}}{m^2+\omega^2\bar{g}^2}.
\end{equation}
Note that this contribution is essentially quantum one as long as the spectral density corresponding to it behaves as \eqref{oscil_corr}. The mention also should be made that this contribution vanishes for $l$ odd and, in particular, for $l=-1$ (the contribution to the coefficient at $\be^{-1}$ in the high-temperature expansion for the bosonic case). The resulting integral over $\omega$ is saturated near the saddle point $\omega=m/\hat{g}$ provided $m\hat{g}/(m_{ge}g_s)\gg1$. Supposing that this inequality is fulfilled, we arrive in the leading order at
\begin{equation}\label{sigma-3}
\begin{split}
    \tilde{\s}^l_\nu\Big|_{III}&\approx e^{i\pi\nu}\sin\frac{\pi (l+1)}{2}\int \frac{d\spx\sqrt{|g|}}{8\pi^{1/2}}\frac{m^{1+l-\nu}}{g_s^{l}}\Big(\frac{g_s}{\hat{g}}\Big)^{l+4-\nu}\frac{m_{gm}m_{ge}^{2-\nu}}{1+\bar{g}^2/\hat{g}^2} \frac{e^{-2m\hat{g}/(m_{ge}g_s)}}{\sin(\pi m_{gm}/m_{ge})\sh^{1/2}\pi}\approx\\
    &\approx e^{i\pi\nu}\sin\frac{\pi (l+1)}{2} \int \frac{d\spx\sqrt{|g|}}{16\pi^{1/2}} \frac{g_s^{-l}m^{1+l-\nu}m_{gm}m_{ge}^{2-\nu}}{\sin(\pi m_{gm}/m_{ge})\sh^{1/2}\pi}e^{-2m/m_{ge}},
\end{split}
\end{equation}
where the last equality has been obtained under the assumption that $\e\ll1$ and the estimations \eqref{estim_weak} are satisfied, i.e., in the weak field limit. In this limit, one should set $g_s=1$. Nevertheless, we have kept the explicit dependence on $g_s$ to save the covariance of the expression under dilatations of the Killing vector: $\xi^\mu\rightarrow\la\xi^\mu$, $\la=const$. Obviously, the expression obtained is finite at $\nu=0$.

In the massive case, it remains to find the contribution of the singular points $\tau=in/(m_{gm}g_s\omega)$, when the inequality \eqref{omega_restr} holds. First of all, we redefine the integration variable $\tau$ as $\tau\rightarrow\omega^{-1}\tau$. Then, the singular points pass into $\tau=in/(m_{gm}g_s)$. The integrals in the vicinity of these singular points will be evaluated by the WKB method, which is justified in the case when the inequality \eqref{UV_strong} is satisfied (see \eqref{S_mag} below). The integration contour is depicted in Fig. \ref{cuts}. Also, in order to simplify the calculations, we assume that $\e\ll1$ and the estimations \eqref{estim_weak} are fulfilled. Then the expression standing in the exponent \eqref{HK_diag_Gauss} is well approximated in the neighbourhood of the singular points by the three leading terms of the Laurent series in the vicinity of these singular points (see Fig. \ref{cuts}). In particular, such a truncated expansion provides a good approximation for the integrand in the neighbourhood of the saddle points where the integral is saturated. The preexponential factor is replaced by the leading term of the Laurent series in the vicinity of the singular point. Thus we have
\begin{equation}\label{S_mag}
    S\approx i\Big(\frac{nm^2}{m_{gm}g_s\omega}-\frac{\omega n\tilde{g}^2}{m_{gm}g_s}-\pi m_{ge}g_s\omega \frac{b^2_{\parallel}}{2\la_2^2}\tnh(\frac{\pi n}{2}\frac{m_{ge}}{m_{gm}}) \Big)+x\Big(\frac{m^2}{\omega}-\omega g^2\Big)+\omega\frac{B^2_n}{x},
\end{equation}
where $x:=\tau-in/(m_{gm}g_s)$,
\begin{equation}
    B^2_n:=\frac{b^2_{\perp}}{2\la_1^2}\big[1-(-1)^n\cos(nH/(m_{gm}g_s))\big],
\end{equation}
and all the terms of the order of $\e$ and higher are thrown away. As seen from this expansion, the variable $x$ is of the order of unity in a neighbourhood of the saddle point. The preexponential factor in $\tilde{\s}^l_\nu$ expanded in a neighbourhood of the singular points reads as
\begin{equation}\label{preexp_mag}
    e^{i\pi\nu}e^{\pm i\pi(3-2\nu)/4}\frac{(-1)^n}{16\pi^2i}\frac{\omega^{3/2+l-\nu}g_s^{3/2-\nu}m_{gm}^{1-\nu}m_{ge}^{1/2}}{x|n|^{1-\nu}\sh^{1/2}(\pi|n|m_{ge}/m_{gm})},
\end{equation}
where the upper sign corresponds to positive $n$'s, the lower sign is for negative ones, and the first term of the Laurent series is only retained. The integral over $x$ is easily evaluated
\begin{equation}\label{Bessel_mag}
    \oint\frac{dx}{x}e^{-x\big(\omega g^2-\frac{m^2}{\omega}\big)+\omega\frac{B^2_n}{x}}=2\pi iJ_0\big(2B_n(g^2\omega^2-m^2)^{1/2}\big).
\end{equation}
As we see, this contribution to $\tilde{\zeta}_+(\nu,\omega)$ is essentially quantum one.

From the formulas \eqref{S_mag}, \eqref{preexp_mag}, and \eqref{Bessel_mag} we deduce that a pair of singular points corresponding to $\pm n$, $n>0$, makes a contribution to $\tilde{\s}^l_\nu$ of the form
\begin{multline}\label{sigma-npm}
    \tilde{\s}^l_\nu\Big|_{\pm n}\approx e^{i\pi\nu} \int \frac{d\spx\sqrt{\bar{g}}}{4\pi} \frac{(-1)^ng_s^{3/2-\nu}m_{gm}^{1-\nu}m_{ge}^{1/2}}{n^{1-\nu}\sh^{1/2}(\pi nm_{ge}/m_{gm})}\int^\infty_{m/\tilde{g}}d\omega\omega^{3/2+l-\nu}J_0\big(2B_n(g^2\omega^2-m^2)^{1/2}\big)\times\\
    \times\sin\Big[\omega\big(\frac{n\tilde{g}^2}{m_{gm}g_s} +\pi m_{ge}g_s \frac{b^2_{\parallel}}{2\la_2^2}\tnh(\frac{\pi n}{2}\frac{m_{ge}}{m_{gm}}) \big) -\frac{nm^2}{m_{gm}g_s\omega} +\frac\pi2\nu -\frac\pi4 \Big].
\end{multline}
Usually $b_\perp$ and, consequently, $B_n$ are rather small (see Eq. (48) of \cite{prop}). Therefore, we shall consider separately two cases, when (a) $B_n=0$, and (b)
\begin{equation}\label{extr_contrib_restr}
    \frac{b^2_\parallel}{2\la_2}+\frac{b_\perp^2}{2\la_1}\Big(1+\frac{2-H^2/\la_1}{\pi^2 n^2}\Big)<0,\qquad \frac{mm_{gm} B_n^2}{n}\gg1.
\end{equation}
Recall that $\la_1<0$. The origin of the conditions \eqref{extr_contrib_restr} becomes clear soon.

\begin{figure}[t]
\centering
\includegraphics*[width=0.4\linewidth]{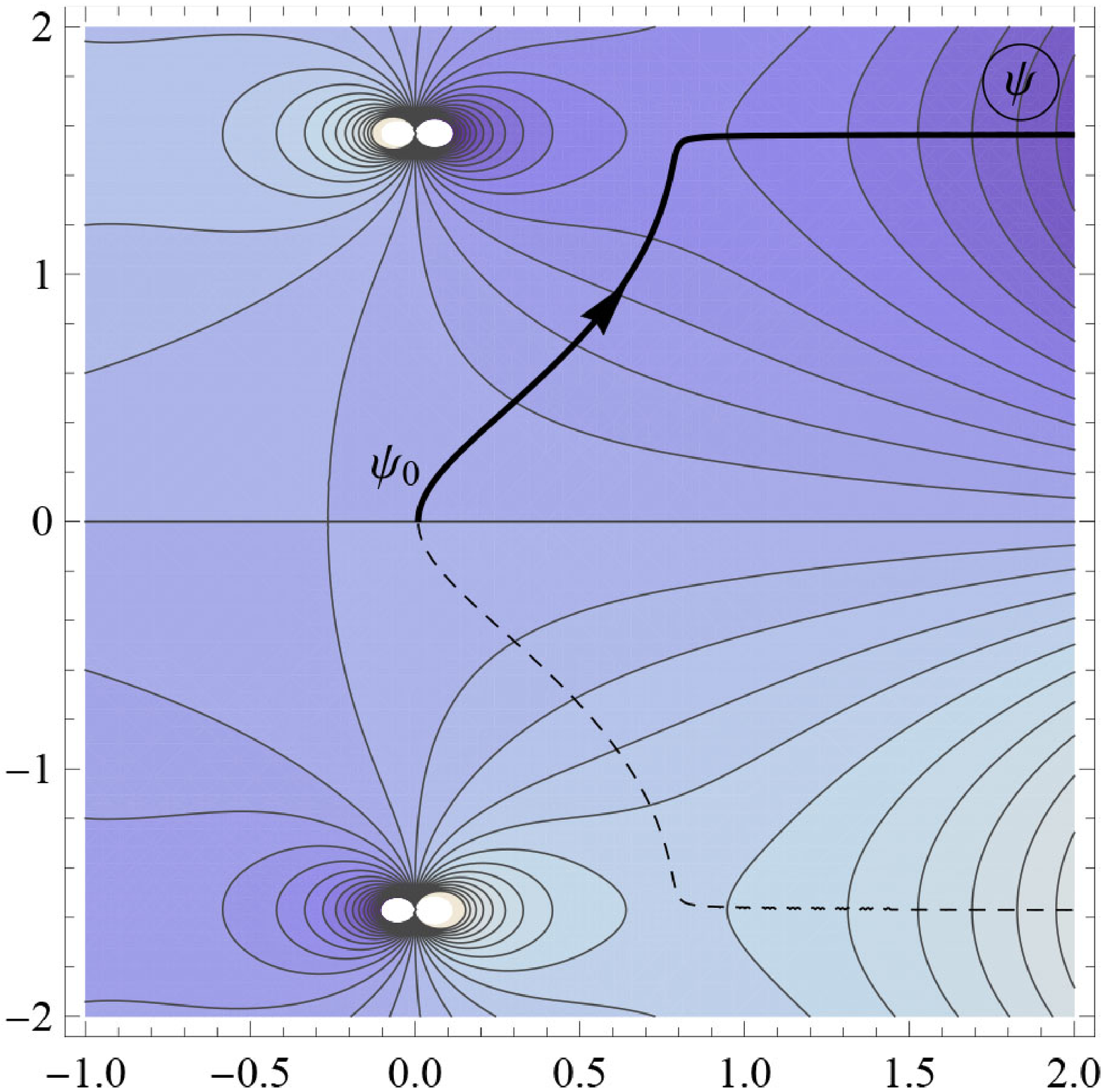}\,
\includegraphics*[width=0.4\linewidth]{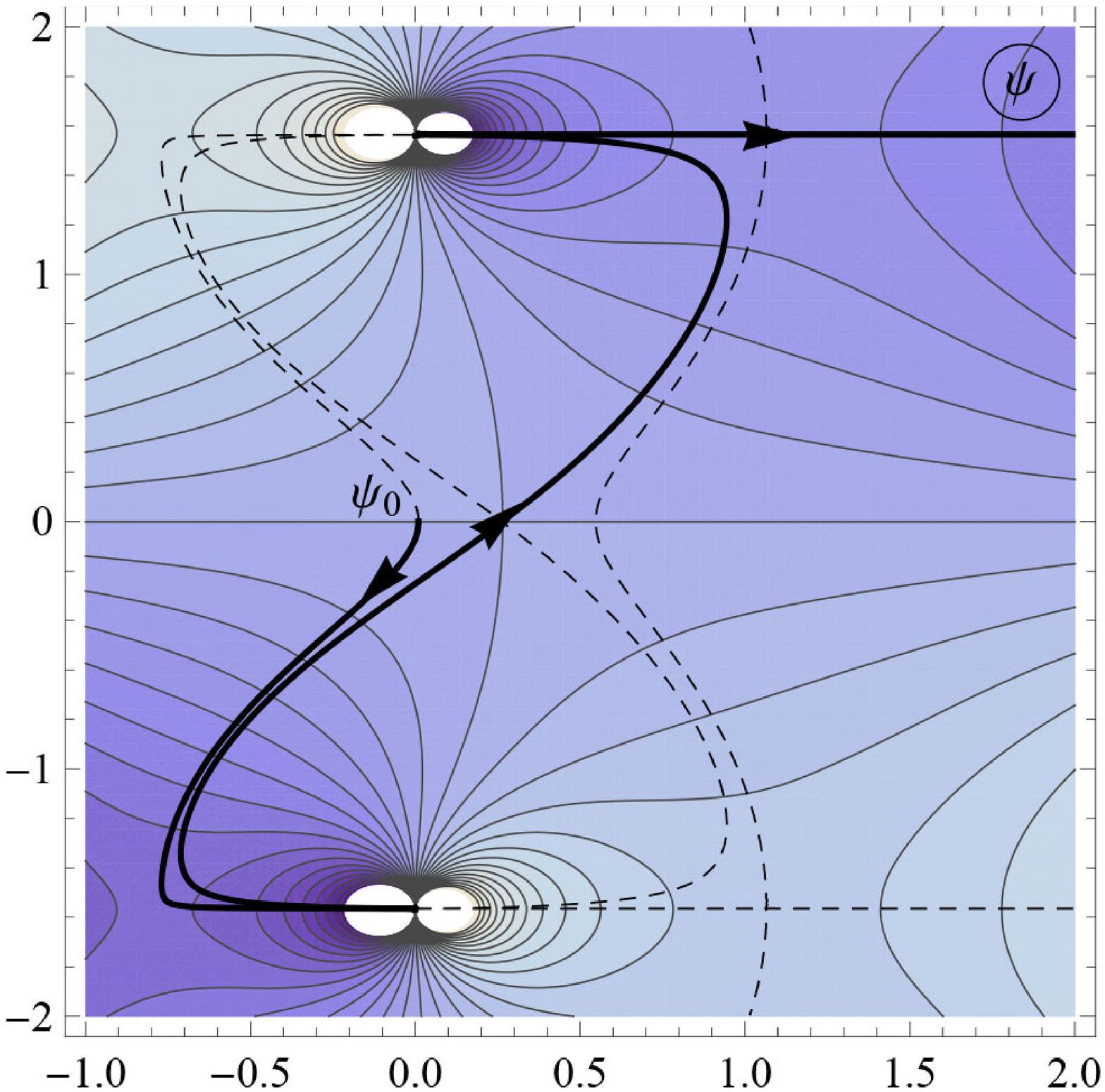}
\caption{{\footnotesize The level lines of the imaginary part of the function \eqref{exp_oscill} at $a_n=1$ and $\psi_0=10^{-2}$. The dashed lines are the lines of the steepest descent. The integration contours go along the lines of the steepest descent. Left panel: The function \eqref{exp_oscill} with the plus sign. Right panel: The function \eqref{exp_oscill} with the minus sign. Remark that (though it is not relevant for our calculations) the integration contour on the left plot does not pass through the saddle point. The integration contour on the right plot does not traverse the saddle point in the left bottom corner either.}}
\label{contours_ps}
\end{figure}

In the case (a), the major contribution to the integral over $\omega$ comes from the boundary point. The sine argument also possesses the stationary point, but its contribution is exponentially suppressed at $m/m_{gm}\gg1$ as compared to the contribution of the boundary point. The derivative of the sine argument calculated at the boundary point takes the form
\begin{equation}
    \frac{2n\tilde{g}^2}{m_{gm}g_s} +\pi mm_{ge}g_s \frac{b^2_{\parallel}}{2\la_2^2}\tnh(\frac{\pi n}{2}\frac{m_{ge}}{m_{gm}})\approx \frac{2n\tilde{g}^2}{m_{gm}g_s}.
\end{equation}
As a result, making use of the standard formula for the leading contribution from a boundary point and neglecting the corrections of the order of $\e$ in the preexponential factor, we have
\begin{equation}\label{sigma_mag_bound}
    \tilde{\s}^l_\nu\Big|_{\pm n}\approx e^{i\pi\nu} \int \frac{d\spx\sqrt{|g|}}{8\pi} \frac{(-1)^nm^{3/2+l-\nu}m_{gm}^{2-\nu}m_{ge}^{1/2}}{n^{2-\nu}\sh^{1/2}(\pi nm_{ge}/m_{gm})g_s^{l}}\cos\Big[\pi mm_{ge}\frac{g_s}{\tilde{g}} \frac{b^2_{\parallel}}{2\la_2^2}\tnh(\frac{\pi n}{2}\frac{m_{ge}}{m_{gm}})+\frac\pi2\nu-\frac\pi4\Big].
\end{equation}
In the case (b), let us employ the asymptotic formula for the Bessel function at large arguments \cite{GrRy}:
\begin{equation}
    J_0(x)\approx\sqrt{\frac{2}{\pi x}}\cos(x-\pi/4).
\end{equation}
Then, substituting $\omega=mg_s^{-1}\ch\psi$, the integral over $\omega$ is reduced to
\begin{equation}\label{int_psi}
    \int_{\psi_0}^\infty d\psi\sh^{1/2}\psi\ch^{3/2+l-\nu}\psi\cos(2B_nm\sh\psi-\pi/4)\sin\Big[\frac{mn}{m_{gm}}(a_n\ch\psi-1/\ch\psi) +\frac\pi2\nu-\frac\pi4\Big],
\end{equation}
where
\begin{equation}\label{an}
    \psi_0:=\arcch(g/\tilde{g}),\qquad a_n:=\frac{\tilde{g}^2}{g^2} +\frac{\pi m_{ge}m_{gm}}{n} \frac{b^2_{\parallel}}{2\la_2^2}\tnh(\frac{\pi n}{2}\frac{m_{ge}}{m_{gm}}).
\end{equation}
It is evident that the trigonometric functions in \eqref{int_psi} are highly oscillating. To analyze the saddle points, we write them via the exponents. This shows that the saddle points are located at the extremum points of the function
\begin{equation}\label{exp_oscill}
    \frac{mn}{m_{gm}}(a_n\ch\psi-1/\ch\psi)\pm2B_nm\sh\psi.
\end{equation}
The plot of this function is given in Fig. \ref{contours_ps}. This function has the extrema at the points
\begin{equation}
    \tnh\psi\approx\pm\sqrt{2},\qquad\psi\approx\pm\frac{m_{ge}B_n}{n}.
\end{equation}
In the latter case the signs are consistent with \eqref{exp_oscill}. It follows from the assumptions \eqref{extr_contrib_restr} that the extremum at the point $\psi=m_{ge}B_n/n$ falls into the integration interval over $\psi$ (the first condition in \eqref{extr_contrib_restr}) and is sufficiently sharp (the second condition in \eqref{extr_contrib_restr}). In that case, the contributions from the boundary point and other saddle points are small in comparison with the contribution of the extremum point $\psi=m_{ge}B_n/n$ (see the integration contour in Fig. \ref{contours_ps}). In particular, the contribution to the integral over $\psi$ from the exponent with \eqref{exp_oscill} (multiplied by an imaginary unit) taken with the plus sign is small as compared to the contribution of the analogous exponent, but with the minus sign in formula \eqref{exp_oscill}.

Developing \eqref{exp_oscill} as a series in $\psi$ up to the second order and keeping only the terms of the order of $\e^{1/2}$ and lower, we come to (the minus sign is taken)
\begin{equation}\label{S_quad}
  \frac{mn}{m_{gm}}\Big[a_n -1 -\frac{B_n^2m_{gm}^2}{n^2} \Big] +\frac{mn}{m_{gm}}(\psi-B_nm_{gm}/n)^2.
\end{equation}
Evaluating the Gaussian integral, the contribution to $\tilde{\s}^l_\nu$ becomes
\begin{equation}\label{sigma_mag_extr}
    \tilde{\s}^l_\nu\Big|_{\pm n}\approx e^{i\pi\nu} \int \frac{d\spx\sqrt{|g|}}{8\pi} \frac{(-1)^nm^{3/2+l-\nu}m_{gm}^{2-\nu}m_{ge}^{1/2}}{n^{2-\nu}\sh^{1/2}(\pi nm_{ge}/m_{gm})g_s^{l}}\sin\Big[\frac{mn}{m_{gm}}\Big(a_n -1 -\frac{B_n^2m_{gm}^2}{n^2} \Big)  +\frac\pi2\nu+\frac\pi4\Big].
\end{equation}
The total contribution of the singular points $\tau=in/(m_{gm}g_s)$ to $\tilde{\s}^l_\nu$ is obtained by summing the above expressions \eqref{sigma_mag_bound} or \eqref{sigma_mag_extr} over all natural $n$. The arising series in $n$,
\begin{equation}\label{sigma-gm}
  \tilde{\s}^l_\nu\Big|_{gm}:=\sum_{n=1}^\infty \tilde{\s}^l_\nu\Big|_{\pm n},
\end{equation}
converges for any $\nu\in \mathbb{C}$. At $\nu=0$ and $m_{ge}/m_{gm}>1$, the series is rapidly convergent and well approximated by its first term. Notice that, in deriving the expressions \eqref{sigma_mag_bound} and \eqref{sigma_mag_extr}, we neglected the exponentially suppressed terms as against the leading contribution. These disregarded terms may be of the order of or even larger than the exponentially suppressed terms in \eqref{sigma-2} and \eqref{sigma-3}. We are only interested in the leading non-perturbative contributions.

The corrections \eqref{a_1_ap}-\eqref{c_ap} can be estimated as follows. In the vicinity of the singular points $\tau=in/(m_{gm}g_s\omega)$, the most relevant contributions come from the terms in the $t$ and $\ell$ tensors that are the most singular at these points. For the tree and one-loop contributions containing $\bar{R}_{(acb)d}$, they are proportional to $x^{-2}$, while for the two-loop contributions with the $V_{2p}$ vertex they are proportional to $x^{-1}$. Hence, we have
\begin{equation}
    \udg{0.7}{\diagai}\sim\frac{\omega\bar{R}}{x^2m_{gm}g_s},\qquad \udg{0.7}{\diagaii}\sim \frac{\bar{R}}{x^2(m_{gm}g_s)^2}, \qquad \udg{0.7}{\diagav}\sim\frac{\bar{R}}{x\omega(m_{gm}g_s)^2}.
\end{equation}
Taking $x$ at the extremum point of \eqref{S_mag}, substituting $\omega=mg_s^{-1}\ch\psi$, and expanding the resulting expression in $\psi$, we find
\begin{equation}\label{alpha2_cntrb}
    \udg{0.7}{\diagai}\sim\frac{mn}{m_{gm}}\frac{\bar{R}\psi^2}{B_n^2n},\qquad \udg{0.7}{\diagaii}\sim\frac{mn}{m_{gm}}\frac{\bar{R}\psi^2}{B_n^2nm_{gm}m}, \qquad \udg{0.7}{\diagav}\sim\frac{\bar{R}\psi}{B_nm_{gm}^2m}.
\end{equation}
where the terms at most quadratic in $\psi$ have been only kept. These contributions should be compared with the terms in \eqref{S_quad} at the same powers of $\psi$. We see that the contributions \eqref{alpha2_cntrb} are negligible in the weak field limit for a large mass $m$. As for the rest contributions of the order $\al^2$ with the vertices $V_{1p}$ and $V_0$, they are much smaller than those we have considered.

\paragraph{High-temperature expansion.}

Thus, $\tilde{\s}^l_\nu$ can be written as
\begin{equation}
    \tilde{\s}^l_\nu=\tilde{\tau}^l_\nu+\tilde{\s}^l_\nu\Big|_{II}+\tilde{\s}^l_\nu\Big|_{III}+\tilde{\s}^l_\nu\Big|_{gm},
\end{equation}
where the items are presented in \eqref{tau_tild}, \eqref{sigma-2}, \eqref{sigma-3}, and \eqref{sigma-gm}. It is only the first term in $\tilde{\tau}^l_\nu$ which possesses singularities in the complex $\nu$ plane. The mention should be made that the non-perturbative contributions \eqref{sigma-2}, \eqref{sigma-3}, and \eqref{sigma-gm} in $\tilde{\s}^l_\nu$ cannot be developed as a Laurent series with a finite principal part in the inverse powers of $m^2$. The point $m^2=\infty$ is a transcendent branch point for these contributions. Having cast out the exponentially suppressed contributions at $\be\rightarrow0$, the high-temperature expansion takes the form
\begin{multline}\label{free_energy_b}
    -F_b=\sum_{k,j,n=0}^\infty \int d\spx\frac{\sqrt{|g|}\bar{a}_k^{(j)}}{(4\pi)^{d/2}} \frac{(-1)^{k}\Ga(D+j-2\nu-2k-2n)\zeta(D+j-2\nu-2k-2n)}{n!\Ga(d/2-\nu-k-n+1)\be_T^{D+j-2\nu-2k-2n}(-m^2)^{-n}e^{-i\pi\nu}}+\\
    +\sum_{l=-1}^\infty\frac{(-1)^{l}\zeta(-l)}{\Ga(l+1)}\tilde{\s}^l_\nu\be^{l},
\end{multline}
where $\be_T:=\sqrt{\xi^2}\beta$ is the Tolman reciprocal temperature, and the limit of a vanishing $\nu$ is implied. In accordance with the general properties of the high-temperature expansion investigated in Sec. \ref{HTE} and the direct calculations \cite{KalKaz}, $F_b$ is an entire function of $\nu$. Besides, as long as $e^{i\pi\nu}$ can be factored out in front of the expression regular at $\nu\rightarrow0$, this factor can be omitted in proceeding to the limit $\nu\rightarrow0$. Also note that the first and second terms in \eqref{free_energy_b} taken separately possess the poles in the $\nu$ plane. However, these divergencies are mutually canceled out. For the ``scalar fermions'' the analogous expansion reads as
\begin{multline}\label{free_energy_f}
    -F_f=\sum_{k,j,n=0}^\infty \int d\spx\frac{\sqrt{|g|}\bar{a}_k^{(j)}}{(4\pi)^{d/2}} (1-2^{2n+2k+2\nu-j-d})\frac{(-1)^{k}\Ga(D+j-2\nu-2k-2n)\zeta(D+j-2\nu-2k-2n)}{n!\Ga(d/2-\nu-k-n+1)\be_T^{D+j-2\nu-2k-2n}(-m^2)^{-n}e^{-i\pi\nu}}+\\
    +\sum_{l=0}^\infty(1-2^{1+l})\frac{(-1)^{l}\zeta(-l)}{\Ga(l+1)}\tilde{\s}^l_\nu\be^{l}.
\end{multline}
In fact, one just needs to add the non-perturbative corrections we found to the high-temperature expansions given in Sec. 7 of \cite{KalKaz}. At that, the formulas of \cite{KalKaz} should be rewritten in terms of $m^2$ rather than the effective mass squared $\tilde{m}^2$. In particular, $\ln\tilde{m}^2$ must be expanded in the inverse powers of the mass squared as
\begin{equation}
    \ln\tilde{m}^2=\ln m^2+O(m^{-2}).
\end{equation}
Then the explicit expressions for the finite and divergent parts of the high-temperature expansion presented in \cite{KalKaz} are left unchanged save $\ln\tilde{m}^2$, which should be replaced by $\ln m^2$.

One of the interesting consequences of the result we have obtained is that we explicitly separated the perturbative and non-perturbative contributions to the effective action at zero temperature, the convergent (in terms of the Feynman diagrams) perturbative corrections standing at the negative powers of the mass squared being independent of the Killing vector and expressed through the metric only. Indeed, the one-loop nonrenormalized contribution of one bosonic degree of freedom to the effective action at zero temperature can be obtained from \eqref{free_energy_f} by the use of the formula \cite{olopq,KalKaz}
\begin{equation}
    \Ga^{(1)}_{1b}/T=-\lim_{\be\rightarrow0}\partial_\be(\be F_f),
\end{equation}
where $\be^{-1}$ plays the role of a cutoff parameter and $T$ is the time interval tending to infinity. Upon cancelation of singularities in the $\nu$ plane, the divergent and finite parts of the high-temperature expansion of $F_f$ for $D$ even have the form (see (42) of \cite{KalKaz})
\begin{multline}\label{F_f}
    -F_f=\int \frac{d\spx\sqrt{|g|}}{(4\pi)^{d/2}}\sum_{k,j=0}^\infty(-1)^k\Big\{\sideset{}{'}\sum_{n=0}^\infty(1-2^{2k+2n-j-d})\frac{\Ga(D+j-2k-2n)\zeta(D+j-2k-2n) \bar{a}_k^{(j)}}{n!\Ga(d/2-k-n+1)\be_T^{D+j-2k-2n}(-m^2)^{-n}}\\
    -\frac{(-m^2)^{(D+j)/2-k}\bar{a}_k^{(j)}}{4((D+j)/2-k)!\Ga((1-j)/2)} \big[\ln\frac{4m^2\be_T^2e^{2\ga}}{\pi^2}-\psi((D+j)/2-k+1)+\psi((1-j)/2)\big]\\
    +\bar{a}_k^{(j)}m^{D+j-2k}\frac{\Ga(k-(D+j)/2)}{4\Ga((1-j)/2)}\Big\}+\frac12\big(\tilde{\s}^0_0\Big|_{II}+\tilde{\s}^0_0\Big|_{III}+\tilde{\s}^0_0\Big|_{gm}\big),
\end{multline}
where $\psi(x):=\Ga'(x)/\Ga(x)$ and $\ga$ is the Euler constant. The prime at the sum over $n$ says that the singular terms are discarded, the second term is zero by definition when the argument of the factorial is negative, and the last term in curly brackets vanishes by definition when the gamma function entering the numerator tends to infinity. In Sec. \ref{DescForm}, we proved that the coefficient in front of $\ln\be$ in \eqref{F_f} and the terms at the negative powers of $m^2$ (the last term in the curly brackets in \eqref{F_f}) do not depend explicitly on the Killing vector and are expressed solely in terms of the metric.

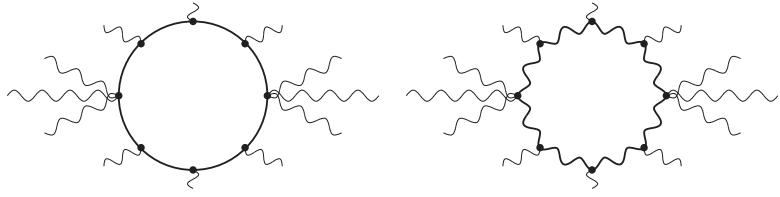
\begin{figure}[t]
\centering
\scalebox{0.7}{\hbox{%
\begin{picture}(200,100)(50,0)
\SetWidth{1}
\CArc(150,50)(40,0,180)
\CArc(150,50)(40,180,360)
\SetWidth{0.5}
\Photon(50,50)(110,50){3}{4}
\Photon(70,70)(110,50){3}{3}
\Photon(70,30)(110,50){-3}{3}
\Vertex(110,50){2}
\Photon(190,50)(250,50){3}{4}
\Photon(190,50)(230,70){-3}{3}
\Photon(190,50)(230,30){3}{3}
\Vertex(190,50){2}
\Photon(150,90)(150,100){3}{1}
\Vertex(150,90){2}
\Photon(150,10)(150,0){3}{1}
\Vertex(150,10){2}
\Photon(178,78)(198,88){3}{2}
\Vertex(178,78){2}
\Photon(122,78)(102,88){-3}{2}
\Vertex(122,78){2}
\Photon(122,22)(102,12){3}{2}
\Vertex(122,22){2}
\Photon(178,22)(198,12){-3}{2}
\Vertex(178,22){2}
\end{picture}}}\quad
\scalebox{0.7}{\hbox{%
\begin{picture}(200,100)(50,0)
\SetWidth{1}
\PhotonArc(150,50)(37,17,197){3}{8}
\PhotonArc(150,50)(37,197,377){3}{8}
\SetWidth{0.5}
\Photon(50,50)(110,50){3}{4}
\Photon(70,70)(110,50){3}{3}
\Photon(70,30)(110,50){-3}{3}
\Vertex(110,50){2}
\Photon(190,50)(250,50){3}{4}
\Photon(190,50)(230,70){-3}{3}
\Photon(190,50)(230,30){3}{3}
\Vertex(190,50){2}
\Photon(150,90)(150,100){3}{1}
\Vertex(150,90){2}
\Photon(150,10)(150,0){-3}{1}
\Vertex(150,10){2}
\Photon(178,78)(198,88){3}{2}
\Vertex(178,78){2}
\Photon(122,78)(102,88){-3}{2}
\Vertex(122,78){2}
\Photon(122,22)(102,12){3}{2}
\Vertex(122,22){2}
\Photon(178,22)(198,12){-3}{2}
\Vertex(178,22){2}
\end{picture}}}
\caption{{\footnotesize The typical diagrams contributing to the one-loop effective action $\Ga[g_{\mu\nu}]$. The external lines are just the labels and do not correspond to propagators. The internal wavy lines depict the graviton propagators, while the solid lines are for the scalar field propagators. The ghosts also contribute to the one-loop effective action. The diagrams for these contributions look like the diagrams with gravitons on the right, but with the replacement of the graviton loop by the ghost one. It is clear from the diagrams that the contributions with the graviton and ghost loops presented in the right figure are independent of $m^2$. Other one-loop contributions are absent for a vanishing background scalar field. Formula \eqref{F_f} contains only the contributions of the diagrams depicted in the left figure.}}
\label{diag_1loop}
\end{figure}

The formula \eqref{F_f} allows us to understand why we do not ``see'' the dependence on the Killing vector when evaluating the effective action using the standard perturbation theory on a flat background. The sum of all the diagrams presented on the left picture in Fig. \ref{diag_1loop} corresponds exactly to the terms in the curly brackets in \eqref{F_f}. Despite the fact that the terms at the nonnegative powers of $m^2$ does explicitly depend on the Killing vector, they can be completely canceled out by the appropriate counterterms added to the initial action (see \cite{KalKaz} for details). In the framework of the dimensional regularization, this cancelation happens automatically, whereas for the other regularization schemes this cancelation is achieved by requiring the fulfillment the Ward identities for the vertex functions. The terms standing at the negative powers of $m^2$ correspond to the convergent contributions of the perturbation theory, are unambiguously evaluated, and can be represented in a general covariant form in terms of the metric. On the other hand, the last three terms in \eqref{F_f} cannot be reproduced at any finite order of the perturbation theory on a flat background. In order to reveal these non-perturbative terms, one has to sum the asymptotic series of perturbation theory in some way. This, in essence, is equivalent to the introduction of the background field. In a certain sense, one may say that the effective action is nonanalytic in the vicinity of the point $g_{\mu\nu}=\eta_{\mu\nu}$.

One could think, keeping in mind the analogy with the vector gauge fields in the presence of their sources, that the components of the metric $g_{0\mu}$ can be expressed through the components of the energy-momentum tensor with the help of the constraints. Obviously, even if such a situation took place, the problem of the explicit dependence on the Killing vector would not be solved since, in this case, we would also need some extra structure distinguishing the time components of the energy-momentum tensor. One could also hope that the non-perturbative contribution found would be canceled out if we took into account the contributions from the graviton sector (see Fig. \ref{diag_1loop}). However, this is not the case in the one-loop approximation. For a vanishing background scalar field, the non-perturbative corrections depend nontrivially on the mass $m$, whereas the one-loop contributions from the graviton sector are independent of the mass of a scalar field. The non-perturbative corrections obtained cannot be canceled by the counterterms either, inasmuch as these corrections describe the actual physical phenomena similar to the Landau oscillations at a nonzero chemical potential \cite{LandOsc,VshiKli,Shoenb,LifshKag,LandLifshStat,LandLifstat} or the vacuum polarization effects \cite{Schwing,HeisEul} in quantum electrodynamics. Also recall that, in the standard model,  $m^2$ is expressed in terms of the $SU(2)$ doublet $\phi$ of the Higgs field. The addition of the counterterm nonpolynomial in $\phi^2$ violates the renormalizability of the Higgs sector (even with the inclusion of the nonminimal term $R|\phi|^2$). Thus we conclude that the effective action at zero temperature (its non-perturbative part, at least) depends explicitly on the Killing vector and cannot be written solely in terms of the metric.

\subsubsection{Massless case}

\paragraph{Main contribution.}

Let us turn now to the massless case $m^2=0$. In this case, the inequality \eqref{omega_restr} is always satisfied and we can deform the integration contour as depicted in Fig. \ref{cuts}
\begin{equation}
    \int_C\frac{d\tau\tau^{\nu-1}}{2\pi i}\Tr e^{-\tau\tilde{H}(\omega)}=\Big[\int_{C_0}\frac{d\tau\tau^{\nu-1}}{2\pi i} +\int_{C_{ge}}\frac{d\tau\tau^{\nu-1}}{2\pi i} +\int_{C_{gm}}\frac{d\tau\tau^{\nu-1}}{2\pi i} \Big]\Tr e^{-\tau\tilde{H}(\omega)}.
\end{equation}
Then it is necessary to evaluate the contributions from the contours $C_0$, $C_{ge}$, and $C_{gm}$. We begin with the main contribution $C_0$. Substituting the heat kernel expansion, we obtain
\begin{equation}
    \tilde{\s}^l_\nu\Big|_{C_0}=\sum_{k=0}^\infty e^{i\pi d/2}\int \frac{d\spx\sqrt{\bar{g}}}{(4\pi)^{d/2}}\int_0^\infty d\omega\omega^l\bar{a}_k(\omega,x)\int_{C_0}\frac{d\tau}{2\pi i}\tau^{k+\nu-d/2-1}e^{-\tau\omega^2 g^2}.
\end{equation}
The integral over $\tau$ is the incomplete gamma function $\ga(\al,x)$. Therefore,
\begin{equation}
    \tilde{\s}^l_\nu\Big|_{C_0}=\sum_{k,j=0}^\infty \int \frac{d\spx\sqrt{\bar{g}}e^{i\pi \nu}}{\pi(4\pi)^{d/2}}\sin\pi(\nu-d/2)\bar{a}_k^{(j)}(x)\int_0^\infty d\omega\omega^l(g_s\omega)^{d-2\nu-2k+j}\gamma(k+\nu-d/2,g_s\omega/m_{ge}),
\end{equation}
where we have expanded $\bar{a}_k(\omega,x)$ in $\omega$ as in formula \eqref{akj}. The integral over $\omega$ converges in the strip of the complex $\nu$ plane
\begin{equation}
    \frac{D+j+l}{2}-k<\re\nu<\frac{d}2+j+l+1-k.
\end{equation}
This allows us to find unambiguously the function $\tilde{\s}^l_\nu|_{C_0}$ in the form of an analytic function of $\nu$. Making use of the formula \cite{GrRy},
\begin{equation}
    \int_0^\infty dxx^{\mu-1}\ga(\nu,x)=-\mu^{-1}\Ga(\mu+\nu),\qquad\re(\mu+\nu)>0,\quad\re\mu<0,
\end{equation}
which is easy to prove integrating by parts, we arrive at
\begin{equation}\label{sigma-0_msls}
    \tilde{\s}^l_\nu\Big|_{C_0}=\sum_{k,j=0}^\infty \int \frac{d\spx\sqrt{|g|}}{(4\pi)^{d/2}} \frac{e^{i\pi \nu}(-1)^{k+l+j+1}g_s^{-l}m_{ge}^{D-2\nu-2k+j+l}\bar{a}_k^{(j)}}{(D-2\nu-2k+j+l)\Ga(k+\nu-d/2-j-l)}.
\end{equation}

\paragraph{Contour $C_{ge}$.}

We shall calculate the contributions of the contours $C_{ge}$ and $C_{gm}$ by the steepest descent method supposing that the estimations \eqref{estim_weak} and the inequality \eqref{UV_strong} are fulfilled. Let us start with the contour $C_{ge}$. Upon stretching the integration variable, $\tau\rightarrow\omega^{-1}\tau$, the expression $S_0$ standing in the exponent in \eqref{HK_diag_Gauss} is written as
\begin{equation}
    S_0=\frac{b^2_\parallel}{\la_2^2x}-xg^2 -\frac{2n+1}{m_{ge}g_s}\tilde{g}^2 +\pi g_sm_{gm}\frac{b^2_\perp}{2\la_1^2}\frac{\ch(\pi(2n+1)m_{gm}/m_{ge})-\ch\big((2n+1)H/(m_{ge}g_s)\big)}{\sh(\pi(2n+1)m_{gm}/m_{ge})},
\end{equation}
in the neighbourhood of the singular points. Here, $x=\tau-(2n+1)/(m_{ge}g_s)$ and we have discarded all the terms of the order $\e$ and higher. The extremum points are now readily found
\begin{equation}
    x_\pm=\pm i\frac{|b_\parallel|}{\la_2 g_s},\qquad S_0''|_{extr}=\pm 2i\omega g_s^3\frac{\la_2}{|b_\parallel|}.
\end{equation}
The value of $x_\pm$ is of the order of $\e^0$. The contribution of the two extremum points $x_\pm$ becomes
\begin{equation}
    \tilde{\s}^l_\nu\Big|_{\pm}=\int \frac{d\spx\sqrt{|g|}}{8\pi} \Big(\frac{m_{ge}g_s}{2n+1}\Big)^{1-\nu}\frac{e^{i\pi\nu}m_{gm}g_s}{\sh(\pi(2n+1)m_{gm}/m_{ge})}\int_0^\infty d\omega\omega^{1+l-\nu}\frac{e^{S_0(\tau_+)-i\pi\nu}-e^{S_0(\tau_-)+i\pi\nu}}{2i},
\end{equation}
where
\begin{equation}
    S_0(\tau_\pm)\approx -g_s\omega\Big(\frac{2n+1}{m_{ge}}\pm\frac{2i|b_\parallel|}{\la_2}\Big),
\end{equation}
and we have only kept the leading contribution in $\e$ in the preexponential factor. As we see, the contribution of the contour $C_{ge}$ to $\tilde{\zeta}_+(\nu,\omega)$ is essentially quantum one and has the form \eqref{oscil_corr}.

The integral over $\omega$ converges for $\re\nu<(l+2)$ and is easily taken. Then
\begin{equation}\label{sigma_pm}
    \tilde{\s}^l_\nu\Big|_{\pm}=\int \frac{d\spx\sqrt{|g|}}{8\pi g_s^l}\Big(\frac{m_{ge}}{2n+1}\Big)^{l+3-\nu}\frac{e^{i\pi\nu}m_{gm}\Ga(l+2-\nu)}{\sh(\pi(2n+1)m_{gm}/m_{ge})}\im\Big[e^{-i\pi\nu} \Big(1+\frac{2im_{ge}|b_\parallel|}{(2n+1)\la_2}\Big)^{\nu-l-2}\Big],
\end{equation}
where $\im$ acts on the expression in the square brackets just as the operation of taking the imaginary part for $\nu$ real. The total contribution of the contour $C_{ge}$ is obtained by summing \eqref{sigma_pm} over all natural numbers $n$:
\begin{equation}
    \tilde{\s}^l_\nu\Big|_{C_{ge}}=\sum_{n=1}^\infty\tilde{\s}^l_\nu\Big|_{\pm}.
\end{equation}
The mention should be made that, in contrast to the massive case, the essentially quantum contribution $\tilde{\s}^l_\nu$ possesses the ``infrared'' poles at $\nu=l+2,l+3,\ldots$. Proceeding to the limit of a vanishing $\nu$ and taking into account that the second term in the round brackets under the $\im$ sign in \eqref{sigma_pm} is small, we derive in the leading order
\begin{equation}\label{sigma-ge_msls}
\begin{split}
    \tilde{\s}^l_0\Big|_{C_{ge}} & = -\sum_{n=1}^\infty \int \frac{d\spx\sqrt{|g|}}{4\pi g_s^l} \frac{|b_\parallel|}{\la_2} \Big(\frac{m_{ge}}{2n+1}\Big)^{l+4}  \frac{m_{gm}(l+2)!}{\sh(\pi(2n+1)m_{gm}/m_{ge})}\\
    & = -(l+2)!\int \frac{d\spx\sqrt{|g|}}{4\pi g_s^l} \frac{|b_\parallel|}{\la_2} m_{gm}m_{ge}^{l+4}\Big[\phi_1\big(l+4,\pi \frac{m_{gm}}{m_{ge}}\big) -2^{-l-4}\phi_1\big(l+4,2\pi \frac{m_{gm}}{m_{ge}}\big)\Big],
\end{split}
\end{equation}
where we have introduced the function
\begin{equation}
    \phi_\mu(\nu,z):=\sum_{n=1}^\infty\frac{1}{n^\nu\sh^\mu(zn)}=2^\mu\sum_{k=0}^\infty\frac{\Ga(\mu+k)}{k!\Ga(\mu)}\Li_\nu(e^{-z(\mu+2k)}).
\end{equation}
The special function $\Li_\nu(z)$ is the polylogarithm which is defined as
\begin{equation}
    \Li_\nu(z)=\frac{z}{\Gamma(\nu)}\int_0^\infty dx\frac{x^{\nu-1}}{e^x-z},\qquad\Li_\nu(z)=\sum_{k=1}^\infty\frac{z^k}{k^\nu},\quad|z|<1.
\end{equation}
Using the Poisson summation formula, it is no difficult to find the asymptotic expansion of $\phi_\mu(\nu,z)$ for $z$ small:
\begin{equation}
\begin{split}
    \phi_\mu(\nu,z) = & \sum_p\frac{\Ga(\mu+\sum_kp_k)}{\Ga(\mu)}\zeta\Big(\mu+\nu-2\sum_kp_kk\Big)z^{-\mu+2\sum_kp_kk}\prod_k\frac{(-1)^{p_k}}{p_k!((2k+1)!)^{p_k}}\\
    &+z^{\nu-1}\int_0^\infty \frac{dxx^{-\nu}}{\sh^\mu x}+O(e^{-2\pi^2/|z|})=\\
    = &  z^{-\mu}\zeta(\mu+\nu) -\frac{\mu}{6}z^{2-\mu}\zeta(\mu+\nu-2)+\cdots +z^{\nu-1}\int_0^\infty \frac{dxx^{-\nu}}{\sh^\mu x}+O(e^{-2\pi^2/|z|}),
\end{split}
\end{equation}
where, in the first line, the summation is carried over all the collections $p=\{p_k\}$, $p_k=\overline{0,\infty}$, $k=\overline{1,\infty}$. The integral over $x$, for those values of $\mu$ and $\nu$ where it diverges, should be understood in the sense of an analytical continuation in these parameters. In particular, for $\mu=1$,
\begin{equation}
    \int_0^\infty\frac{dx x^{-\nu}}{\sh x}=(2-2^\nu)\Ga(1-\nu)\zeta(1-\nu)=\pi^{1-\nu}\frac{2^{1-\nu}-1}{\sin(\pi\nu/2)}\zeta(\nu).
\end{equation}

\paragraph{Contour $C_{gm}$.}

It is left to evaluate the contribution to $\tilde{\s}^l_\nu$ from the contour $C_{gm}$. From formula \eqref{sigma-npm} with $m^2=0$, we have for the contribution of a pair of singular points
\begin{multline}\label{sigma-pmn_msls}
    \tilde{\s}^l_\nu\Big|_{\pm n}\approx e^{i\pi\nu} \int \frac{d\spx\sqrt{\bar{g}}}{4\pi} \frac{(-1)^ng_s^{3/2-\nu}m_{gm}^{1-\nu}m_{ge}^{1/2}}{n^{1-\nu}\sh^{1/2}(\pi nm_{ge}/m_{gm})}\int^\infty_{0}d\omega\omega^{3/2+l-\nu}J_0(2B_ng_s\omega)\times\\
    \times\sin\Big[\omega g_s\big(\frac{n\tilde{g}^2}{m_{gm}g_s^2} +\pi m_{ge} \frac{b^2_{\parallel}}{2\la_2^2}\tnh(\frac{\pi n}{2}\frac{m_{ge}}{m_{gm}}) \big)  +\frac\pi2\nu -\frac\pi4 \Big].
\end{multline}
This contribution is evidently quantum one. The integral over $\omega$ is reduced to the Gauss hypergeometric function (\cite{GrRy}, 6.699)
\begin{equation}
    \int_0^\infty dxx^{\nu-1}\sin(ax+\vf)J_\la(bx)=\Big(\frac{b}2\Big)^\la a^{-\la-\nu}\frac{\Ga(\la+\nu)}{\Ga(\la+1)}\sin\big(\frac\pi2(\la+\nu)+\vf\big) F\big(\frac{\la+\nu}2,\frac{1+\la+\nu}2;\la+1;\frac{b^2}{a^2}\big),
\end{equation}
for $0<b<a$ and $-\re\la<\re\nu<3/2$. In our case, the integral over $\omega$ on the right-hand side of \eqref{sigma-pmn_msls} equals
\begin{equation}
    -\Ga(5/2+l-\nu)\sin\frac{\pi l}{2}\Big(\frac{m_{gm}}{g_sna_n}\Big)^{\nu-5/2-l}F\big(\frac{5+2l-2\nu}{4},\frac{7+2l-2\nu}{4};1;\frac{4m_{gm}^2B_n^2}{n^2a_n^2}\big),
\end{equation}
where $a_n$ is given in \eqref{an}. We see that the integral vanishes at $\nu=0$ and $l$ even. In particular, the contribution of the contour $C_{gm}$ to the finite part of the effective action ($l=0$, $\nu=0$) is equal to zero. The vanishing of this contribution at even $l$ is the exact property of the heat kernel \eqref{HK_diag_Gauss} for $m^2=0$, i.e., it holds not only for the weak field limit that we are considering now. Further, we shall be interested only in the leading contribution in $\e$. In this limit, $a_n$ and the hypergeometric function equal to unity. Then, \eqref{sigma-pmn_msls} is written as
\begin{equation}
    \tilde{\s}^l_\nu\Big|_{\pm n}\approx e^{i\pi\nu}\Ga(5/2+l-\nu)\sin\frac{\pi l}{2}\int \frac{d\spx\sqrt{|g|}}{4\pi g_s^l} \Big(\frac{m_{gm}}{n}\Big)^{7/2+l-2\nu} \frac{(-1)^{n+1}m_{ge}^{1/2}}{\sh^{1/2}(\pi nm_{ge}/m_{gm})},
\end{equation}
and the contribution of all the singular points on the imaginary axis of the $\tau$ plane takes the form
\begin{multline}\label{sigma-gm_msls}
    \tilde{\s}^l_\nu\Big|_{C_{gm}} = \sum_{n=1}^\infty\tilde{\s}^l_\nu\Big|_{\pm n}
    \approx e^{i\pi\nu}\Ga(5/2+l-\nu)\sin\frac{\pi l}{2}\times\\
    \times \int \frac{d\spx\sqrt{|g|}}{4\pi g_s^l} m_{ge}^{1/2}m_{gm}^{7/2+l-2\nu}\Big[\phi_{1/2}\big(l+\frac72-2\nu,\pi \frac{m_{ge}}{m_{gm}}\big)-2^{2\nu-l-5/2}\phi_{1/2}\big(l+\frac72-2\nu,2\pi\frac{m_{ge}}{m_{gm}}\big) \Big].
\end{multline}
Notice that the oscillations are not present in $\tilde{\s}^l_\nu$. This is a characteristic feature of a massless case at zero chemical potential (see, e.g., \cite{olopq}).

\paragraph{High-temperature expansion.}

In order to obtain the high-temperature expansion, we just need to substitute the expressions \eqref{sigma-0_msls}, \eqref{sigma-ge_msls}, and \eqref{sigma-gm_msls} to the general formula \eqref{free_energy_b} or \eqref{free_energy_f} and cancel the poles at $\nu\rightarrow0$, if they arise. For even $D$ such singularities do appear in the first terms of  \eqref{free_energy_b} and \eqref{free_energy_f} and in the contribution \eqref{sigma-0_msls} of the contour $C_0$ to $\tilde{\s}^l_\nu$. Consider separately these two contributions to the free energy. Bearing in mind that $\bar{a}_k^{(j)}=0$ for $j$ even and $j\leq[4k/3]$, we get for the divergent and finite parts of the high-temperature expansion in the bosonic case
\begin{equation}\label{F0b}
\begin{split}
    -F^0_b = & \int\frac{d\spx\sqrt{|g|}}{(4\pi)^{d/2}}\sideset{}{'}\sum_{k,j=0}^\infty(-1)^k\bar{a}_k^{(j)}\Big[\frac{\Ga(D+j-2k)\zeta(D+j-2k)}{\Ga(d/2-k+1)\be_T^{D+j-2k}} +\frac{m_{ge}^{d+j-2k}\be_T^{-1}}{(d+j-2k)\Ga(k-j-d/2+1)} \Big]+\\
    & +\int\frac{d\spx\sqrt{|g|}}{(4\pi)^{d/2}}\sum_{k=D/2}^{3D/2} \frac{(-1)^k\bar{a}_k^{(j)}}{4\Ga\big((1-j)/2\big)} \Big[\ln\frac{m_{ge}^2\be_T^2}{64\pi^2}+2\big(\psi((1-j)/2)+\ga+\ln4\big)\Big]_{j=2k-D}-\\
    & -(-1)^{D/2}\int\frac{d\spx\sqrt{|g|}}{(4\pi)^{d/2}}\sideset{}{'}\sum_{s=-D/2}^\infty\frac{m_{ge}^{-2s}}{4\pi s}\sum_{n=0}^{2s+D}\Ga(1/2-s+n)\bar{a}^{(2n)}_{s+D/2+n}+\cdots,
\end{split}
\end{equation}
where the omission points denote the terms at $\be$ in higher powers. The primes at the sum signs recall that all the singular terms arising at certain values of the indices $k$, $j$, and $s$ are thrown away. The term in the second line is deliberately written in such a form to make a comparison of \eqref{F0b} with (41) of \cite{KalKaz} simpler (the values of the digamma functions are presented in (43) of \cite{KalKaz}). Matching \eqref{F0b} with formula (41) of \cite{KalKaz}, it is easy to see that the first term in \eqref{F0b} is the same as the first term in (41) of \cite{KalKaz} expressed through $m^2$ and then $m^2$ set to zero. The coefficient at the logarithm in \eqref{F0b} coincides with the same coefficient in (41) of \cite{KalKaz} at $m^2=0$. The expression in the argument of the logarithm is obtained by the replacement $\tilde{m}^2\rightarrow m^2_{ge}e^{-2\ga}/4$. The second term in the second line of \eqref{F0b} equals two times the expression in the third line in (44) of \cite{KalKaz} expressed via $m^2$ (see (B21) of \cite{KalKaz}) at $m^2=0$, the terms at the negative powers of $m^2$ being absent here. The coefficient at $\be^{-1}_T$ and the last term in \eqref{F0b} are the expansions in derivatives of the metric and the Killing vector field. These expansions do not coincide with the analogous terms in (41) of \cite{KalKaz} on identifying $m^2_{ge}\rightarrow m^2$ (it is not surprising, of course). In particular, the coefficients at $m^2_{ge}$ in the last term of \eqref{F0b} are not expressible solely in terms of the metric. In the fermionic case, the formula analogous to \eqref{F0b} can be cast into the form
\begin{equation}\label{F0f}
\begin{split}
    -F^0_f = & \int\frac{d\spx\sqrt{|g|}}{(4\pi)^{d/2}}\sideset{}{'}\sum_{k,j=0}^\infty(-1)^k\bar{a}_k^{(j)}(1-2^{2k-j-d})\frac{\Ga(D+j-2k)\zeta(D+j-2k)}{\Ga(d/2-k+1)\be_T^{D+j-2k}} -\\
    & -\int\frac{d\spx\sqrt{|g|}}{(4\pi)^{d/2}}\sum_{k=D/2}^{3D/2} \frac{(-1)^k\bar{a}_k^{(j)}}{4\Ga\big((1-j)/2\big)} \Big[\ln\frac{m_{ge}^2\be_T^2}{4\pi^2}+2\big(\psi((1-j)/2)+\ga+\ln4\big)\Big]_{j=2k-D}+\\
    & +(-1)^{D/2}\int\frac{d\spx\sqrt{|g|}}{(4\pi)^{d/2}}\sideset{}{'}\sum_{s=-D/2}^\infty\frac{m_{ge}^{-2s}}{4\pi s}\sum_{n=0}^{2s+D}\Ga(1/2-s+n)\bar{a}^{(2n)}_{s+D/2+n}+\cdots.
\end{split}
\end{equation}
Recall that the coefficients $\bar{a}_k^{(j)}$ are expressed through the coefficients $\tilde{a}_k^{(j)}$ found in \cite{KalKaz} (see \eqref{ak_tild_ak_bar}). The total high-temperature expansion without the exponentially suppressed terms at $\be\rightarrow0$ becomes
\begin{equation}
    F_b=F_b^0-\sum_{l=-1}^\infty\frac{(-1)^l\zeta(-l)}{\Ga(l+1)}(\tilde{\s}^l_0\big|_{C_{ge}}+\tilde{\s}^l_0\big|_{C_{gm}})\be^l,\qquad F_f=F_f^0-\sum_{l=0}^\infty(1-2^{1+l})\frac{(-1)^l\zeta(-l)}{\Ga(l+1)}(\tilde{\s}^l_0\big|_{C_{ge}}+\tilde{\s}^l_0\big|_{C_{gm}})\be^l,
\end{equation}
where the non-perturbative contributions are presented in \eqref{sigma-ge_msls} and \eqref{sigma-gm_msls}.

In conclusion, we outline how the procedure above is changed when $m_{ge}<m_{gm}$. In that case, the expansion in $\tau$ of the heat kernel $G(\tau)$ is convergent in the disc $|\tau|<1/(m_{gm}g_s\omega)$. Therefore, it is convenient for both massive and massless cases to part the integration contour $C$ in \eqref{zeta+_nu} into two: $C_1$ passing from the singular point $\tau=i/(m_{gm}g_s\omega)$ to $\tau=-i/(m_{gm}g_s\omega)$, and the contour $\bar{C}_1$ being the rest part of the contour $C$ (see the structure of singularities in Fig. \ref{cuts}). The integral along the contour $C_1$ is evaluated in the same way as we did for the integrals along the contours $C_1$ in the massive case and $C_0$ for the massless one. The integration contour $\bar{C}_1$ is rotated in an appropriate way subject to the sign of the expression standing on the left-hand side of \eqref{omega_restr}. Then the integral over $\tau$ is calculated by the steepest descent method just as we did for the contour $\bar{C}_1$ and the contributions of the singular points on the imaginary axis of the $\tau$ plane. A detailed study of the expressions resulting from this procedure will be given elsewhere.

\section{Discussion}

The results that we obtained are valid within the bounds of the approximations made, when the Klein-Gordon equation on a curved background holds and the basic principles of QFT are kept. Using the background field gauge \cite{DeWGAQFT,BuchOdinShap}, one may expect on the basis of the perturbative analysis on a flat background that, in the infrared limit, the effective action $\Ga[g_{\mu\nu}]$ is expressed in a covariant form in terms of the metric, its curvature, and covariant derivatives constructed with the help of this metric. However, as we saw, this can be done only for the perturbative contributions after a suitable renormalization procedure (see the details of this procedure in \cite{DeWQFTcspt,KalKaz}). Non-perturbative corrections depend explicitly on the Killing vector, which defines the vacuum state and, consequently, the representation of the algebra of observables according to the GNS construction. This dependence is just a manifestation of the dependence of averages on the state with respect to which they are calculated. Of course, there could be a situation that the unique vacuum state is singled out by a certain natural local structure made of the metric $g_{\mu\nu}$ alone. In particular, for a stationary gravitational background this would mean that the Killing vector, or some its analog, should be constructed in terms of the spacetime curvature and its covariant derivatives. However, there is no such a natural construction (see, e.g., the discussion in \cite{Isham}). The Killing vector field is not expressed through the metric in a local form. Furthermore, if one tries to construct QFT on the background metric, which is ``produced'' by some extended massive object (say, a star) that at the initial moment moved uniformly and rectilinearly, then was accelerated by some external force, and after that became moving uniformly and rectilinearly, but with other velocity, then one comes almost inevitably to the conclusion that, under the assumption of locality and causality of QFT, the field $\xi^\mu$ must be dynamical.


The non-perturbative corrections we found are extremely small far from the ergosphere. For example, the leading non-perturbative contribution to the energy density in the massive case \eqref{sigma_mag_extr}, \eqref{sigma-gm} is of the order of
\begin{equation}
    \tilde{\s}^0_0\Big|_{gm}/V\sim5.5 \Big(\frac{m}{\text{GeV}}\Big)^{3/2}\times10^{-70}\text{GeV}^4,
\end{equation}
on the Earth surface. For comparison, the ``vacuum energy density'' associated with the cosmological constant is approximately equal to $10^{-47}$ GeV${}^4$. Nevertheless, the fact that the non-perturbative corrections cannot be expressed via the metric in a covariant way is very important. For comparison, one may recollect the non-perturbative contributions to the effective action of quantum electrodynamics. The Heisenberg-Euler effective action is invariant with respect to the $U(1)$ gauge transformations and expressible in terms of curvatures. On the other hand, there is another suggestive example in QFT. It is quantum chromodynamics (QCD) with the spontaneously broken chiral symmetry. At the present moment, the well established proof (starting from the QCD Lagrangian) of the existence of spontaneous breaking of the chiral symmetry and the description of the quark condensate properties are not given. However, the very assumption that this symmetry is spontaneously broken leads to a great number of nontrivial phenomenological consequences formalized in chiral perturbation theory (see, e.g., \cite{DonGolHol}). In particular, as long as the non-perturbative corrections to the effective action for gravity depend explicitly on the Killing vector, this vector field (or $g_\mu=\xi_\mu/\xi^2$) and its derivatives must appear in the initial Lagrangian in accordance with the general prescriptions of the effective field theories \cite{Weinb,DonGolHol} and renormalization theory (see, e.g., \cite{BogolShir,Collins}). Therefore, a complete cancelation of the perturbative quantum contributions to the effective action depending on the Killing vector by the appropriate counterterms is unnatural, at least.

At the same time, one should bear in mind that an immediate transfer of reasonings used in constructing chiral perturbation theory is illicit. For instance, the Goldstone theorem does not apply in the case we consider since it rests on the assumption that an open system tends to a minimum energy state (the ground state). It is clear that this principle does not work in general relativity so long as the energy operator is an ill-defined concept till the symmetry is broken, i.e., until a certain external structure defining the representation of the algebra of observables and the energy operator is introduced.

In general relativity it is natural to demand the fulfillment of the principle of general covariance from the in-in effective action $\Ga[g^\pm_{\mu\nu},\Phi^\pm,\ldots]$ (in the background field gauge) for all the fields including those defining the vacuum state and the representation of the algebra of observables. The omission points in the argument of the effective action denote these fields. The calculations presented in this paper and many others \cite{gmse,KalKaz,prop,FrZel,AndHisSam,BrOtPa,Page,GriMaMos,Howard,DeWQFTcspt,MamMostStar} suggest that the role of such fields is played by some quantum timelike vector field $\xi^\mu$, or the Tolman temperature one-form $g_\mu$ constructed in terms of this vector field. For stationary spacetimes, the average of this field should coincide with (or be close to) the Killing vector field. The requirement of general covariance appears to be sufficient to determine the dynamics of this field \cite{gmse}. In this case, the so-called background independence is nothing but the existence of the in-in effective action $\Ga[g^\pm_{\mu\nu},\Phi^\pm,g^\pm_\mu]$ satisfying the Ward identities following from the general covariance:
\begin{equation}\label{Ward_ident}
    \nabla^\mu_+\frac{\de\Ga[g^\pm_{\mu\nu},\Phi^\pm,g^\pm_\mu]}{\sqrt{|g_+|}\de g^+_{\mu\nu}}\approx0,
\end{equation}
where the approximate equality means the fulfilment of this equality on the solution to the equations of motion for the fields $\Phi^+$ and $g_\mu^+$. The same equality holds for the ``minus'' fields as well. The vacuum state of quantum fields and all their correlators are restored from $\Ga$. The dynamics of the fields $g_\mu^\pm$ are found from the Ward identities \eqref{Ward_ident}, i.e., in fact, from the self-consistency condition for the quantum Einstein equations. In particular, the equations of motion for the average field $g_\mu$ look like \eqref{Ward_ident}, but with the ``plus'' and ``minus'' fields identified upon variation. Note that, for a stationary spacetime, this last equation possesses the solution coinciding with the Killing vector \cite{gmse}, although other solutions to \eqref{Ward_ident} may also exist in this case. Therefore, the Killing vector field, which is usually taken to define the vacuum state, does not violate the Ward identities for the average field (see, e.g., \cite{FrZel}). The higher Ward identities, i.e., \eqref{Ward_ident} without identification of the ``plus'' and ``minus'' fields, can be satisfied only if the field $g_\mu$ is quantized.

As shown in \cite{gmse}, the equations \eqref{Ward_ident} for the field $g_\mu$ are the equations of a hydrodynamical type. It is interesting to note that the sound speed in this ``fluid'', i.e., the speed of propagation of small disturbances of the field $g_\mu$, is determined by the asymptotics of the energy-momentum tensor in the weak field limit \cite{gmse,prop,KalKaz} and independent of the details of the model. Of course, inasmuch as the field $g_\mu$ interacts with the metric $g_{\mu\nu}$, whose perturbations propagate with the velocity of light, the disturbances of the metric induce the perturbations of the field $g_\mu$ that propagate with the velocity of light too, as for the usual fluid in a gravitational field. The quantization of the field $g_\mu$ can be constructed employing the well-known relativistic Lagrangian for an Eulerian fluid (see, e.g., \cite{FockB,Taub,Schutz,Brown,HajKij,JNPP,rrmm}).

It would be interesting to investigate the quantum dynamics of the field $g_\mu$ in the weak field limit, when the dispersion law for its small perturbations is known, and obtain possible phenomenological restrictions on its vertices. The immediate generalization of the results of the paper is, of course, the development of a procedure, similar to the one presented in the paper, for higher spin fields at a nonzero chemical potential including the interaction with the background fields other than metric.

\paragraph{Acknowledgments.}

The work is supported in part by the Tomsk State University Competitiveness Program and by the RFBR grant No. 13-02-00551.

\appendix
\section{The proof of $\zeta_+(\nu,0)=0$}\label{Zeta0}

Let us show that if a vacuum is stable \eqref{vac_stab} then $\zeta_+(\nu,0)=0$, i.e., $H(0)$ does not possess positive eigenvalues. Let $a$ be a real parameter characterizing the background fields entering $H(\omega,a)$ such that at $a=0$ the operator $H(\omega,0)$ ceases to depend on the background fields and becomes the standard relativistic wave operator in the Minkowski space, while at $a=1$ we have the initial operator $H(\omega)\equiv H(\omega,1)$. The operator $H(0,0)$ has no positive eigenvalues. It is proven by the standard means (see \cite{Newton_scat}, Ch. 9) that $\e_k(\omega,a)$ is an analytic function of $\omega$ in the neighbourhood of the real axis and for complex $\omega$ may possess the branch points $\omega_0$, when
\begin{equation}
    \e_k(\omega_0,a)=\e_l(\omega_0,a),\qquad \psi_k(\omega_0,a)=\psi_l(\omega_0,a),
\end{equation}
where $\psi_k$ and $\psi_l$ are the eigenfunctions of the operator $H$ corresponding to the eigenvalues $\e_k$ and $\e_l$, respectively. In other words, the resolvent $(\e-H)^{-1}$ has multiple poles at these points. The analytic function $\e_k(\omega,a)$ obeys the Schwarz symmetry principle
\begin{equation}\label{Schwar_princ}
    \e^*_k(\omega,a)=\e_k(\omega^*,a).
\end{equation}

We shall be interested in the behaviour of the eigenvalues $\e_k(0,a)$ as functions of $a$. At that, we suppose that $\e_k(\omega,a)$ is a smooth function of $a$. Let the eigenvalue $\e_k(0,a)$ change its sign at $a=a_0\in(0,1)$:
\begin{equation}
    \e_k(0,a_0)=0,\qquad \partial_a\e_k(0,a)|_{a=a_0}>0, \qquad \partial_\omega\e_k(\omega,a_0)|_{\omega=0}=0,
\end{equation}
where the last equality follows from \eqref{vac_stab}. In a small vicinity of the point $a_0$, we deduce that
\begin{equation}\label{point_a_0}
    \omega_k \approx\pm i\sqrt{2(a-a_0)\partial_{a_0}\e_k(\omega,a_0)/\partial^2_{\omega_0}\e_k(\omega_0,a_0)},\qquad\partial^2_{\omega_0}\e_k(\omega_0,a_0)>0,
\end{equation}
where $\omega_k$ is the solution to the equation $\e_k(\omega,a)=0$. As we see, the vacuum becomes unstable at $a>a_0$ and $\e_k(0,a)>0$.

Now we have to trace the subsequent ``dynamics'' of the roots $\omega^{1,2}_k(a)$ of the equation $\e_k(\omega,a)=0$ with $a$ changing from $a_0$ to $1$. In virtue of the Schwarz symmetry, the complex root $\omega_k(a)$ always comes with its complex conjugate $\omega^*_k(a)$. Consequently, the roots $\omega_k^1(a)$, $\omega_k^2(a)$ leave the real axis colliding and scattering to the opposite directions along the line parallel to the imaginary axis. They return to the real axis in the reverse manner moving perpendicularly to the real axis (in its small neighbourhood) from the opposite sides, colliding, and scattering to the opposite directions along the real axis. Further, we assume that these two roots do not tend to infinity in increasing $a$ from $a_0$ to $1$ and also other roots of the equation $\e_k(\omega,a)=0$ do not appear in the complex $\omega$ plane. Otherwise, we call that the two operators (at $a=0$ and at $a=1$) are not smoothly deformable to each other.

In increasing $a$ from $a_0$ to $1$, different situations may occur. The roots $\omega^{1,2}_k(a)$ may stay complex, or they may come back to the real axis, or they may come back to the real axis and then leave it and so on. Let us distinguish the following cases, which are realized in increasing $a$ from $a_0$ to $1$:
\begin{enumerate}
  \item $\omega_k(1),\omega_k^*(1)\in \mathbb{C}$;
  \item $\omega_k^1(1),\omega_k^2(1)\in \mathbb{R}$ and the roots $\omega_k^1$, $\omega_k^2$ cross the point $\omega=0$ even times moving along the real axis;
  \item $\omega_k^1(1),\omega_k^2(1)\in \mathbb{R}$ and the roots $\omega_k^1$, $\omega_k^2$ cross the point $\omega=0$ odd times moving along the real axis.
\end{enumerate}
Denote as $N$ the number of crossings of the point $\omega=0$ by the roots $\omega_k^1$, $\omega_k^2$ moving along the real axis in increasing $a$ from $a_0$ to $1$. In the case, when the roots collide at the point $\omega=0$ and disperse along the imaginary axis just as at $a=a_0$ \eqref{point_a_0}, we count this as one crossing.

It follows from \eqref{Schwar_princ} that the quantity $\e_k(0,a)$ is real. The function $\e_k(0,a)$ changes its sign when one of the roots $\omega^{1,2}_k(a)$, or both at once, traverses the origin of the $\omega$ plane. Therefore,
\begin{equation}
    \sgn(\e_k(0,1))=(-1)^N.
\end{equation}
In the case $1$, the system is unstable at $a=1$ (since $\omega_k(1)$ is complex) and $N$ is an even number. Indeed, if the roots remain complex for $a\in(a_0,1]$ then $N=0$. If they returned to the real axis and then go out to the complex plane then, as follows from the properties of $\omega_{k}^{1,2}(a)$ discussed above, $N$ must be even. In the case $2$, the system is unstable since $\omega_k^1$ and $\omega_k^2$ lie on the one side from the point $\omega=0$ and $\e'(\omega_k^1)\e'(\omega_k^2)<0$, which contradicts \eqref{vac_stab}. The case $3$ is only left. Consequently, $N$ is an odd number provided the vacuum is stable at $a=1$, and so $\e_k(0,1)<0$ for any $k$. Thus we have shown that $\zeta_+(\nu,0)=0$ subject to the above mentioned assumption on a smooth deformability of $H(\omega,0)$ into $H(\omega,1)$ and the vacuum stability at $a=1$.

\section{Perturbation theory}\label{Pert_Theor}
\subsection{General formulas}

We shall use the condensed notations for the fields $\phi$:
\begin{equation}
    \mu:=(t,i),\qquad i:=(\spx,a),\qquad\de^i_j\equiv\de^a_b\de(\spx-\spy),
\end{equation}
where $a$ is a set of tensor indices, viz.,
\begin{equation}
    \phi^\mu\equiv\phi^i(t)\equiv\phi^a(x).
\end{equation}
In our case, the set $(x^i(\tau),p_j(\tau))$ plays the role of $\phi^\mu$. The field operators in different representations are labeled as
\begin{equation}
    \phi^i(t) - \text{Heisenberg},\qquad\phi^i_I(t) - \text{interaction picture},\qquad\phi^i - \text{Schr\"{o}dinger}.
\end{equation}
The relations among the representations ($H=H_0(t,\phi)+V(t,\phi)$):
\begin{equation}
\begin{gathered}
    \phi^i(t):=U_{0,t}\phi^iU_{t,0},\qquad\phi_I^i(t)=U^0_{0,t}\phi^iU^0_{t,0},\qquad\phi^i(t')=U_{0,t'}U_{t,0}\phi^i(t)U_{0,t}U_{t',0},\\
    \phi^i(t)=U_{0,t}U^0_{t,0}\phi_I^i(t)U^0_{0,t}U_{t,0}=S_{0,t}\phi_I^i(t)S_{t,0},
\end{gathered}
\end{equation}
where $U^0$ is the evolution operator corresponding to $H_0$ and the $S$-matrix is defined as
\begin{equation}
\begin{gathered}
    S_{t_2,t_1}=U^0_{0,t_2}U_{t_2,t_1}U^0_{t_1,0},\qquad U_{t_2,t_1}=U^0_{t_2,0}S_{t_2,t_1}U^0_{0,t_1},\\
    S_{t_3,t_2}S_{t_2,t_1}=S_{t_3,t_1},\qquad S^\dag_{t_2,t_1}S_{t_2,t_1}=S_{t_2,t_1}S^\dag_{t_2,t_1}=id.
\end{gathered}
\end{equation}
This operator satisfies the equation
\begin{equation}
    i\partial_t S_{t,t'}=V_I(t)S_{t,t'},\qquad S_{t,t}=id,\qquad S_{t,t'}=\Texp\Big[-i\int_{t'}^td\tau V_I(\tau)\Big],
\end{equation}
where $V_I(t)=V(t,\phi_I(t))$. The instant of time $t=0$ is arbitrary. In general, for each its choice there will be its own interaction and Heisenberg representations. All of them are related by a (formal) unitary transform.

Let us given the eigenvectors of the Schr\"{o}dinger Hamiltonian taken at the initial and final times
\begin{equation}
    H(t_{in})|in,t_{in}\ran=E_{in}|in,t_{in}\ran,\qquad H(t_{out})|out,t_{out}\ran=E_{out}|out,t_{out}\ran.
\end{equation}
Such a definition of the initial and final states is typical for QFT. Generally, the states $|in,t_{in}\ran$ and $|out,t_{out}\ran$ can be determined by any other way. All the subsequent formulas are left intact in this case. For example, in case of the perturbation theory for the matrix element of the evolution operator in the $x$ representation, the initial and final states are specified by the relations
\begin{equation}\label{xy_inout}
    \hat{x}^i|in,t_{in}\ran=y^i|in,t_{in}\ran,\qquad \hat{x}^i|out,t_{out}\ran=x^i|out,t_{out}\ran.
\end{equation}
Define the in- and out-states in the Heisenberg representation and the interaction picture as
\begin{equation}\label{vacua}
\begin{aligned}
    |in\ran:&=U_{0,t_{in}}|in,t_{in}\ran,&\qquad|out\ran:&=U_{0,t_{out}}|out,t_{out}\ran,\\
    |\overline{in}\ran:&=S_{t_{in},0}|in\ran=U^0_{0,t_{in}}|in,t_{in}\ran,&\qquad|\overline{out}\ran:&=S_{t_{out},0}|out\ran=U^0_{0,t_{out}}|out,t_{out}\ran.
\end{aligned}
\end{equation}
Then
\begin{equation}\label{evol_oper}
    \lan out,t_{out}|U_{t_{out},t_{in}}|in,t_{in}\ran=\lan \overline{out}|S_{t_{out},t_{in}}|\overline{in}\ran=\lan out|in\ran,
\end{equation}
and the Heisenberg averages calculated in the standard in-out perturbation theory take the form
\begin{equation}
    \lan out|T\{\phi(t_n)\cdots\phi(t_1)\}|in\ran=\lan \overline{out}|T\{\phi_I(t_n)\cdots\phi_I(t_1)S_{t_{out},t_{in}}\}|\overline{in}\ran.
\end{equation}

In order to construct the perturbation theory, let us define the generating functional of the free Green functions
\begin{multline}\label{gener_func_free}
    Z_0(J)=e^{iW_0(J)}=\lan \overline{out}|\Texp\Big\{i\int_{t_{in}}^{t_{out}}d\tau J_i(\tau)\phi^i_I(\tau)\Big\}|\overline{in}\ran=\\
    =\lan out,t_{out}|\Texp\Big\{-i\int_{t_{in}}^{t_{out}}d\tau[H_0(\tau,\phi)-J_i(\tau)\phi^i]\Big\}|in,t_{in}\ran.
\end{multline}
If $H_0(t,\phi)$ is quadratic in the fields $\phi^\mu$ (this will be assumed henceforward), the evolution operator in the last line can be explicitly calculated. It is not difficult to show that
\begin{equation}\label{Wick_thm}
    \frac{\de^3 W_0(J)}{\de J_\mu\de J_\nu\de J_\rho}\equiv0.
\end{equation}
Therefore, varying \eqref{gener_func_free}, we obtain
\begin{equation}\label{gener_func_free_expl}
    e^{iW_0(J)}=\lan \overline{out}|\overline{in}\ran e^{-\frac{i}2 J_\mu D^{\mu\nu} J_\nu+i\bar{\phi}^\mu J_\mu},
\end{equation}
where
\begin{equation}\label{phi_prop}
    \lan \overline{out}|\overline{in}\ran\bar{\phi}^\mu:=\lan\overline{out}|\phi_I^i(t)|\overline{in}\ran,\qquad\lan \overline{out}|\overline{in}\ran D^{\mu\nu}:=-i\lan\overline{out}|T\{(\phi_I^i(t_1)-\bar{\phi}^i(t_1))(\phi_I^j(t_2)-\bar{\phi}^j(t_2))\}|\overline{in}\ran.
\end{equation}
Using \eqref{gener_func_free_expl} and \eqref{evol_oper}, we derive for the matrix element of the evolution operator
\begin{equation}\label{evol_oper_pt}
\begin{split}
    \frac{\lan out|in\ran}{\lan \overline{out}|\overline{in}\ran} &= e^{-i\int_{t_{in}}^{t_{out}}d\tau V(\tau,-i\frac{\de}{\de J_i(\tau)}) } e^{-\frac{i}2 J_\mu D^{\mu\nu} J_\nu+i\bar{\phi}^\mu J_\mu}\Big|_{J=0}=\\
    &= e^{\frac{i}{2}\frac{\de}{\de\phi^\mu}D^{\mu\nu}\frac{\de}{\de\phi^\nu}+\bar{\phi}^\mu\frac{\de}{\de\phi^\mu}}e^{-i\int_{t_{in}}^{t_{out}}d\tau V(\tau,\phi(\tau))}\Big|_{\phi=0} = e^{\frac{i}{2}\frac{\de}{\de\phi^\mu}D^{\mu\nu}\frac{\de}{\de\phi^\nu}}e^{-i\int_{t_{in}}^{t_{out}}d\tau V(\tau,\phi(\tau))}\Big|_{\phi=\bar{\phi}}.
\end{split}
\end{equation}
One of the simplest ways to take the noncommutativity of the operators in the vertex $V(\phi)$ into account is to shift their arguments by infinitesimal quantities so that under the $T$-ordering they would stand in the proper order. This resolves the ambiguity of the expressions like $D^{ij}(t,t)$ arising in the perturbative computations.

The Feynman diagram technique is just a graphic representation of the right-hand side of \eqref{evol_oper_pt}. The easiest way to get it is to use the last formula in \eqref{evol_oper_pt}. From now on, we follow \cite{Vasil}. The following relation holds
\begin{equation}\label{Fd_formul}
    F\Big(\frac{\partial}{\partial\phi}\Big)\prod_{k=1}^nF_k(\phi)=F\Big(\sum_{k=1}^n\frac{\partial}{\partial\phi_k}\Big)\prod_{k=1}^nF_k(\phi_k)\Big|_{\phi_k=\phi}.
\end{equation}
This formula is obviously valid for the functions $F_k=\exp(p_k\phi_k)$ (no summation). All the other functions that can be Taylor expanded are obtained by a differentiation of $\prod_k\exp(p_k\phi_k)$ with respect to the parameters $p_k$, these parameters vanishing after a differentiation. The differentiations with respect to $\phi_k$ and $p_k$ commute.

The formula \eqref{Fd_formul} implies that
\begin{equation}
    e^{\frac{i}2\frac{\de}{\de\phi^\mu}D^{\mu\nu}\frac{\de}{\de\phi^\nu}}S_{int}^n(\phi)=\exp\Big\{\sum_{k=1}^n\frac{iD_{kk}}{2}+\sum_{k<l}iD_{kl}\Big\}S_{int}(\phi_1)\cdots S_{int}(\phi_n)\Big|_{\phi^\mu_k=\phi^\mu},
\end{equation}
where
\begin{equation}
    S_{int}:=-\int_{t_{in}}^{t_{out}}d\tau V(\tau,\phi(\tau)),\qquad D_{kl}:= \frac{\de}{\de\phi^\mu_k}D^{\mu\nu}\frac{\de}{\de\phi^\nu_l}.
\end{equation}
Developing the exponent as a series, we come to
\begin{equation}
    \exp\Big\{\sum_{k=1}^n\frac{iD_{kk}}{2}+\sum_{k<l}iD_{kl}\Big\}S^{int}(\phi_1)\cdots S^{int}(\phi_n)=\sum_\pi\prod_k\Big[\frac{(iD_{kk})^{\pi_{kk}}}{2^{\pi_{kk}}\pi_{kk}!}\Big] \prod_{k<l}\Big[\frac{(iD_{kl})^{\pi_{kl}}}{\pi_{kl}!}\Big]S^{int}_1\cdots S^{int}_n,
\end{equation}
where $\pi=\{\pi_{kl}, k\leq l\}$, $\pi_{kl}=\overline{0,\infty}$, $k,l=\overline{1,\infty}$, and $S^{int}_k\equiv S_{int}(\phi_k)$. The right-hand side of the last formula can be represented graphically as a sum of all possible graphs with $n$ numbered vertices, where $\pi_{kl}$ defines the number of lines\footnote{The graph can be identified with a symmetric matrix with nonnegative integer elements $\pi_{kl}$, the so-called adjacency matrix. Two numbered graphs are equal if their adjacency matrices are equal.} coming from the vertex $k$ to the vertex $l$, the vertices with $L$ lines correspond to the expressions
\begin{equation}\label{vertices}
    \frac{\de^L S^{int}_k}{\de\phi^{\mu_1}_k\cdots\de\phi^{\mu_L}_k},
\end{equation}
and the lines correspond to
\begin{equation}
    iD^{\mu\nu},
\end{equation}
the summation over all the identical Greek indices is implied. The coefficient at the diagram equals
\begin{equation}\label{coeff_diag}
    \Big[\prod_k\big(2^{\pi_{kk}}\pi_{kk}!\big) \prod_{k<l}\pi_{kl}!\Big]^{-1}.
\end{equation}
It is easy to see that this quantity does not depend on the way of numeration of the vertices. On identifying $\phi_k=\phi$, the vertices \eqref{vertices} become independent of the number $k$. Collecting all the identical diagrams with unnumbered vertices, we see that the coefficient at every the unnumbered diagram is given by \eqref{coeff_diag} multiplied by $n!/s$, where $n!$ is the number of permutations of vertices in the diagram and $s$ is the number of permutations of the numbered vertices in the graph that do not affect it (the symmetry factor). Thus, we have the following Feynman rules for the right-hand side of \eqref{evol_oper_pt}:
\begin{equation}\label{Feyn_rule}
    \text{line}=iD^{\mu\nu},\qquad\text{vertex}=\frac{i\de^L S_{int}}{\de\phi^{\mu_1}\cdots\de\phi^{\mu_L}}\Big|_{\phi=\bar{\phi}},\qquad\text{coefficient}=\Big[s\prod_k\big(2^{\pi_{kk}}\pi_{kk}!\big) \prod_{k<l}\pi_{kl}!\Big]^{-1},
\end{equation}
where $\pi_{kl}$, $k\leq l$, is the number of lines coming from the vertex $k$ to the vertex $l$ for some fixed (arbitrary) numeration of the diagram vertices.

The computation of higher orders of the perturbation theory can be simplified if we observe that
\begin{equation}\label{con_part}
    \ln\frac{\lan out|in\ran}{\lan \overline{out}|\overline{in}\ran}=\text{conn.part} \frac{\lan out|in\ran}{\lan \overline{out}|\overline{in}\ran},
\end{equation}
where ``conn.part'' means that only the connected diagrams with their coefficients are left in the expression. The connected diagram is a graph where one can pass moving along the lines from any vertex to any other vertex. A graph consisting of a single vertex is connected by definition. Let us prove the formula \eqref{con_part}. If
\begin{equation}
    \ln\frac{\lan out|in\ran}{\lan \overline{out}|\overline{in}\ran}=\sum_{a=1}^\infty c_aW_a,
\end{equation}
where $c_a$ is the numeric coefficient \eqref{Feyn_rule} of the connected diagram $W_a$, then
\begin{equation}\label{exp_con}
    \frac{\lan out|in\ran}{\lan \overline{out}|\overline{in}\ran}=e^{\sum_a c_aW_a}=\prod_a\sum_{m_a=0}^\infty \frac{(c_aW_a)^{m_a}}{m_a!}=\sum_{m}\prod_a \frac{c_a^{m_a}}{m_a!}W_a^{m_a},
\end{equation}
where the sum is performed over all the possible collections $m=\{m_a, a=\overline{1,\infty}\}$, $m_a=\overline{0,\infty}$. On the other hand, let us gather all the identical disconnected diagrams in the expansion of the right-hand side of \eqref{evol_oper_pt} containing the diagrams $W_1$, $\ldots$, $W_n$ as the connected parts that are replicated with the multiplicities $m_1$, $\ldots$, $m_n$. The symmetry factor for such a diagram is
\begin{equation}
    s=\prod_{a=1}^n m_a! \,s^{m_a}_{W_a},
\end{equation}
where $s_{W_a}$ is the symmetry factor for the graph $W_a$. The factorial in this formula corresponds to the number of permutations of the numbered vertices in $m_a$ copies of the graph $W_a$ (all the vertices of one copy are permuted with all the vertices of the other). Other factors in the coefficient \eqref{Feyn_rule} at the diagram fall into the product of multipliers corresponding to the connected components $W_a$. Therefore, the coefficient at such a diagram is equal to
\begin{equation}
    \prod_{a=1}^n\frac{c_a^{m_a}}{m_a!},
\end{equation}
where $c_a$ is the coefficient \eqref{Feyn_rule} of the diagram $W_a$. Comparing the derived expression with \eqref{exp_con}, we arrive at \eqref{con_part}.

Sometimes it is useful to make certain components of the fields $\phi^\mu\rightarrow\phi_a^i(t)$ explicit, where the index $a$ numbers the components. Then the Feynman rules can be cast into the form
\begin{equation}\label{Feyn_rule_multi}
\begin{gathered}
    \text{line}_{ab}=iD^{\mu\nu}_{ab},\qquad\text{vertex}=\frac{i\de^L S_{int}}{\de\phi^{\mu_1}_{a_1}\cdots\de\phi^{\mu_L}_{a_L}}\Big|_{\phi=\bar{\phi}},\\
    \text{coefficient}=\Big[s\prod_k\prod_a\big(2^{\pi^{aa}_{kk}}\pi^{aa}_{kk}!\big) \prod_{k}\prod_{a<b}\pi^{ab}_{kk}!\prod_{k<l}\prod_{a,b}\pi^{ab}_{kl}!\Big]^{-1},
\end{gathered}
\end{equation}
where $\pi^{ab}_{kl}$ is the number of lines of the type $ab$ connecting (directly) the vertices $k$ and $l$. The symmetry factor $s$ is calculated in the same way, but taking into account that the lines are of a different type now. The statement about the connectedness of the diagrams defining \eqref{con_part} remains valid. Also, in some cases, it is convenient to single out the ``mean field'' $\bar{\phi}$ explicitly and expand all the expression in it. It follows from the first formula in the second line of \eqref{evol_oper_pt} that, in this case, the additional ingredient in the Feynman rules arises
\begin{equation}
    \text{external line}=\bar{\phi}^\mu,
\end{equation}
and the coefficient at the diagram acquires the extra factor
\begin{equation}\label{coef_ext_line}
    \Big(\prod_k\pi_k!\Big)^{-1},
\end{equation}
where $\pi_k$ is the number of external lines joining to the vertex $k$. With this definition, the free endpoints of the external lines must not be considered as vertices in calculating the symmetry factor $s$. If the free endpoints of the external lines are regarded as vertices, the factor \eqref{coef_ext_line} ought not to be placed. It is reproduced automatically in evaluating $s$. It is clear that the statement about the connectedness of \eqref{con_part} holds true even after the inclusion of external lines.

\subsection{Correction of order $\al^2$}

Let us obtain, using the above general theory, the correction of order $\al^2$ to the diagonal matrix element of the evolution operator \eqref{heat_kern}. The Hamiltonian $H_0$ defined in \eqref{Ham_func} with the coefficients satisfying the relations \eqref{E_f} determines the averages $\bar{x}(\tau)$, $\bar{p}(\tau)$, the propagators, and the matrix element $\lan \overline{out}|\overline{in}\ran$, where the states $|\overline{in}\ran$ and $|\overline{out}\ran$ are specified in \eqref{xy_inout} and \eqref{vacua}. Solving the Heisenberg equations for the Hamiltonian $H_0$, which are equivalent to the corresponding Hamilton equations in this case, it is not difficult to find (see \cite{prop} for details)
\begin{equation}
\begin{split}
    x_I(\tau)=&\frac{\sin(\tau\omega\ka)}{\sin(s\omega\ka)}e^{(s-\tau)\omega f}(x_0+x)+\frac{\sin((s-\tau)\omega\ka)}{\sin(s\omega\ka)}e^{-\tau\omega f}(x_0+y)-x_0,\\
    p_I(\tau)=\frac12(\dot{x}_I+fx_I)=&\frac{\omega\ka}2\frac{\cos(\tau\omega\ka)}{\sin(s\omega\ka)}e^{(s-\tau)\omega f}(x_0+x)-\frac{\omega\ka}2\frac{\cos((s-\tau)\omega\ka)}{\sin(s\omega\ka)}e^{-\tau\omega f}(x_0+y)-\frac12\omega fx_0,
\end{split}
\end{equation}
where $\varkappa:=\sqrt{-f^2-2E}$, $x_0:=E^{-1}b$, $x=x_I(s)$, $y=x_I(0)$, and
\begin{equation}
    [y_i,x_j]=\Big(\frac{2i}{\omega\ka}\sin(s\omega\ka)e^{s\omega f}\Big)_{ij}.
\end{equation}
Consequently, we have for the averages and propagators \eqref{phi_prop}:
\begin{equation}\label{aver_ap}
\begin{split}
    \bar{x}(\tau)=&\Big[\frac{\sin(\tau\bar{\ka}/s)}{\sin\bar{\ka}}e^{(1-\tau/s)\bar{f}}+\frac{\sin((1-\tau/s)\bar{\ka})}{\sin\bar{\ka}}e^{-\tau\bar{f}/s}-1\Big]x_0=\\
    =&\Big[\frac{\sin b_+}{\sin\bar{\ka}}e^{ib_-\s}-\frac{\sin b_-}{\sin\bar{\ka}}e^{ib_+\s}-1\Big]x_0,\\
    \bar{p}(\tau)=&\Big[\bar{\ka}\frac{\cos(\tau\bar{\ka}/s)}{\sin\bar{\ka}}e^{(1-\tau/s)\bar{f}}-\bar{\ka}\frac{\cos((1-\tau/s)\bar{\ka})}{\sin\bar{\ka}}e^{-\tau\bar{ f}/s} -\bar{f}\Big]\frac{x_0}{2s}=\\
    =&\Big[\frac{\bar{\ka}\sin b_+}{i\sin\bar{\ka}}e^{ib_-\s}+\frac{\bar{\ka}\sin b_-}{i\sin\bar{\ka}}e^{ib_+\s} -\bar{f}\Big]\frac{x_0}{2s},\\
\end{split}
\end{equation}
\begin{equation}\label{props_ap}
\begin{split}
    D_{xx}(\s_1,\s_2)=&\frac{2s}{\bar{\ka}}\Big[\theta(\s_1-\s_2)\frac{\sin((1-\s_1)\bar{\ka}/2)\sin((1+\s_2)\bar{\ka}/2)}{\sin\bar{\ka}}+\\ &+\theta(\s_2-\s_1)\frac{\sin((1+\s_1)\bar{\ka}/2)\sin((1-\s_2)\bar{\ka}/2)}{\sin\bar{\ka}}\Big]e^{\bar{f}(\s_2-\s_1)/2},\\
    D_{xp}(\s_1,\s_2)=&\Big[\theta(\s_1-\s_2)\frac{\sin((1-\s_1)\bar{\ka}/2)\cos((1+\s_2)\bar{\ka}/2)}{\sin\bar{\ka}}-\\ &-\theta(\s_2-\s_1)\frac{\sin((1+\s_1)\bar{\ka}/2)\cos((1-\s_2)\bar{\ka}/2)}{\sin\bar{\ka}}\Big]e^{\bar{f}(\s_2-\s_1)/2},\\
    D_{pp}(\s_1,\s_2)=&-\frac{\bar{\ka}}{2s}\Big[\theta(\s_1-\s_2)\frac{\cos((1-\s_1)\bar{\ka}/2)\cos((1+\s_2)\bar{\ka}/2)}{\sin\bar{\ka}}+\\ &+\theta(\s_2-\s_1)\frac{\cos((1+\s_1)\bar{\ka}/2)\cos((1-\s_2)\bar{\ka}/2)}{\sin\bar{\ka}}\Big]e^{\bar{f}(\s_2-\s_1)/2},
\end{split}
\end{equation}
where we have taken into account that the action of the operator $\hat{x}$ on the left state and $\hat{y}$ on the right one vanishes for the matrix element $\lan \overline{out}|\overline{in}\ran$. The general expression (out of the diagonal) for the averages and the propagators is presented in \eqref{aver_props}. There was introduced the notation in \eqref{aver_ap} and \eqref{props_ap}:
\begin{equation}\label{bb_kb_fb}
  \s=2\tau/s-1,\qquad b_\pm=\frac{s\omega}{2}(if\pm\ka),\qquad \bar{\ka}=s\omega\ka,\qquad \bar{f}=s\omega f.
\end{equation}
The variable $\s$ ranges between $-1$ and $1$. In particular, it follows from \eqref{props_ap} that
\begin{equation}\label{prop_coinc_ap}
\begin{gathered}
  D_{xx}(\s,\s)=s\frac{\cos(\bar{\ka}\s)-\cos\bar{\ka}}{\bar{\ka}\sin\bar{\ka}},\qquad D_{pp}(\s,\s)=-\frac{\bar{\ka}}{4s}\frac{\cos(\bar{\ka}\s)+\cos\bar{\ka}}{\sin\bar{\ka}},\\
  D_{xp}(\s+0,\s)=\frac{\sin\bar{\ka}-\sin(\bar{\ka}\s)}{2\sin\bar{\ka}},\qquad D_{xp}(\s,\s+0)=-\frac{\sin\bar{\ka}+\sin(\bar{\ka}\s)}{2\sin\bar{\ka}},
\end{gathered}
\end{equation}
The representation of the averages \eqref{aver_ap} and the propagators \eqref{props_ap}, \eqref{prop_coinc_ap} in terms of $\s$ appears to be more useful for the calculation of integrals in vertices. The expressions for the propagators \eqref{props_ap} are resolved into factors depending only on $\s_1$ or $\s_2$, with the exception of the $\theta$ functions; this simplifies the integration.

The potential defining the perturbation of the free Hamiltonian $H_0$ at the order $\al^2$ takes on the form (see \eqref{Hamiltonian2}, \eqref{gi}, and \eqref{grav_pot})
\begin{equation}\label{V2}
    V_2=\frac13\bar{R}_{ikjl}p_ix_kx_lp_j+\frac{\omega}{3}\partial_jf_{ki}(p_ix_jx_k+x_kx_jp_i)+\frac12\bnabla_ih_i+\frac14 h_ih_i-\frac16\bar{R}.
\end{equation}
Whence, we get for the vertices of the type $V_{2p}$, $V_{1p}$, and $V_0$ at the order $\al^2$ (see the notation in \eqref{vertices_type}),
\begin{equation}\label{vertices_a2_ap}
\begin{split}
    \udg{0.7}{\diagViipw} = &-\frac{i}{3}\bar{R}_{i'k'j'l'}\int_0^s d\tau \big[\de_{i'i}(\tau+0-\tau_1)\de_{j'j}(\tau-0-\tau_4)+\de_{i'j}(\tau+0-\tau_4)\de_{j'i}(\tau-0-\tau_1)\big]\times\\ &\times\big[\de_{k'k}(\tau-\tau_2)\de_{l'l}(\tau-\tau_3)+\de_{k'l}(\tau-\tau_3)\de_{l'k}(\tau-\tau_2)\big],\\
    \udg{0.7}{\diagVipw} = &-\frac{i\omega}{3}\partial_{j'}f_{k'i'}\int_0^s d\tau \big[\de_{i'i}(\tau+0-\tau_1)\de_{j'j}(\tau-\tau_2)\de_{k'k}(\tau-\tau_3) +\de_{i'i}(\tau+0-\tau_1)\de_{j'k}(\tau-\tau_3)\de_{k'j}(\tau-\tau_2)\\
    &+\de_{i'i}(\tau-0-\tau_1)\de_{j'j}(\tau-\tau_2)\de_{k'k}(\tau-\tau_3) +\de_{i'i}(\tau-0-\tau_1)\de_{j'k}(\tau-\tau_3)\de_{k'j}(\tau-\tau_2)\big],\\
    \udg{0.7}{\diagc} =\; &is\big(\frac16\bar{R}-\frac12\bnabla_ih_i-\frac14 h_ih_i\big),
\end{split}
\end{equation}
respectively. The shifts of time arguments by the infinitesimal quantities are responsible for the ordering of the operators in \eqref{V2}. Similar expressions can be derived for the vertices of higher order in $\al$.

In order to simplify the subsequent formulas, it is convenient to use the basis specified by a tetrad
\begin{equation}
    e^i_a:=\{\ups_1^i,\bar{\ups}^i_1,\ups^i_2\},
\end{equation}
where its vectors are defined in \eqref{E_f}. It is useful to introduce the following tensors in this basis
\begin{equation}\label{t_tens_ap}
    t^{a_1\cdots a_{n+k}}_{x\cdots xp\cdots p}:=(-i)^k\bar{\ka}_{a_{n+1}}\cdots\bar{\ka}_{a_{n+k}}\sum_{\s}\s(1)\cdots\s(n)\frac{\sin\sum_{r=1}^{n+k}b_{\bs(r)}^{a_r}}{\sum_{r=1}^{n+k}b_{\bs(r)}^{a_r}} \prod_{r=1}^{n+k}\frac{\sin b_{\s(r)}^{a_r}}{\sin\bar{\ka}_{a_r}},
\end{equation}
where $\s(r)=\pm$, $r=\overline{1,n+k}$, the sum over $\s$ denotes the sum over $2^{n+k}$ combinations of $\s(r)$, the bar over $\s(r)$ implies the sign change of $\s(r)$, $n$ is the number of indices of $x$, and $k$ is the number of indices of $p$. The quantities $b^a_\pm$, $\bar{\ka}_a$, and $\bar{f}_a$ are the eigenvalues of the matrices \eqref{bb_kb_fb}. The tensor \eqref{t_tens_ap} is recovered in the initial basis as
\begin{equation}\label{basis_conver}
    t^{i_1j_1\cdots i_{n+k}j_{n+k}}=\sum_{a_1,\ldots,a_{n+k}}t^{a_1\cdots a_{n+k}}_{x\cdots xp\cdots p} e^{i_1}_{a_1}\bar{e}^{i_1}_{a_1}\cdots e^{i_{n+k}}_{a_{n+k}}\bar{e}^{i_{n+k}}_{a_{n+k}}.
\end{equation}
The tensor $t$ arises in calculations of the tree one-vertex diagrams. In fact, this tensor equals the half of the integral over $\s$ from $-1$ to $1$ of the product of a corresponding number of the ``first'' terms (containing $\exp(ib_\pm\s)$) in the expressions for $\bar{x}$ and $\bar{p}$ in \eqref{aver_ap}.

For the one-vertex diagrams involving tadpoles, the integral over $\s$ will contain the product of several sines and cosines coming from the propagators with the coinciding arguments \eqref{prop_coinc_ap}, in addition to the factors from $\bar{x}$ and $\bar{p}$. Therefore, it is also useful to introduce the tensors
\begin{multline}\label{l_tens_ap}
    \ell^{a_1\cdots a_{n+k}b_1\cdots b_{m+l}}_{x\cdots xp\cdots ps\cdots sc\cdots c}:=\frac{(-i)^{k+m}}{2^{m+l}}\frac{\bar{\ka}_{a_{n+1}}\cdots\bar{\ka}_{a_{n+k}}}{\sin\bar{\ka}_{b_1}\cdots\sin\bar{\ka}_{b_{m+l}}}\times\\ \times\sum_{\s,\rho}\s(1)\cdots\s(n)\rho(1)\cdots\rho(m)\frac{\sin\big(\sum_{r=1}^{n+k}b_{\bs(r)}^{a_r}+\sum_{q=1}^{m+l}\bar{\ka}^{b_q}_{\rho(q)}\big)}{\sum_{r=1}^{n+k}b_{\bs(r)}^{a_r}+\sum_{q=1}^{m+l}\bar{\ka}^{b_q}_{\rho(q)}} \prod_{r=1}^{n+k}\frac{\sin b_{\s(r)}^{a_r}}{\sin\bar{\ka}_{a_r}},
\end{multline}
where $\rho(q)=\pm$, $q=\overline{1,m+l}$, the sums are over all $2^{n+k+m+l}$ combinations of $(\s(r),\rho(q))$, $m$ is the number of indices $s$, and  $l$ is the number of indices $c$. Furthermore, $\bar{\ka}_\pm:=\pm\bar{\ka}$. In the initial basis, the tensor is constructed as given in \eqref{basis_conver}. The obvious properties hold for the $\ell$ tensor
\begin{equation}
    \ell^{a_1\cdots a_{n+k}}_{x\cdots xp\cdots p}=t^{a_1\cdots a_{n+k}}_{x\cdots xp\cdots p},\qquad \ell^{b_1\cdots b_{2m+1}c_1\cdots c_l}_{\ s\cdots s\ \ \ \ \ \ c\cdots c}=0.
\end{equation}
The tensors $t$ and $\ell$ are symmetric with respect to permutations of the indices in each of the group $x$, $p$, $s$, and $c$. Here are some specific values:
\begin{equation}
    t^a_x=\Big(\frac1{b_-^a}-\frac1{b_+^a}\Big)\frac{\sin b_-^a\sin b_+ ^a}{\sin\bar{\ka}_a},\qquad t^a_p=\Big(\frac{\bar{\ka}_a}{b_-^a}+\frac{\bar{\ka}_a}{b_+^a}\Big)\frac{\sin b_-^a\sin b_+ ^a}{2i\sin\bar{\ka}_a},\qquad \ell^a_c=\frac1{\bar{\ka}_a},\qquad \ell^a_s=0.
\end{equation}

With the help of these tensors it is not difficult to write all contributions of the perturbation theory of the order $\al^2$. Exploiting \eqref{aver_ap}, \eqref{prop_coinc_ap}, and \eqref{vertices_a2_ap}, we arrive at
\begin{flalign}\label{a_1_ap}
    \frac14 \udg{0.7}{\diagai}&=-\frac{s}{4}\frac{2i}{3}\bar{R}_{(acb)d}\big[t^{acdb}_{pxxp}-t^{acb}_{pxp}-t^{adb}_{pxp} +t^{ab}_{pp} -\bar{f}^{(a}(t^{cdb)}_{xxp}-t^{cb)}_{xp}+t^{b)}_p) +\bar{f}^a (t^{cd}_{xx}-t^c_x-t^d_x+1) \bar{f}^b\big]\frac{x_0^ax_0^cx_0^dx_0^b}{4s^2}&\nonumber\\
    &=-\frac{i\bar{R}_{(acb)d}}{24s}\big[t^{acdb}_{pxxp}-t^{acb}_{pxp}-t^{adb}_{pxp} +t^{ab}_{pp} -\bar{f}^{(a}(t^{cdb)}_{xxp}-t^{cb)}_{xp}+t^{b)}_p) +\bar{f}^a (t^{cd}_{xx}-t^c_x-t^d_x+1)
    \bar{f}^b\big]x_0^ax_0^cx_0^dx_0^b,&
\end{flalign}
\begin{flalign}
    \frac14 \udg{0.7}{\diagaii}&=-\frac{is^2}{4}\frac{2i}{3}\bar{R}_{(acb)d}\frac{\de_{cd}}{\bar{\ka}_c}\big[\ell^{abc}_{ppc}-\bar{f}^{(a}\ell^{b)c}_{px} +\bar{f}^a\bar{f}^b\ell^c_c-\ctg\bar{\ka}_c(t^{ab}_{pp}-\bar{f}^{(a}t^{b)}_p+\bar{f}^a\bar{f}^b ) \big]\frac{x_0^ax_0^b}{4s^2}&\nonumber\\
    &=\frac{1}{24}\bar{R}_{(acb)d}\frac{\de_{cd}}{\bar{\ka}_c}\big[\ell^{abc}_{ppc}-\bar{f}^{(a}\ell^{b)c}_{px} +\bar{f}^a\bar{f}^b\ell^c_c-\ctg\bar{\ka}_c(t^{ab}_{pp}-\bar{f}^{(a}t^{b)}_p+\bar{f}^a\bar{f}^b ) \big]x_0^ax_0^b,&
\end{flalign}
\begin{flalign}
    \frac14 \udg{0.7}{\diagaiii}&=\frac{is}{4}\frac{2i}{3}\bar{R}_{(acb)d}\de_{ab}\frac{\bar{\ka}_a}{4s}\big[\ell^{acd}_{cxx} -\ell^{ac}_{cx}-\ell^{ad}_{cx}+\ell^a_c +\ctg\bar{\ka}_a(t^{cd}_{xx}-t^c_x-t^d_x+1) \big]x_0^cx_0^d&\nonumber\\
    &=-\frac{1}{24}\bar{R}_{(acb)d}\de_{ab}\bar{\ka}_a\big[\ell^{acd}_{cxx} -\ell^{ac}_{cx}-\ell^{ad}_{cx}+\ell^a_c +\ctg\bar{\ka}_a(t^{cd}_{xx}-t^c_x-t^d_x+1) \big]x_0^cx_0^d,&
\end{flalign}
\begin{flalign}
    \udg{0.7}{\diagaiv}&=-\frac{is}{2}\frac{i}{3}\bar{R}_{acbd}\big[\de_{cb}(t^{ad}_{px}-\bar{f}^at^b_x-t^a_p+\bar{f}^a-\ell^{cad}_{spx}+\bar{f}^a\ell^{cd}_{sx}+\ell^{cd}_{sp})\frac{x_0^ax_0^d}{2s}-&\nonumber\\
    &-\de_{ad}(t^{cb}_{xp}-t^b_p-t^c_x\bar{f}^b+\bar{f}^b +\ell^{acb}_{sxp} -\ell^{ab}_{sp} -\ell^{ac}_{sx}\bar{f}^b)\frac{x_0^bx_0^c}{2s}\big]&\nonumber\\
    &=-\frac{1}{6}\bar{R}_{acbd}\de_{ad}(\ell^{acb}_{spx}-\bar{f}^c\ell^{ab}_{sx}-\ell^{ac}_{sp})x_0^bx_0^c,&
\end{flalign}
\begin{flalign}
    \frac14 \udg{0.7}{\diagav}&=-\frac{1}{4}\frac{2i}{3}\bar{R}_{(acb)d}\de_{ab}\de_{cd}\frac{\bar{\ka}_a}{4\bar{\ka}_c}(\ell^{ac}_{cc}-\ctg\bar{\ka}_c\ell^a_c+\ctg\bar{\ka}_a\ell^c_c-\ctg\bar{\ka}_a\ctg\bar{\ka}_c)&\nonumber\\
    &=-\frac{is}{24}\bar{R}_{(acb)d}\de_{ab}\de_{cd}\frac{\bar{\ka}_a}{\bar{\ka}_c}(\ell^{ac}_{cc}-\ctg\bar{\ka}_c\ell^a_c+\ctg\bar{\ka}_a\ell^c_c-\ctg\bar{\ka}_a\ctg\bar{\ka}_c),&
\end{flalign}
\begin{flalign}
    \udg{0.7}{\diagavi}&=-s\frac{i}{3}\bar{R}_{acbd}\frac{2}{4}\de_{ad}\de_{bc}(1-\ell^{ab}_{ss})
    =\frac{is}{6}\bar{R}_{acbd}\de_{ad}\de_{bc}(\ell^{ab}_{ss}-1),&
\end{flalign}
\begin{flalign}
    \frac12 \udg{0.7}{\diagbi}&=-\frac{s}{2}\frac{i\omega}{3}2\partial_{(b}f_{c)a}\Big[t^{abc}_{pxp}-t^{ac}_{px}-t^{ab}_{px}+t^a_p-\bar{f}^a(t^{bc}_{xx}-t^b_x-t^c_x+1) \Big]\frac{x_0^ax_0^bx_0^c}{2s}&\nonumber\\
    &=-\frac{i\omega}{6}\partial_{(b}f_{c)a}\Big[t^{abc}_{pxp}-t^{ac}_{px}-t^{ab}_{px}+t^a_p-\bar{f}^a(t^{bc}_{xx}-t^b_x-t^c_x+1) \Big]x_0^ax_0^bx_0^c,&
\end{flalign}
\begin{flalign}
    \frac12 \udg{0.7}{\diagbii}&=-\frac{is^2}2\frac{i\omega}{3}2\partial_{(b}f_{c)a}\frac{\de_{bc}}{\bar{\ka}_b}\Big[\ell^{ba}_{cp}-\bar{f}^a\ell^b_c-\ctg\bar{\ka}_bt^a_p+\ctg\bar{\ka}_b\bar{f}_a\Big]\frac{x_0^a}{2s}&\nonumber\\
    &=\frac{\omega s}{6}\partial_{(b}f_{c)a}\frac{\de_{bc}}{\bar{\ka}_b}\Big[\ell^{ba}_{cp}-\bar{f}^a\ell^b_c-t^a_p\ctg\bar{\ka}_b+\bar{f}_a\ctg\bar{\ka}_b\Big]x_0^a,
\end{flalign}
\begin{flalign}
    \udg{0.7}{\diagbiii}&=-is\frac{i\omega}{3}\partial_{b}f_{ca}(-\de_{ab}\ell^{ac}_{sx})x_0^c
    =-\frac{\omega s}{3}\partial_{b}f_{ca}\de_{ab}\ell^{ac}_{sx}x_0^c,&
\end{flalign}
\begin{flalign}\label{c_ap}
    \udg{0.7}{\diagc}&=is(\frac16\bar{R}-\frac12\bnabla_ih_i-\frac14 h_ih_i).&
\end{flalign}
The coefficients at the diagrams are calculated making use of the formulas \eqref{Feyn_rule_multi}. The summation is implied over all the tetrad indices. Note that four terms (see \eqref{vertices_a2_ap}) correspond to each of the diagram on the left apart from the last one. However, in virtue of the symmetry properties of the tensors $\bar{R}_{acbd}$ and $\partial_bf_{ca}$, some terms are similar or vanish. For instance, the contraction of the fields with the indices $a,c$ or $b,d$ for $V_{2p}$ and the indices $a,c$ for $V_{1p}$ vanishes in the loop diagrams. The total contribution to the logarithm of the diagonal of the evolution operator $G(\omega,s;\spx,\spx)$ of the order $\al^2$ is given by the sum of the expressions \eqref{a_1_ap}-\eqref{c_ap}.

\end{document}